# Structured Event Representation and Stock Return Predictability[*]

Gang Li[†]    Dandan Qiao[‡]    Mingxuan Zheng[§]

December 22, 2025

**Abstract**

We find that event features extracted by large language models (LLMs) are effective for text-based stock return prediction. Using a pre-trained LLM to extract event features from news articles, we propose a novel deep learning model based on structured event representation (SER) and attention mechanisms to predict stock returns in the cross-section. Our SER-based model provides superior performance compared with other existing text-driven models to forecast stock returns out of sample and offers highly interpretable feature structures to examine the mechanisms underlying the stock return predictability. We further provide various implications based on SER and highlight the crucial benefit of structured model inputs in stock return predictability.

---
[*]We are grateful to Allaudeen Hameed, Johan Sulaeman, as well as participants at NUS Workshop (2025), SCECR (2025), INFORMS International Conference (2025), IC2S2 (2025) for many helpful comments.
[†]The Chinese University of Hong Kong, CUHK Business School, `gang.li@cuhk.edu.hk`
[‡]National University of Singapore, School of Computing, `qiaodd@nus.edu.sg`
[§]National University of Singapore, School of Computing, `mingxuanzheng@u.nus.edu`

# 1 Introduction

Text data has become an increasingly important source of information in finance. Unlike structured data, text data contains rich and high-dimensional information that reflects both the current states of assets and expectations about future investment opportunities (Gentzkow, Kelly, and Taddy, 2019; Hassan et al., 2025). The typical method to extract textual information is to transform text data either into sentiment features representing emotional polarity (e.g., Tetlock, 2007; Tetlock, Saar-Tsechansky, and Macskassy, 2008; Tetlock, 2010; Loughran and McDonald, 2011; Jegadeesh and Wu, 2013; Hassan et al., 2019; Ke, Kelly, and Xiu, 2022), or into low-dimensional semantic vectors using pretrained language models (e.g., Chen, Kelly, and Xiu, 2022; Bybee, Kelly, and Su, 2023; Tan, Wu, and Zhang, 2023; Cong et al., 2025; Jha, Liu, and Manela, 2025). These derived representations are then linked to asset returns to capture latent state variables embedded in financial language that help explain cross-sectional and time-series variations in asset prices.

The existing approaches, however, are limited to interpretability. The process of mapping an article to a numerical vector is inherently opaque, and leaves it unclear how the model derives such representations or what the vectors signify. This opacity poses a particular challenge for text-driven return prediction, where it is essential to understand not only the predictive outcomes but also the underlying mechanisms linking textual content to expected returns. Without interpretability, it becomes difficult to evaluate the reliability of the extracted information and to justify subsequent decisions based on these model outputs.

To address this limitation, in this paper, we propose an alternative approach beyond traditional semantic representations: **structured event representation** (SER). Our key innovation lies in using large language models (LLMs) not merely as embedding tools but as controllable extractors that identify key events from unstructured text data. Specifically, we integrate task-specific prompt design, in-context exemplars, external knowledge graphs, and post-extraction validation to build an LLM-based event-identification framework. With this framework, we are able to transform an article into multiple event triplets (i.e., *subject-action-object*) and treat them as the basic features for model input. By examining these human-readable events, we can learn how they are mapped to stock returns and conduct attribution analyses for our model.

Taking advantage of the SER structure, we propose a novel modeling architecture that can effectively link SERs to stock returns. First, we introduce a stock-specific representation which encodes firm-level characteristics relevant to stock returns. More importantly, such a stock-level representation can interact with event representation in our model, so that we can account for heterogeneous event effects, reflecting the fact that identical events often generate asymmetric market reactions across firms. For example, Donald Trump's election victory may benefit traditional-energy companies while hurt clean-energy stocks. Second, because multiple events can occur at the same time with varying levels of importance, we allow the model to differentially weight events to capture their heterogeneous effects on stock returns. This is achieved through an attention mechanism that dynamically assigns weights to events (Vaswani et al., 2017), thereby enabling the model to place greater emphasis on events that are more informative for return estimation. Finally, we use the aggregated event representations to predict stock returns through a multi-layer perceptron (MLP), optimized using mean squared error (MSE) loss.

We empirically validate our model on a return prediction task for U.S. stocks using newspaper data with full texts. In particular, we first process all newspaper data into events (i.e., each SER with subject, action, and object), and then train the designed attention model with SERs as inputs to forecast stock returns. Out-of-sample results consistently demonstrate that structuring textual information into event representations yields markedly stronger predictive performance across both daily and weekly horizons. The model delivers an annualized return of 10.93% with a Sharpe ratio of 0.78 for daily predictions and 5.23% with a Sharpe ratio of 0.63 for weekly predictions, substantially outperforming both sentiment-based and embedding-based benchmarks. Moreover, the model generates a significantly positive Fama–French five-factor $\alpha$, indicating that its predictive gains are not subsumed by any standard risk premia. The portfolio sorting results highlight the economic value of transforming unstructured text into event-level signals that encode explicit economic semantics.

Cross-sectional Fama–MacBeth regressions further confirm the robustness of these out-of-sample findings, as our SER model forecasts exhibit significantly positive predictive coefficients across frequencies. More importantly, across multiple benchmark models, our approach consistently outperforms the existing methods based on sentiment features (e.g., sentiment scores) and semantic embeddings (e.g., BERT and FinBERT models), underscoring the predictive value of event-based representations in financial text analysis. Our method thus provides a new angle to improve model performance with respect to general asset return predictability in finance. While some recent studies have advanced financial prediction through increasingly powerful AI methods, our study highlights the importance of how such advanced methods need to be adapted and deployed in domain-specific settings. In particular, by constructing clean and structured event-based representations, our SER approach enables large language models to be used more effectively and efficiently in financial applications, mitigating the well-known "garbage in, garbage out" problem and improving asset return predictability.

In addition to the improvement of stock return prediction, our method provides a highly interpretable output. Unlike using bag of words or word embeddings, we address the black-box nature of traditional text-based models by using LLMs to extract human-readable event structures (i.e., the *subject-action-object* triplet) and construct a novel event-driven prediction framework. To explain what the model has learned, we further implement a gradient-based attribution method that quantifies each event's or event entity's (subject or object of the event) marginal contribution to predicted returns. This allows us to rank informational drivers and link return predictability to specific economic mechanisms (i.e., particular events in our case) in a fully transparent manner. Many of the extracted high-importance events and entities are directly related to core financial activities, such as earnings announcements, policy interventions, capital flows, and corporate actions, making it possible to clearly identify which types of information drive future returns. Furthermore, our approach reveals rich heterogeneity in the informativeness of financial events and entities heterogeneity that, to our knowledge, has not been documented in prior work. Existing studies typically treat textual features as homogeneous inputs (i.e., consistently either positively or negatively forecast stock returns), whereas we provide the first systematic evidence that even within the same event type or involving the same entity, the predictive effect can differ substantially depending on event context and firm specificity. We visualize and analyze these important events and entities from different aspects, and further compare differences in marginal contributions to return predictability among various horizons (e.g., daily and weekly). By organizing events and entities into various topics, we show that the drivers of daily and weekly return predictability are different. This highlights an important form of information-set heterogeneity: different categories of news event and roles inform stock returns over different horizons.

Furthermore, our model design allows us to extend our analysis to examine whether entities extracted by the LLM give rise to different degrees of entity-based stock comovement, and whether more important entities induce stronger comovement. In particular, we identify firm linkages through shared textual exposures, specifically when different firms appear in news events associated with the same entity. This approach builds on the literature on information-based return comovement (Cohen and Frazzini, 2008; Anton and Polk, 2014; Hameed et al., 2015; Chinco, Clark-Joseph, and Ye, 2019), which suggests that firms connected through common information channels tend to exhibit correlated price movements. We then group entities by their model-assigned importance scores and construct portfolios of firms with the highest historical exposure to each entity, and measure the comovement of realized returns around news events involving that entity. We find that while some baseline level of comovement exists, consistent with the connected-stocks framework (Anton and Polk, 2014), the effect becomes substantially stronger for entities with higher importance scores. Taken together, this comovement analysis not only extends the connected-stocks literature by showing that information linkages can be directly inferred from shared textual entity exposures, which are extracted from structured events decomposed from news articles, but also demonstrates that entities assigned higher importance by the model induce stronger realized comovement, providing evidence that the model captures economically meaningful information channels. Furthermore, this exercise serves as a strong example illustrating that our SER-based framework offers a promising event-study approach for linking model interpretability outputs to actual market behavior and for systematically exploring firm-level exposures to specific informational triggers.

Our research is closely related to the literature in text-driven asset pricing. There are primarily two approaches to incorporating text data into asset pricing models. The first approach focuses on analyzing the correlation between document word characteristics and expected returns, such as sentimental polarity (Tetlock, 2007; Tetlock, Saar-Tsechansky, and Macskassy, 2008; Tetlock, 2010; Loughran and McDonald, 2011; Jegadeesh and Wu, 2013; Jeon, McCurdy, and Zhao, 2022). Sentiment variables are documented to explain stock market responses to text data, although their performance is limited in recent periods. The word-level sentiment annotations also rely on human subjective judgment and therefore could be prone to bias. Furthermore, the sentiments expressed by different media outlets regarding the same event may vary significantly, introducing substantial noise and potential bias (Ashtiani and Raahemi, 2023; Goldman, Gupta, and Israelsen, 2024).

Another way to extract word characteristics in text data is topic modeling, which clusters texts into different topic categories through their word distributions.[1] For instance, Bybee et al. (2024) employ latent Dirichlet allocation to segment news articles into 180 topics and subsequently model the correlation between these topics and stock returns. Zhou, Fan, and Xue (2024) utilize high-dimensional matrices of word frequencies to represent each article and apply principal component analysis to decompose the matrix into latent factors and residuals, which are then used for factor modeling in asset pricing. Hirshleifer, Mai, and Pukthuanthong (2025) use a semi-supervised topic model on news articles to evaluate market valuation of disaster risk. While these methods provide a structured approach to capturing information from text, they are limited by the high dimensionality of the data and the difficulty of interpreting the latent factors. Moreover, words can carry entirely different meanings depending on context, an ambiguity that word-level analyses cannot easily resolve.

---

[1] Recent studies include: Manela and Moreira (2017), Cao et al. (2022), Sautner et al. (2023a), Sautner et al. (2023b), among others.

The second approach focuses on semantic features, which often use pretrained LLMs to convert individual words into dense numerical vectors (i.e., word embeddings) that offer stronger contextual understanding. These models can generate embeddings not only for words but also for entire sentences or documents, therefore providing context-aware and document-level representations that better preserve the overall meaning and narrative structure of text data. Such document-level embeddings have been increasingly adopted in finance research, particularly those models derived from the BERT family (Devlin et al., 2018) that leverages deep bidirectional contextual understanding to capture nuanced semantics in financial texts. This type of research often uses document-level embeddings generated by LLMs to fit expected stock returns through linear models or neural networks, such as Chen, Kelly, and Xiu (2022), Bybee, Kelly, and Su (2023), Tan, Wu, and Zhang (2023), Cong et al. (2025), Jha, Liu, and Manela (2025), and Siano (2025). In addition, there is a growing stream of research employing LLMs as generative forecasting models (Bybee, 2023; Lopez-Lira and Tang, 2023; Jha et al., 2025). Rather than first converting text into embeddings for a separate downstream predictor, these frameworks prompt the LLM to read financial news and generate return forecasts directly.

While these semantic feature-based approaches have achieved promising predictive performance, they often suffer from limited interpretability, because words or entire articles are transformed into dense vector representations and are thus inherently opaque. Such a transformation process abstracts away the original linguistic and economic meaning of the text, making it difficult to directly link model inputs to outcomes (e.g., stock returns). Even when some generative LLMs can be instructed to provide explanations, the resulting descriptions constitute post-hoc rationalizations rather than faithful accounts of the model's actual inference (Parcalabescu and Frank, 2024). Moreover, Bybee (2023) shows that LLM-generated expectations exhibit systematic belief distortions akin to well-known biases in human expectation formation, thereby limiting their reliability as direct forecasting tools.

Taken together, these limitations underscore the need for an approach that preserves the rich information contained in textual data while also ensuring transparency in how predictive signals are generated. Motivated by this gap, our study proposes a framework that incorporates both interpretable features and an interpretable predictive model. Our method extracts structured and event-level features with clear economic semantics (i.e., SER), integrates them into a novel model architecture and uses a gradient-based interpretability method to link predicted outcomes directly to the most influential events. In summary, we propose a new event-driven framework that combines LLM-extracted structured events with deep learning prediction and gradient-based interpretation. The study bridges the gap between high-dimensional textual data and interpretable financial forecasting, highlighting the value of event-centric analysis in asset pricing.

The remainder of the paper is organized as follows. Section 2 introduces the methodological framework and the detailed construction of our SER model. Section 3 describes the empirical design, data, key variables, and main empirical results of stock return predictability by the SER model. Section 4 investigates various economic interpretations and implications of the SER model's forecasting capacity. Section 5 concludes the paper.

# 2 Methodology

## 2.1 Event Extraction through Large Language Model

Our goal is to develop a modeling architecture that can effectively predict asset returns through structured event information from text data. The first step is to design a reliable method for extracting events from unstructured text and transforming them into structured representations suitable for prediction. In this section, we outline an LLM-based framework for event extraction in Figure 1. LLMs have been extensively applied to diverse text understanding and generation tasks (Bybee, 2023; Jha et al., 2024; Zhou, 2025). By inputting raw text, such as news, firm announcements, reports, or other information disclosures, into an LLM, we can use a simple prompt like "please extract the events in this article" to obtain relevant outputs. However, such naïve prompting is inadequate to produce consistent and structured representations of events for effective downstream prediction.

Figure 1 about here

There are several challenges in employing LLMs to extract targeted event features. First, it's essential to clearly specify the format of the event representation. This format should be concise, structured, and informative, while also being suitable for use as input to the subsequent stock prediction model. Second, the same event may be scattered in different places, and the associated entities are often expressed inconsistently across multiple articles (e.g., 'Donald Trump' and 'President Trump', 'stock growth' and 'stock profit'). If these semantically identical mentions are treated as different inputs, the same economic information becomes fragmented across surface forms, weakening the model's ability to learn meaningful patterns. Consolidating these variants into unified entity representations is therefore essential to preserve signal strength and ensure comparability across time and firms. Finally, despite detailed prompt specifications, there remains the risk that LLMs produce content unrelated to the prompt instruction, a phenomenon known as hallucination (Bai et al., 2024).

To address these challenges and design effective prompts, we draw on insights from research in prompt engineering, an emerging research area in computer science that explores strategies for improving LLM performance through prompt design. Following this spirit, we design a prompt template tailored to each challenge in order to obtain structured and effective event representations. To address the first challenge, defining a structured format for event representation, we require the LLM to express each extracted event as a triplet consisting of (*subject-action-object*). This triplet structure allows the model to retain the core elements of an event, while filtering out narrative details that introduce noise but rarely contribute predictive value. By compressing linguistically diverse descriptions into a unified and comparable representation, the triplet format enables economically similar events to be treated consistently across articles and over time, which is crucial for downstream prediction and analysis (Deng et al., 2019; Lee and Ram, 2024). Recent research demonstrates that LLMs, when guided by carefully crafted prompts, can be steered to generate outputs that adhere to this triplet-based representation rather than producing unstructured free text (Iga and Silaghi, 2024; Kommineni, König-Ries, and Samuel, 2024). Building on this insight, we design the following format requirements in our prompts to obtain structured representations of event features:

*"The output should be a JSON-formatted sequence of events. Specifically, each event should include the following keys with corresponding values: 'subject', 'action', 'object', and 'context'. The 'subject' key must contain the text of the extracted subject entity. The 'action' key must include the verb or action connecting the subject to the object. The 'object' key should represent the text of the entity which is the object of the action. The 'context' key must provide additional context information about the event in the original article."*

Specifically, the first three elements ('subject', 'action', and 'object') form a triplet representing an event, and 'context' provides the location information of the event in the original text, which can facilitate a post-hoc check of the quality of extracted events as well as interpretable analysis.

For the second challenge, we need to ensure that event entities referring to the same real-world object are marked consistently, even when expressed in different surface forms. To implement this requirement, we draw on traditional entity linking techniques (Baigang and Yi, 2023), which commonly rely on open-source knowledge graphs to assign canonical URLs as identifiers. Among these resources, DBpedia is one of the most widely used knowledge bases for linking textual mentions to standardized entity entries (Deng et al., 2019).[2] Accordingly, we require the LLM to return such DBpedia URL for every subject and object, thereby providing a standardized identifier across instances. This design helps mitigate inconsistency issues: even when different words are used, the same entity will be mapped to the same DBpedia link. A further advantage is that the same event reported across multiple articles can be represented in a uniform way, which facilitates the later merging of duplicates. At this stage, the final event representation format is established, as illustrated in Figure 1. Each event is represented through six attributes: *subject, subject link, action, object, object link,* and *context.*

Finally, to mitigate potential hallucinations, we include selected examples into the prompt, pairing raw textual materials as inputs with their corresponding structured event representations as outputs. This process, known as in-context learning, enables LLMs to grasp task patterns through analogy, which has been shown to effectively improve the quality of the output (Dong et al., 2022). Additionally, we implement a post-processing validation to ensure the reliability of the generated results. This validation is conducted through the LLM-based judge, which evaluates each output against predefined requirements, such as whether the output conforms to the required subject–action–object triplet structure, whether all necessary event fields are present, and whether the extracted content is semantically consistent with the input text. Outputs that fail this validation are either automatically corrected or re-generated by resubmitting the same input. This procedure accounts for the probabilistic nature of LLM outputs as well as occasional network or API-related errors. To prevent the inclusion of unreliable data, cases that fail validation after a fixed number of attempts are discarded.

---

[2]DBpedia is a large-scale, structured knowledge base extracted from Wikipedia; see https://www.dbpedia.org.

By addressing the challenges outlined above, we are able to develop a carefully tailored prompt that adapts to the financial domain, enabling the effective transformation of textual signals into structured event representations (as illustrated in Figure 1). We also include the full prompt template and a representative sample output in the Online Appendix OA.2. This event extraction pipeline serves as the foundation for our subsequent return prediction. For implementation, we employ the commercial GPT series APIs provided by OpenAI. These state-of-the-art LLMs are widely recognized for their advanced capabilities in text understanding and generation and have been successfully applied to a broad range of complex natural language processing tasks (e.g., Iga and Silaghi, 2024 and Kommineni, König-Ries, and Samuel, 2024).

## 2.2 Stock Return Prediction through LLM-extracted Events

Having outlined the procedure for extracting structured events from textual signals, we now turn to the central task of predicting stock returns using these representations. Our objective in this section is to construct a deep learning framework that links event-based features to expected returns. Specifically, we consider two related prediction tasks, daily return prediction and weekly return prediction. While both tasks share a unified formulation, model architecture, and training strategy, they differ slightly in the design of the feature aggregation module. These differences, together with the common framework, are described in detail below.

### 2.2.1 Task Formulation

Given the set of events $\{x_{event}\}_{N,i,t}$ associated with stock $i$ during time window $t$, our objective is to predict the expected return $\hat{y}_{i,t+1}$ in the subsequent time window $t+1$. In the daily return prediction task, $t$ corresponds to a single trading day, while in the weekly return prediction task, $t$ denotes one trading week. Accordingly, $t+1$ represents the next day or the next week in the two tasks, respectively. The set $\{event\}_{N,i,t}$ contains $N$ events linked to stock $i$ within time window $t$, with each event represented in the structured format illustrated in Figure 1.

In addition to event features, we introduce a stock-identifier representation as part of the model input. This input provides a compact numerical encoding of each stock (e.g., a ticker or PERMNO), allowing the model to process a large universe of stocks within a unified latent space.[3] It enables the model to learn latent firm-level characteristics, such as differences in business exposure, industry positioning, or other firm-level patterns that are not fully captured by observable firm features. More importantly, existing text-based return prediction studies typically rely on textual features alone and implicitly assume homogeneous responses across stocks. In practice, however, identical events often have markedly different implications for different firms. For example, the same event may benefit some firms while adversely affecting others, depending on their underlying exposures and connections to the entities involved. With the stock-level representation, our model can allow for such stock-specific representations to interact with event representations, thus accommodating heterogeneous event responses and enabling the same event signal to translate into different return effects across stocks.

Formally, we include a stock identifier, $stock\_ID_i$, as an auxiliary input to indicate the target stock, which is then mapped into a continuous embedding space (in Section 2.2.2).

---

[3]This stock-specific representation draws inspiration from the user representation learning in literature of recommender systems (Li and Zhao, 2020), where discrete entities (e.g., users or items) are mapped into low-dimensional latent spaces to capture unobserved heterogeneity. The use of a shared latent space also enables parameter efficiency in the modeling.

The stock embedding interacts with event representations in a personalized manner, enabling the model to capture both shared structure and idiosyncratic behavior across stocks (in Section 2.2.3). The identifier $stock\_ID_i$ corresponds to a unique attribute of stock $i$ in the dataset, such as a ticker symbol or CRSP permanent number (PERMNO).

Mathematically, our task can be expressed as:

$$\hat{y}_{i,t+1} = f(x) = f(\{event\}_{N,i,t},\ stock\_ID_i), \tag{1}$$

where $f(x)$ denotes the predictive mapping from inputs to expected return. The function $f(x)$ is implemented using a deep learning framework (as shown in Figure 2), which comprises three main components: feature embedding, feature aggregation, and return mapping. These components are introduced in the following subsections.

<div style="text-align:center;">Figure 2 about here</div>

### 2.2.2 Feature Embedding

To map the inputs into expected returns, the first step is to transform both textual and identifier features into vector representations. This step consists of two modules: the event encoder (denoted as ① in Figure 2) and the stock ID encoder (denoted as ② in Figure 2), which vectorize the two types of inputs, respectively.

For the vectorization of events $\{event\}_{N,i,t}$, we adopt the TransE method, which is a widely used model for entity-relation learning (Bordes et al., 2013; Deng et al., 2019). Specifically, it transforms entities and relations into a shared low-dimensional embedding space, where relationships are simulated through simple geometric operations such as vector addition. Unlike generic text embeddings, TransE preserves the directional and relational semantics of a triplet. Although it is originally developed for knowledge-graph completion, this relational modeling makes it naturally suited for encoding our SER triplets in a scalable and structure-preserving manner. Building on the work of Deng et al. (2019), we use

TransE to convert the events extracted by LLM into embedding representations. Prior to the embedding process, we standardize event representations in order to facilitate efficient and consistent modeling, including merging entities or actions with identical references and removing duplicates. An entity (subject or object) or action is then converted into a token, so that each event triplet (*subject-action-object*) is transformed into a concise triplet (*entity token, action token, entity token*). Every token is then represented by a trainable $M$-dimensional vector $\mathbf{e} \in \mathbb{R}^M$. Consequently, each event is transformed into a triplet of vectors: $\{event\}_{N,c,t} \to \{(e_{\text{subject}}, e_{\text{action}}, e_{\text{object}})\}_{N,c,t}$. These vectorized event representations are subsequently processed by the TransE model, which learns embeddings by optimizing spatial relationships such that the sum of the subject and action vectors is close to the object vector in the embedding space, preserving the semantic meaning of events.

$$\mathbf{z}_{\text{event}} = (e_{\text{subject}} + e_{\text{action}} - e_{\text{object}}). \tag{2}$$

Through this process, the structured textual inputs $\{event\}_{N,i,t}$, are encoded into event embeddings $\{\mathbf{z}_{\text{event}}\}_{N,i,t}$. These event embeddings provide a compact representation that captures the semantic meaning and relational structure of each event, serving as a key input for subsequent modules to learn how different events influence stock movements.

For the stock-level vectorization, we encode each stock identifier $stock\_ID_i$ as a trainable $M$-dimensional vector: $stock\_ID_i \to z_{\text{stock\_ID},i}$, where $z_{\text{stock\_ID},i} \in \mathbb{R}^M$. This encoding transforms discrete stock identifiers into meaningful vector representations, allowing the model to explicitly incorporate stock-specific information into the prediction process and also capture their heterogeneous interactions with events.

### 2.2.3 Feature Aggregation

After constructing event-level representations $\{\mathbf{z}_{\text{event}}\}_{N,i,t}$ and stock-specific representations $z_{\text{stock\_ID},i}$ through the embedding processes, the next step is to jointly model and utilize such encoded representations for stock return prediction. This requires an aggregation step that combines multiple event signals with stock-level information. Crucially, the aggregation design must account for the heterogeneous responses of different stocks to the same event. For this purpose, we implement this aggregation via the multi-head self-attention mechanism (Vaswani et al., 2017; Devlin et al., 2018; Cong et al., 2021). Attention mechanisms have garnered significant attention in the field of natural language processing due to their effectiveness across various tasks. Intuitively, an attention mechanism provides a data-driven way to assign different weights to multiple inputs based on their relevance, so that a compact and informative representation can be derived for such inputs. Specific to our context, the attention mechanism allows our model, in the process of further aggregating event signals and stock information, to weight the events differentially to account for their heterogeneous importance across stocks.

The mathematical details of the attention mechanism are explained below. We first concatenate the encoded event embeddings and the stock ID embedding in this way:

$$\mathbf{z}_{i,t}^0 = [\mathbf{z}_{[CLS]}, \{\mathbf{z}_{\text{event}}\}_{N,i,t}, \mathbf{z}_{\text{stock\_ID},i}] \in \mathbb{R}^{(N+2)\times M}. \tag{3}$$

Here, $\mathbf{z}_{[\text{CLS}]} \in \mathbb{R}^M$ is a learnable summary token that is prepended to the combined embeddings.[4] Using a dedicated summary token is a commonly adopted practice in attention-based models, as it provides a flexible and efficient way to aggregate information from all tokens into a single representation via self-attention, achieving information fusion and dimensionality reduction. In our setting, the [CLS] token serves as a summary node that gathers contextual signals from all event embeddings and the stock embedding after self-attention, analogous to how an investor synthesizes information from multiple firm-specific

---
[4]The name "CLS" originates from early attention-based language models (Devlin et al., 2018) as a special token that provides a fixed-length summary representation for downstream classification (CLS) tasks. The same idea, however, is widely applicable beyond classification and can be used for regression or other prediction settings. In our model, we retain the same name for convenience, although the token serves as a general summary embedding for a regression task rather than for classification.

events together with firm characteristics to form an overall assessment of the stock.

We then apply a multi-head attention mechanism to $\mathbf{z}_{i,t}^0$. The concept behind the multi-head mechanism is based on the notion that there may exist multiple aspects that need to be accounted for in linking the input (i.e., event representations and the stock representation) and the output (i.e., the final aggregated representation) (Vaswani et al., 2017). Multi-head attention hence deploys multiple heads to capture the diverse aspects of the input features. Each head independently computes attention weights and contextualized representations. Such representations derived from multiple heads are concatenated and linearly projected to produce the final output. Multi-head attention mechanism has been widely employed in prior studies: for example, BERT employs it to capture cross-word dependencies in natural language (Devlin et al., 2018), while Cong et al. (2021) apply it to learn cross-stock effects on expected returns. In our study, the attention design focuses on capturing cross-event relations, allowing the model to differentiate the relative importance of multiple events to a stock's expected return. Noted that the attention mechanism design for daily return prediction and weekly return prediction is slightly different.

For the daily return prediction task, we directly implement the multi-head self-attention mechanism. Formally, let the self-attention module consist of $L$ layers. At the $l$-th layer, the transformation is defined as:

$$\mathbf{z}_{i,t}^l = [\text{head}_1, \ \text{head}_2, \ \ldots, \ \text{head}_H] \mathbf{W}^O \in \mathbb{R}^{(N+2) \times M}, \tag{4}$$

where $H$ is the number of attention heads, and $\mathbf{W}^O \in \mathbb{R}^{Hd_v \times M}$ is the output projection matrix, with $d_v$ denoting the value dimension of each head. Each attention head is computed as:

$$\text{head}_h = \text{Attention}(\mathbf{Q}_h, \mathbf{K}_h, \mathbf{V}_h) = \text{softmax}\left(\frac{\mathbf{Q}_h \mathbf{K}_h^\top}{\sqrt{d_k}}\right) \mathbf{V}_h, \tag{5}$$

where

$$\mathbf{Q}_h = \mathbf{z}_{i,t}^{l-1}\mathbf{W}_h^Q, \quad \mathbf{K}_h = \mathbf{z}_{i,t}^{l-1}\mathbf{W}_h^K, \quad \mathbf{V}_h = \mathbf{z}_{i,t}^{l-1}\mathbf{W}_h^V.$$

Here, $\mathbf{W}_h^Q, \mathbf{W}_h^K \in \mathbb{R}^{M \times d_k}$ and $\mathbf{W}_h^V \in \mathbb{R}^{M \times d_v}$ are learnable projection matrices for the $h$-th head, where $d_k$ denotes the dimensionality used for the attention scaling factor. After $L$ layers, the aggregated daily representation is obtained from the [CLS] token:

$$\mathbf{Z}_{i,t}^{\text{Daily}} = \mathbf{z}_{i,t}^L[0] \in \mathbb{R}^M. \tag{6}$$

Here, $\mathbf{z}_{i,t}^L \in \mathbb{R}^{(N+2) \times M}$ denotes the output of the final layer ($L$-th layer) and $[0]$ indicates the first token, namely the summary token [CLS]. This vector serves as a global summary of all input features associated with stock $i$ on day $t$. Panel A in Figure 3 illustrates the process of this daily attention algorithm.

Figure 3 about here

For the weekly return prediction task, unlike the daily task, where all events are confined to a single trading day and temporal ordering is less relevant, the weekly task spans up to five consecutive trading days. Over this longer horizon, temporal dynamics, such as how earlier events interact with later ones within the same week, are more likely to influence stock movements. In this way, applying a single flat attention layer to all weekly events would obscure these within-day versus across-day structures and significantly increase computational complexity. To address these limitations, we adopt a hierarchical attention mechanism that first aggregates events within each day and then aggregates across days at the weekly level. This design makes the temporal organization of weekly news explicit, enabling the model to disentangle how intra-day and inter-day informational patterns jointly affect stock movements more effectively.

We show this weekly attention algorithm in Panel B in Figure 3. Formally, we first apply daily-level self-attention independently to each day in week $t$: $\mathbf{Z}_{i,t}^{\text{Daily}_1}$, $\mathbf{Z}_{i,t}^{\text{Daily}_2}$, ..., $\mathbf{Z}_{i,t}^{\text{Daily}_D}$ where $Daily_d$ means the $d$-th day of the week and $D = 5$. These daily vectors are concatenated with a learnable weekly [CLS] token:

$$\mathbf{z}_{i,t}^{\text{Weekly},0} = \left[\mathbf{z}_{[\text{Weekly-CLS}]},\ \mathbf{Z}_{i,t}^{\text{Daily}_1},\ \ldots,\ \mathbf{Z}_{i,t}^{\text{Daily}_D}\right] \in \mathbb{R}^{(D+1) \times M}. \tag{7}$$

Next, we apply a $P$-layer of self-attention to $\mathbf{z}_{i,t}^{\text{Weekly},0}$ at the weekly-level, and take the output corresponding to the [Weekly-CLS] token as the weekly representation:

$$\mathbf{Z}_{i,t}^{\text{Weekly}} = \mathbf{z}_{i,t}^{\text{Weekly},P}[0] \in \mathbb{R}^M, \tag{8}$$

where $\mathbf{z}_{i,t}^{\text{Weekly},P} \in \mathbb{R}^{(D+1) \times M}$ denotes the final output of the weekly attention module after $P$ layers of multi-head self-attention.

Through the mechanisms above, we obtain the aggregated event representation $\mathbf{Z}_{i,t} \in \mathbb{R}^M$, where

$$\mathbf{Z}_{i,t} = \begin{cases} \mathbf{Z}_{i,t}^{\text{Daily}}, & \text{for daily return prediction}, \\ \mathbf{Z}_{i,t}^{\text{Weekly}}, & \text{for weekly return prediction}. \end{cases} \tag{9}$$

### 2.2.4 Return Mapping

The final step is to map the aggregated representation $\mathbf{Z}_{i,t}$ into the predicted stock return. To capture potentially nonlinear relations between event features and returns, we employ a multi-layer perceptron (MLP) with $S$ layers (denoted as ④ in Figure 2):

$$\hat{y}_{i,t+1} = \text{MLP}(\mathbf{Z}_{i,t})^S \in \mathbb{R}. \tag{10}$$

For each layer $s = 1, \ldots, S$, the transformation is defined as:

$$\mathbf{h}^{(s)} = \begin{cases} \sigma\left(\mathbf{W}^{(s)}\mathbf{Z}_{i,t} + \mathbf{b}^{(s)}\right), & \text{if } s = 1, \\ \sigma\left(\mathbf{W}^{(s)}\mathbf{h}^{(s-1)} + \mathbf{b}^{(s)}\right), & \text{if } 1 < s < S, \\ \mathbf{W}^{(S)}\mathbf{h}^{(S-1)} + \mathbf{b}^{(S)}, & \text{if } s = S, \end{cases} \quad (11)$$

where $\sigma(\cdot)$ denotes a non-linear activation function (e.g., ReLU), and $\mathbf{W}^{(s)}$ and $\mathbf{b}^{(s)}$ are learnable parameters of the $s$-th layer. The output layer produces a scalar value $\hat{y}_{i,t+1}$, representing the predicted return for stock $i$ in period $t+1$.

### 2.2.5 Model Training

The preceding subsections describe the forward inference process of the prediction model, that is, how the input features are transformed into the expected return of the target stock. This process involves numerous learnable parameters, including token embeddings in the encoders, projection weights in the attention module, and weights and biases in the MLP. In this subsection, we outline how these parameters are optimized through model training.

We adopt an end-to-end training strategy in which all components of the model are jointly optimized via backpropagation. Model parameters are initialized from a normal distribution. The optimization objective is to minimize the mean squared error (MSE) between the predicted return and the realized return. Let each training sample $n$ correspond to a specific stock–date observation $(i, t)$, with target return $y_{i,t+1}$. For notational simplicity, we denote this target as $y_n$. The loss function can hence be formalized as:

$$\mathcal{L}_{\text{MSE}} = \frac{1}{N_{\text{train}}} \sum_{n=1}^{N_{\text{train}}} (\hat{y}_n - y_n)^2, \quad (12)$$

where $N_{\text{train}}$ denotes the number of training samples.

This concludes the description of our proposed structured event representation (SER)

model framework for event-driven return prediction. For ease of reference, we refer to our framework as the SER model in the remainder of the paper. Details regarding hyperparameter settings and optimization strategies are provided in the following empirical section.

## 3 Empirical Evidence

### 3.1 Data

We evaluate the proposed framework using a large-scale dataset constructed from real U.S. stock market data. To ensure sufficient coverage and robustness of the results, we consider 20 years from January 2003 to December 2022. The stock data is obtained from the Center for Research in Security Prices (CRSP) dataset. We download return data for all stocks listed on the CRSP Stock Security Dataset within the specified time window. Weekly returns are then computed from daily returns through each week by the following formulation.

$$R_{\text{week}} = \exp\left(\sum_{i=1}^{5} \ln(1 + r_i)\right) - 1, \tag{13}$$

where $r_i$ denotes the daily return on trading day $i$ within a given week, and the summation is taken over all valid trading days in that week.

We use news articles as the textual signals for constructing event representations and linking them to firm performance. The news data are collected primarily from ProQuest and complemented with Factiva. We focus on major U.S. newspapers, including the Wall Street Journal, New York Times, Washington Post, and Financial Times, covering the same 20-year period. To link stock data with news texts, we construct a list of company names from CRSP and apply a text fuzzy matching algorithm to search for occurrences of any company names in the articles (at least twice). Text fuzzy matching (e.g., Jaro-Winkler distance or Levenshtein distance) is used to define the similarity between specific strings and target strings. Each article is then assigned to the corresponding company based on the best match. To mitigate potential mismatches, we exclude firms with ambiguous names (e.g., those containing common words).

After linking stock and news data, each observation consists of (i) the list of news texts, (ii) publication date, (iii) associated stock identifier, and (iv) the realized return of the following time window. We employ the OpenAI GPT API (`gpt-3.5-turbo-0125`) to extract structured event representations from the news texts, following the methods we proposed in Section 2.1. To validate the quality of our data processing, we randomly select multiple subsamples of extracted events and manually compare with the context in their original articles to ensure that the LLM-extracted events accurately reflect the underlying news content. We then further preprocess the dataset by standardizing event representations, removing redundant or low-information entities, and eliminating repeated actions. The detailed processing procedures are documented in the Online Appendix OA.1. After these processing steps, the final dataset contains 301,409 firm-day observations, involving 4,355 unique stocks, 6,157,014 events, 1,650,954 entities, and 164,449 actions.

Table 1 about here

The summary statistics in Table 1 provide an overview of the dataset characteristics at both daily and weekly frequencies. The average market capitalization exceeds 9 billion USD, indicating that the sample primarily consists of large publicly listed firms. Besides return and capitalization measures, the table also reports multi-level summary statistics of the dataset, including news articles, extracted events, and the associated entities and actions. This diversity provides a comprehensive basis for modeling cross-sectional return predictability through structured event representations.

## 3.2 Model Setting

We adopt a rolling window strategy to construct our dataset splits. Specifically, within 20 years, the data is divided into 15 sequential subsets. Each subset spans six years, with the first five years serving as the training window and the sixth year as the testing window. For instance, the first subset uses 2003–2007 for training and 2008 for testing, the second uses 2004–2008 for training and 2009 for testing, and so on. The final subset uses 2017–2021 for training and 2022 for testing.[5] This procedure results in a 15-year out-of-sample evaluation period covering 2008–2022. For each subset, the model is trained on the corresponding training (in-sample, IS) window and evaluated on its testing (out-of-sample, OOS) window.

Consistent with standard deep learning practice, our SER-based framework (see Figure 2) organizes event inputs into a fixed-size structure ($N$ events per firm-day) to enable uniform and efficient batch processing. In practice, however, the number of news events associated with each stock varies across days. Based on the data distribution summarized in Table 1, we set the maximum number of daily events per stock to 30. When the number of events exceeds this threshold, excess events are truncated, whereas if fewer events are available, zero-value padding is applied to maintain a consistent input dimension. Given that the average number of events per stock is approximately 20 for the daily task and 14 for the weekly task, this threshold comfortably covers the majority of firm-period observations in our dataset. Such truncation and padding procedures are standard in deep learning models. Importantly, all embedding layers (entity and action embeddings) are re-initialized and trained from scratch within each training window, without relying on any pretrained parameters or data from post-training window. This design prevents information leakage across rolling windows.

For other hyperparameters of our SER model, such as the learning rate, batch size, and the number of layers in each module, we adopt a grouped hyperparameter-transfer procedure following standard practices in hyperparameter optimization (Hutter, Kotthoff, and Vanschoren, 2019). Specifically, the 15 rolling subsets are organized into three non-overlapping groups (subsets 1–5, 6–10, and 11–15). Within each group, we only conduct five-fold cross-validation on the train window of the first subset to select the optimal hyperparameters. These hyperparameters are then held fixed when formally estimating the five models in the group, each trained on its respective training window and evaluated on its corresponding out-of-sample (OOS) test window. This design avoids using the years immediately before the test window as a validation set. The motivation is that some entities, such as newly established organizations or recently appointed officials, often appear in the year before the OOS period. If this year was used for validation rather than training, embeddings of these new entities would not be adequately trained, leading to unreliable validation outcomes and weaker generalization in the OOS evaluation. The grouped strategy ensures the model has sufficient exposure to these kinds of entities during training before being evaluated out of sample. Finally, to reduce the risk of overfitting, we apply an L2 regularization penalty to the model parameters during training.

---

[5]Since the amount of data between 2018 and 2022 is considerably smaller than that of earlier periods, we adopt an extended rolling window approach for the final group. For example, the 12th subset still uses 2019 as the testing year, but its training window is expanded to include the preceding six years, and this pattern continues accordingly.

To better evaluate the performance of the proposed method, we construct a benchmark comprising several representative methods. These methods can be grouped into two categories:

**Sentiment Analysis-based Methods:** We include two widely used sentiment-based methods, *RavenPack* and *Loughran-MacDonald*. RavenPack is a commercial dataset that extracts sentiment signals from market-moving news and social media content. In our implementation, the RavenPack sentiment is applied with the CSS75 setting. Loughran and McDonald (2011), on the other hand, propose a dictionary-based approach that calculates sentiment using the formula sentiment $= \frac{p-n}{p+n}$, where $p$ and $n$ are the counts of positive and negative words in an article, respectively.

**Word Embedding-based Methods:** We employ pre-trained BERT-series models (Devlin et al., 2018), which have been extensively applied in prior studies (Chen, Kelly, and Xiu, 2022; Bybee, Kelly, and Su, 2023; Cong et al., 2025; Siano, 2025). These models adopt a Transformer-based architecture and are pre-trained on large-scale text corpora to learn general-purpose language representations, with the quality of the learned embeddings largely determined by the characteristics of the pre-training corpus. To assess the impact of domain-specific pre-training, we select two representative backbones: a general-purpose model ("bert-base-uncased", hereafter BERT) and a domain-specific model trained on financial text ("ProsusAI/finbert", proposed by Araci (2019), hereafter FinBERT). Both models output 768-dimensional embeddings, and we apply the same single-layer regression mapping to ensure comparability.

For fairness, all methods are evaluated on the same stock universe, within the same time window, and under the same optimization strategy. We perform portfolio sorting experiments and Fama-Macbeth regression analysis. Results are reported across the 15-year testing horizon (2008–2022) to provide consistent out-of-sample comparisons. The details of these evaluation experiments are described in the following sections.

## 3.3 Portfolio Sorting

For portfolio sorting, at the beginning of each holding period, we sort all eligible stocks with available news observations into quintiles based on the cross-sectional distribution of predicted returns from each method. Within each quintile, we compute equal-weighted returns and then construct a long–short portfolio by taking the difference between the top and bottom quintiles. This procedure evaluates the ability of different signals to separate winners from losers in the cross-section of returns.

To measure portfolio performance, we report three standard metrics: (i) annualized return (AR) along with its $t$-statistic, (ii) Sharpe ratio (SR), and (iii) annualized Fama–French five-factor alpha ($\alpha$) with its corresponding $t$-statistic. Table 2 presents the results for both daily and weekly prediction horizons.

Table 2 about here

Table 2 shows that our SER model achieves the strongest performance across both horizons. At the daily frequency, the long–short portfolio delivers an annualized return of 10.93%, a $t$-statistic of 3.31, and a Sharpe ratio of 0.78. Performance remains robust at the weekly horizon, with 5.23% annualized return, a $t$-statistic of 3.03, and a Sharpe ratio of 0.63. In addition, the positive and statistically significant five-factor $\alpha$ further indicates that the model's predictive gains are not explained by standard risk premia.

Among the baselines, the sentiment-based methods, Loughran–McDonald and RavenPack, perform the worst, generating negligible or negative returns. This suggests that coarse sentiment signals alone provide limited incremental information for predicting cross-sectional stock movements. For embedding-based models, BERT produces limited predictive gains (2.12% daily and 3.56% weekly), suggesting that general-purpose language representations capture little domain-specific financial context. FinBERT, trained on financial corpora, significantly improves these results (8.59% daily and 3.78% weekly), indicating that domain-adaptive pretraining enhances the model's sensitivity to financially relevant semantics. Nevertheless, both embedding-based approaches still underperform compared with our SER model. Taken together, these results demonstrate that our event-based approach generally outperforms baselines in terms of profitability, risk-adjusted returns, and statistical significance.

Table 3 about here

Table 3 examines the delayed implementation of our event-based strategy. Compared with immediate execution (+1T), once execution is delayed by one trading interval (+2T), performance drops sharply into negative territory for both daily and weekly annualized returns (−3.1% and −0.07%), reflecting a short-term reversal effect that temporarily offsets the predictive signal. Importantly, profitability does not vanish entirely. At both daily and weekly horizons, returns gradually recover with moderate delays (+3T to +5T), though they fade again at longer horizons. This suggests that event-based signals may be effective at capturing short-term market mispricing caused by slow information diffusion.

## 3.4 Fama-MacBeth Regression

The Fama and MacBeth (1973) regression provides a complementary evaluation of the predictive power of different models in the cross-section of stock returns. Specifically, for each prediction horizon (daily or weekly), we run the following cross-sectional regression at time $t$:

$$R_{i,t} = \alpha_t + \sum_{m=1}^{M} \beta_{m,t} \hat{y}_{i,t-1}^{(m)} + \sum_{k=1}^{K} \gamma_{k,t} X_{i,t-1}^{(k)} + \epsilon_{i,t}, \quad i = 1, \ldots, N_t, \tag{14}$$

where $R_{i,t}$ denotes the realized return of stock $i$ at time $t$, $\hat{y}_{i,t-1}^{(m)}$ represents the predicted return signal from model $m$, and $X_{i,t-1}^{(k)}$ denotes the control variables. The controls include the past 1-week, 2-week, and 1-month (for daily regressions) or the past 1-month, 3-month, and 6-month (for weekly regressions) average realized returns of each stock, aggregated from daily returns to capture short- to medium-horizon momentum and reversal effects. This regression is estimated separately for each time $t$, yielding a time series of slope coefficients $\{\beta_{m,t}\}$.

Following the standard Fama–MacBeth procedure, the average slope coefficient $\bar{\beta}_m$ across all time periods is taken as the estimate of the predictive effect of model $m$. The statistical significance is evaluated using $t$-statistics computed as

$$t\text{-stat} = \frac{\bar{\beta}_m}{s(\beta_{m,t})/\sqrt{T}}, \tag{15}$$

where $s(\beta_{m,t})$ is the time-series standard deviation of the estimated coefficients and $T$ is the number of time periods. To ensure robust inference, we compute heteroskedasticity- and autocorrelation-consistent standard errors (Newey and West, 1987) with 11 lags. All evaluations are performed on the out-of-sample dataset to avoid overfitting and to ensure that the reported $t$-values reflect genuine predictive power rather than in-sample fit. The resulting average coefficients and corresponding $t$-statistics for each prediction method are summarized in Table 4.

<div style="text-align:center">Table 4 about here</div>

Table 4 reports the out-of-sample Fama–MacBeth regression results for daily and weekly horizons, respectively. Consistent with the portfolio-sorting evidence, the *event-based* approaches exhibit the strongest explanatory power across both frequencies. In particular, across all settings, the SER model delivers the largest and most stable coefficients, with economically meaningful magnitudes and strong statistical significance. For the daily horizon, the SER coefficients range from 0.209 to 0.223, with corresponding $t$-statistics between 2.54 and 2.90. For the weekly horizon, the estimated coefficients are substantially larger, approximately 0.430 to 0.493, and remain statistically significant across all settings, with $t$-statistics reaching up to 2.73. This superiority remains stable when controlling for short- to medium-horizon momentum and reversal effects derived from aggregated past returns, indicating that the predictive value of event signals is not subsumed by conventional technical patterns.

In contrast, both *sentiment-based* predictors (Loughran–McDonald and RavenPack) and *embedding-based* models (BERT and FinBERT) produce coefficients that are positive but modest (or slightly negative in the case of RavenPack) and statistically insignificant. For example, the Loughran-McDonald sentiment measure yields coefficients below 0.13 with $t$-statistics below 1.0, while BERT and FinBERT show similarly muted effects. These results suggest that, after including our SER method, sentiment-based and embedding-based approaches offer only limited incremental predictive power once evaluated in a cross-sectional return framework with standard controls. Across horizons, the event-based signals remain consistently significant. These results reinforce that structured event representations pro-vide distinctive and valuable information for return predictability beyond what is reflected in sentiment tone or pre-trained embeddings.

# 4 Economic Interpretation of the Predictive Power by LLM

Beyond delivering promising predictive performance, our proposed method also demonstrates a notable degree of interpretability. The LLM-extracted event features are inherently meaningful, as they capture structured and human-understandable information in the form of (*subject-action-object*) triplets, while remaining compatible with deep learning architectures. Leveraging this property in our approach, we employ gradient-based interpretability techniques to quantify the contribution of each event and entity, together with its polarity, to predicted stock returns (Section 4.1). This gradient-based approach serves as the foundation for all subsequent analyses in this section.

Building on this foundation, we proceed along several complementary directions to demonstrate the effectiveness and interpretability of our approach. First, we analyze event-level importance (Section 4.2) by examining which exact events (defined by identical subjects, actions, and objects) consistently exert the strongest influence on predictions. This analysis also reveals that the same event can produce different directional impacts depending on its contextual appearance, illustrating a core strength of SER: its ability to preserve fine-grained economic semantics that enable event-level interpretability. Second, we aggregate these events into broader event topics (Section 4.3) to uncover higher-level patterns in influential news themes and the systematic differences in their predictive effects. Third, we investigate entity-level importance heterogeneity (Section 4.4), leveraging SER's structured subject/object fields to identify which firms, individuals, or economic entities systematically drive return predictability. Finally, we examine event-based stock comovement (Section 4.5), evaluating whether stocks exposed to highly influential entities exhibit stronger comovement patterns, thereby linking micro-level event signals to broader market dynamics.

Together, these analyses form a coherent interpretability framework that starts from local gradient-based attributions for exact events and scales upward to topics, entities, and cross-stock relationships. All empirical analyses in this section are conducted using the last sub-dataset, with a training window from 2013 to 2021 and the corresponding trained SER model, which together serve as the representative configuration for interpretation and reporting.

## 4.1 The Gradient-based Interpretable Methods

To quantify the contribution of each event and its underlying entities to the predicted stock returns, we adopt a gradient-based interpretability approach (Shrikumar, Greenside, and Kundaje, 2017) that attributes the model's output to its input representations. This method provides a local measure of importance by evaluating how sensitive the model's prediction is to small perturbations in each feature and weighting this sensitivity by the feature's own magnitude.

Formally, given a sample input $x$ with $I$ features and a model $M$ with prediction function $f(x)$, the importance score of the $i$-th feature $x_i$ is defined as:

$$\text{Importance}_i = x_i \times \frac{\partial f(x)}{\partial x_i}, \tag{16}$$

where $x_i$ denotes the embedding of feature $i$, $f(x)$ represents the model's prediction function, and $\frac{\partial f(x)}{\partial x_i}$ is the gradient of the prediction with respect to $x_i$. The resulting attribution score, $\text{Importance}_i$, captures both the magnitude (how strongly the feature is present) and sensitivity (how much the prediction changes with respect to that feature). Importantly, the sign of the importance score indicates the direction of the feature's influence: a positive value implies that an increase in this feature raises the predicted return (a return-enhancing effect), whereas a negative value suggests that the feature contributes to lowering the predicted return (a return-depressing effect). This polarity allows us to distinguish between event features or entities associated with favorable versus unfavorable market implications, which becomes crucial for subsequent analyses of entity- and event-level heterogeneity.

Specifically, we apply this gradient-based attribution at two distinct levels of representation:

- Event-level importance: when $x_i$ represents the *event embedding*, constructed from the overall (*subject-action-object*) triplet, the resulting importance score measures how much each event as a whole contributes to the predicted return.

- Entity-level importance: when $x_i$ corresponds to the embedding of the *subject* or *object* within the triplet, the attribution isolates how much each entity independently drives the prediction.

At the entity level, the importance scores derived from the embeddings of subjects and objects are first computed for each occurrence of an entity across the dataset. We then average these scores across all instances to derive the entity's overall contribution to the model's predictions. This score reflects how strongly a specific entity influences the model's output. To capture both direction and strength, we consider two variants: (1) the absolute importance, which measures the overall magnitude of influence regardless of sign, and (2) the polarity importance, which distinguishes between positive and negative effects on predicted returns. These entity-level importance measures form the basis for the subsequent analysis of entity-heterogeneity in Section 4.4.

At the event level, we follow an analogous procedure but leverage the additional structural granularity offered by SER. Because each event is expressed as an exact triplet (*subject-action-object*), we can compute gradient-based importance scores for fully identical events, preserving fine-grained economic semantics that conventional text representations cannot capture. For each exact event, we aggregate its local importance scores across all occurrences to obtain both absolute and polarity-based event importance. These event-level importance measures form the basis for the subsequent analysis of event-heterogeneity in Section 4.2. In addition, we further group events into event topic types using the topic modeling procedure described in Section 4.3. Within each topic cluster, we aggregate the importance scores of all constituent events to assess the contribution of broader news themes. This two-level analysis allows us to examine (i) which specific events exert the strongest predictive influence and (ii) how different categories of news events differ systematically in their return implications.

## 4.2 Event-Heterogeneity Effect

The first dimension of interpretability examines event-level importance, which quantifies the predictive contribution of individual exact events. Using the gradient-based attribution procedure described in Section 4.1, we compute an importance score for each event by aggregating local attributions across all occurrences of the same (*subject-action-object*) triplet. This yields a global measure that reflects both the magnitude and the polarity of each event's influence on predicted returns.

Building on these event-level scores, we can identify which exact events exert the strongest impact in the representative dataset. To ensure reliability, we restrict the analysis to high frequency events ($frequency > 20$) in the dataset.[6] Tables 5–6 report the top-ranked events for the daily and weekly horizons, including their aggregated importance, polarity composition, and frequency of appearance. These results illustrate the events to which the model assigns disproportionate predictive weight and highlight the interpretive advantages of structured event representation.

Table 5 about here

The ranked event list in Table 5 illustrates three key insights enabled by SER. First, SER's explicit (*subject-action-object*) format allows us to interpret events at both macro and micro levels in a unified manner. Macro-level signals, such as (Donald_Trump–sign–Executive_order), capture broad shifts in market sentiment or sector momentum, while micro-level, firm-specific disclosures, such as (IBM–report–Revenue) and (Macy's–report–Sales), highlight precise corporate actions that directly shape expectations about cash flows and performance. This multi-resolution interpretability is enabled entirely by SER's structured encoding, which preserves economic meaning inside each event triplet.

---

[6]We evaluate alternative minimum-frequency thresholds (5–50) and find that importance estimates stabilize around 20, so we adopt 20 as a conservative choice.

Second, the event-level gradients reveal substantial event-heterogeneity, where identical events exhibit both positive and negative contributions depending on context. For example, (Walmart–raise–Wage) typically carries a positive polarity, which is often interpreted as a sign of strong labor demand or operational confidence, yet also displays nontrivial negative contributions, consistent with investor concerns about rising labor costs. Similarly, (Share–fell_with–Percentage) appears with mixed polarity: although fundamentally negative, the positive share reflects contexts where price drops may trigger contrarian expectations, short-covering, or signals of market overreaction. This coexistence of positive and negative polarity for the same event underscores how SER allows the model to condition on surrounding text and market conditions rather than assigning uniform sentiment labels.

Third, the ranked events reveal meaningful behavioral patterns consistent with investor underreaction to bad news and overreaction to good news (Frank and Sanati, 2018). Events tied to positive performance signals, such as (Revenue–raise–Percentage), (Apple.Inc.–sell–IPhone), and (Netflix–add–Subscriber), exhibit overwhelmingly positive polarity and high importance scores, indicating that markets disproportionately emphasize favorable news. By contrast, events that convey negative developments, such as (Stock–fell_with–Percentage), (Apple.Inc.–decline–Comment), and (Macy's–report–Sales), show relatively weaker negative contributions despite their adverse implications. The muted negative weights on events, such as (European_Central_Bank–cut–Interest_rate) and (Federal_Reserve_System–raise–Interest _rate), further suggest that monetary policy shocks, while informative, may produce gradual rather than immediate price adjustments.

Table 6 about here

The ranked events in Table 6 indicate that many of the patterns observed in the daily model persist at the weekly horizon. Macro-financial developments, such as (United States Department of the Treasury–schedule–Finance) and (Federal Reserve System–raise–Interest rate), remain among the most influential predictors. These events reflect monetary, fiscal, and credit-related actions whose market effects unfold gradually, making their predictive power particularly pronounced when returns are measured over multi-day windows. At the same time, micro-level corporate information continues to play a meaningful role: events like (Apple.Inc.–fell with–Percentage), (Apple.Inc.–decline–Comment), and (Epic.Games - boycott-Apple.Inc.) demonstrate that firm-specific price moves, and platform-policy disputes remain important signals at the weekly frequency.

Within this broader mix, several weekly-specific patterns become particularly salient. One illustrative example is the joint appearance of (Investor–pour–United_States_dollar) and (Investor–pull–United_States_dollar) among the top-ranked events. These two events represent opposite forms of capital movement, net inflows versus withdrawals, and the model not only ranks them jointly as influential but also assigns sharply contrasting polarity compositions. (Investor–pour–United_States_dollar) is dominated by positive polarity, suggesting risk-on reallocations toward dollar-denominated stocks, whereas (Investor–pull–United_States _dollar) carries predominantly negative polarity, reflecting risk-off repositioning and capital flight. Their simultaneous prominence, combined with their opposite polarity profiles, indicates that multi-day return windows enable the model to distinguish between these two distinct investor behaviors, which is a less visible asymmetry at the daily horizon. A second horizon-specific pattern emerges in the polarity associated with (Federal_Reserve_System–raise –Interest_rate). At the daily horizon, this event exhibits mixed and relatively more positive contributions, suggesting short-run reversals, initial optimism about economic strength, or temporary overreactions around policy announcements. However, at the weekly horizon, the polarity becomes predominantly negative. This shift aligns with well-established behavioral and informational dynamics: the adverse implications of rate hikes, higher borrowing costs, tighter financial conditions, and slower growth expectations, typically materialize with a delay, and investors may take several days to fully update their beliefs.

Together, the daily and weekly results demonstrate that SER provides a clear and interpretable view of how informative signals operate across different horizons and news categories. By capturing news' roles and their exact behaviors in the events, SER allows us to uncover both stable economic drivers and horizon-specific patterns. The next section moves beyond individual exact events and groups them into broader event topics, enabling us to examine more general patterns in how different categories of news events contribute to return predictability.

## 4.3 Event-Topic-Heterogeneity Effect

Following the individual event analysis, we next examine how different types of news events exert distinct impacts on stock returns. Building on the gradient based attribution framework introduced in Section 4.1, we extend the same interpretability method to the event level, where each event's contribution to the model's prediction is quantified through its gradient based importance score and then aggregated by topic type to evaluate the influence of different event categories. Unlike conventional article level topic modeling (e.g., Bybee et al., 2024), which clusters entire news articles based on overall word distributions, our approach conducts **topic modeling at the event level**, namely grouping individual events extracted from news articles into latent thematic categories and providing a more fine grained level of analysis of textual information. This event-centered analysis enables us to identify which categories of events, such as financial disclosures, executive changes, or policy actions, carry the greatest predictive weight and how their informational effects differ across time horizons, thereby linking textual event structures to interpretable market dynamics.

As a first step of the analysis, we categorize each event into a latent topic type, enabling us to analyze importance patterns across broader thematic groupings of news events. Each news event extracted by the LLM contains a context sentence that encapsulates the original news fragment. We apply latent Dirichlet allocation (LDA) to cluster these event contexts into interpretable topics.

LDA is a probabilistic generative model that represents each document (i.e., an event's context sentence in our case) as a mixture of latent topics, where each topic is characterized by a distribution over words. The model intuitively captures the notion that different documents may touch upon multiple themes, and that certain terms tend to co-occur within the same thematic context. This makes LDA particularly suitable for modeling news texts, which often blend various angles (e.g., policy, finance, corporate behavior). A critical hyperparameter of LDA is the number of topics. To select this parameter systematically, we follow the approach of Bybee et al. (2024) and choose the number of topics that maximizes the Bayes factor within a predefined range. The Bayes factor evaluates how well a model with a given number of topics explains the data relative to simpler models, balancing model fit against overfitting. Based on this criterion, we select 250 topics.

After training the LDA model, we manually assign human-readable labels to each topic based on its most probable terms. While we take inspiration from the label conventions used in Bybee et al. (2024), our data exhibits a different topical structure due to differences in corpus composition and market context. Therefore, we refine and adapt labels accordingly, in some cases creating entirely new labels to capture distinctive themes in our dataset. Full details of the topic selection process and resulting topic distributions are provided in the Online Appendix OA.4.

Based on the obtained topic labels for each event, we then evaluate which event topics exert the greatest influence on the model's predictions. To this end, we build on the gradient based attribution method introduced in Section 4.1. Specifically, we aggregate the local importance scores of all events belonging to each LDA-derived topic, thereby quantifying the relative contribution of each event type to predicted returns. Formally, the analysis proceeds in three steps:

- First, for each input sample, we compute the gradient-based importance scores for individual events, capturing how much each event contributes to the model's predicted return.

- Second, since every event has been assigned to a topic label by LDA based on its context sentence, we aggregate the importance scores across all samples by topic type and compute the average score of events within each topic.

- Finally, the aggregated topic-level importance scores are used to identify and rank the most influential event topics in both daily and weekly prediction settings, allowing for a direct comparison of how temporal horizons affect the model's informational focus (e.g., financial disclosures, executive moves, regulatory actions, and consumer sentiment).

Table 7 about here

Figure 6 about here

Based on the aforementioned approach, we obtain the global importance score for each event topic in both the daily and weekly prediction models. Table 7 reports the top 20 event topics ranked by their total importance scores for the two models, respectively. Each topic entry includes its absolute importance, the proportions of positive and negative polarity contributions, and representative high-impact events associated with that topic, providing a comprehensive view of how different categories of news shape the model's predictions. Detailed examples of individual high-importance events within each topic are provided in the Online Appendix OA.3 for reference. Figure 6 further compares these results side by side, highlighting both the shared and distinctive patterns of topic importance across the two temporal horizons. Several clear patterns emerge from this comparative analysis.

For the daily return prediction model, the most influential event topics are primarily centered on short-term financial shocks and sentiment-driven market triggers. As shown in the left panel of Table 7, topics like "Bankruptcy," "Control Stakes," "Collaboration," and "Key Role" dominate the ranking, reflecting the model's sensitivity to immediate and tangible firm-level behaviors related to financial distress, ownership restructuring, and executive actions. Other highly ranked topics, such as "Natural Gas Industry," "Pharmaceutical Industry," and "Commercial Airlines", indicate that the model also responds to industry-specific developments, particularly those linked to supply disruptions or regulatory uncertainty. Importantly, these leading topics exhibit clear polarity heterogeneity: some are dominated by return-depressing contributions (e.g., "Bankruptcy"), while others tilt more toward return-enhancing signals (e.g., "Collaboration" and "Casinos"), with several topics showing a more balanced mix (e.g., "Movie industry" and "Global Exchange Strategies"). Overall, the daily model highlights a mix of corporate actions, key role movements, and sectoral signals, but with a stronger emphasis on emotionally charged or sentiment-sensitive events, especially bankruptcy-related news that triggers abrupt market reactions and same-day price adjustments.

In contrast, the weekly return prediction model places greater emphasis on structural, strategic, and relational topics that unfold over longer horizons. As indicated in the right panel of Table 7, leading topics, such as "Pharmaceutical Industry," "Control Stakes," "Shareholder Payouts," "Corporate Bond Risk," and "Earning Forecast", reflect the model's focus on fundamental corporate and macroeconomic drivers. These topics reflect enduring market narratives involving capital structure management, investor compensation, and the overall financial health of sectors, which affect returns through gradual information diffusion.

The prominence of "Corporate Governance," "Takeovers," and "Retail" further underscores the weekly model's sensitivity to inter-firm dynamics, strategic decision-making, and consumer demand trends that evolve more gradually. Moreover, the leading weekly topics also exhibit clear polarity heterogeneity. Some predominantly contribute in a return-depressing direction (e.g., "Corporate Bond Risk" and "Corporate Governance"), whereas others skew toward return-enhancing contributions (e.g., "Earning Performance" and "Auto Manufacturing"). In sum, while both models emphasize corporate actions, executive behavior, investment activities, and industry-specific signals, the daily model prioritizes sentiment- and event-driven shocks, whereas the weekly model highlights persistent, sectoral, and structurally embedded informational patterns.

Figure 6 provides a side-by-side comparison of the top 20 event topics across the two horizons. Both models exhibit several recurring themes, including "Control Stakes," "Pharmaceutical Industry," and "Key Role," suggesting that corporate control, health-related developments, and executive changes play a persistent role in shaping return predictability across different timescales. According to Table 7, these recurring topics differ in both relative ranking and directional contributions. For instance, "Control Stakes" and "Pharmaceutical Industry" remain consistently near the top of both lists and retain a mildly return-enhancing tilt across horizons, whereas "Key Role" ranks high in the daily model but drops substantially in the weekly model, despite exhibiting broadly similar near-balanced polarity. Moreover, some overlapping topics show horizon-dependent sign shifts, for example, "Global Exchange Strategies" becomes less emphasized at the weekly horizon and tilts more return-depressing relative to its daily profile, highlighting that the same thematic news category can operate through different channels across horizons.

Overall, both models highlight economically meaningful categories of information, including corporate actions, executive movements, investment behaviors, and industry-specific developments, demonstrating the LLM's capacity to capture a wide range of market-relevant signals. However, their focal tendencies and distributional patterns differ with temporal granularity. The daily model shows a sharper and more focused attention to sentiment-driven and event-triggered shocks, particularly news related to bankruptcy or corporate restructuring, which tend to elicit immediate market reactions. In contrast, the weekly model distributes importance more evenly across topics, highlighting sectoral and structural themes that evolve gradually through sustained information diffusion.

## 4.4 Entity-Heterogeneity Effect

Building on the event-level analysis, the next dimension of interpretability investigates the entity-heterogeneity effect, which reveals how different entities embedded in event structures, namely subjects, and objects, contribute unequally to stock return predictions. This analysis provides economic insight into which firms, individuals, locations, or financial terms carry the strongest informational weight in driving market reactions, thereby linking the model's internal representations to interpretable economic drivers of stock prices. Using the LLM extracted (*subject-action-object*) triplets and the gradient based attribution method introduced in Section 4.1, we obtain an importance score for each entity in the representative dataset. These scores quantify how strongly and in which direction a certain entity influences the model's output through its embedding representation.

To identify the entities that most strongly drive the model's return forecasts, we visualize their global importance across the two prediction horizons, daily and weekly returns. For each task, we aggregate the importance scores of all entity occurrences and generate word clouds to highlight those with the highest average importance. To ensure reliability, we restrict the analysis to entities that appear more than 700 times in the dataset, thereby avoiding sparsity-induced noise in importance estimation.[7] Figures 4 and 5 present the resulting visualizations, each following a two-panel layout. The upper panels are based on the absolute importance scores, where the size of each word reflects the overall magnitude of an entity's contribution to the model's predictions, and colors denote semantic categories (e.g., organizations, persons, locations, and economic terms). The lower panels correspond to the polarity importance scores, which further distinguish entities by their directional effects: entities with positive importance (on the left side of the figure) are associated with return-enhancing signals, whereas those with negative importance (on the right side of the figure) indicate return-depressing effects. In addition to the visual analysis, Table 8 summarizes the top 20 entities ranked by their total importance scores for both the daily and weekly models, along with their polarity proportions. This table provides a compact quantitative reference, facilitating direct comparison of entity-level importance across temporal horizons and corroborating the visual patterns observed in the word cloud representations. Detailed examples of high-impact events associated with each entity are provided in the Online Appendix OA.3 for reference.

---

[7]We evaluate alternative minimum-frequency thresholds (100–1000) and find that importance estimates stabilize around 700 for both prediction horizons, so we adopt 700 as a conservative choice.

Figure 4 about here

Table 8 about here

For the daily return prediction task, as shown in the top panel of Figure 4, the most prominent entities include both standard financial concepts and broader contextual signals. A substantial share of high-importance entities are financial and accounting terms (highlighted in orange), such as "Bankruptcy," "Bond," "Financial_statement," "Revenue," and "Price." Their consistently high importance values indicate that the model places considerable weight on information related to firms' financial conditions, credit risk, and valuation fundamentals. In addition, entities linked to social, regulatory, or public-attention dimensions, such as "Vaccine," "Challenge," and "Fine,", also receive meaningful importance scores. This pattern suggests that the model systematically incorporates signals associated with regulatory scrutiny, operational disruptions, and shifts in public sentiment, all of which appear in news with sufficient regularity to influence the predictor's decisions. Prominent geopolitical and regional entities, such as "California," "Washington, D.C.," "United Kingdom," and "New York City", further indicate that location-specific events carry predictive content. Such entities often serve as markers for region-related policy environments, legal disputes, or local economic conditions that affect multiple firms. Their high importance therefore reflects the model's ability to utilize geographic cues that correlate with cross-firm exposures and sectoral sensitivities.

The bottom panel further decomposes entity importance by polarity. The left figure displays entities that, on average, contribute positively to predicted returns, while the right figure highlights those exerting negative influences. Positively associated entities include "Financial_statement," "United Kingdom," "California," and "Cost." Such entities often appear in favorable contexts, such as strong earnings reports or policy stability. Conversely, negatively associated entities include "Bankruptcy," "Fine," "Tariff," "Sales," and "Vaccine." These entities typically correspond to events with adverse economic implications. This visualization shows that the daily return prediction model identifies a wide range of interpretable and economically meaningful signals, spanning financial fundamentals, affective framing, and geographical context.

Figure 5 about here

For the weekly return prediction task, as shown in Figure 5, the results follow the same two-panel visualization template introduced earlier. Unlike the daily return model, which emphasized financial disclosures and emotionally charged terms, the weekly model highlights industry-specific concepts and major corporations (top panel in Figure 5). This distinction likely reflects the temporal granularity of information diffusion: while daily returns react to sharp news shocks, weekly returns respond to slower-moving strategic developments and inter-firm dynamics. For instance, "Consumer" emerges as one of the most influential entities. It appears in news related to consumer sentiment and policy reactions, especially in sectors where customer behavior is central to performance, such as fast food (e.g., McDonald's reassurance about customer retention) and electronics (e.g., Amazon or Apple introduc-ing new consumer features). The importance of this entity reflects the model's sensitivity to consumer feedback loops, which play out gradually and are crucial for weekly return adjustments. Similarly, the frequent appearance of major companies like "Tesla, Inc.," "Microsoft," "ExxonMobil," and "Royal_Dutch_Shell" further indicates that the weekly model relies on signals tied to inter-firm linkages and corporate decision-making. These entities hold structurally important positions in their industries and frequently act as bellwethers, with information about them influencing assessments of strategic direction, competitive conditions, and risk outlooks across the sector and occasionally the broader market. Additional high-importance entities, such as "Retail," "Mortgage_loan," "Advertising," and "Letter", indicate that long-horizon investor attention tends to coalesce around broad sectoral signals, consumer lending conditions, marketing strategies, and even political or executive communications, which are topics whose market impact develops gradually over time rather than immediately. Overall, the visualization suggests that the weekly model captures structural and relational information associated with sustained shifts in corporate strategies, industry dynamics, and macroeconomic sentiment.

The bottom panel of Figure 5 further decomposes entity importance by polarity. The left figure shows entities that, on average, contribute positively to predicting weekly returns, while the right figure highlights those exerting negative effects. In the positive polarity group (left panel), entities like "Microsoft," "Verizon_Communications," "Deal," "Business," and "Fine" dominate. These entities typically occur in favorable business contexts, such as expansionary agreements or competitive gains. On the negative side (right panel), top entities include "Tesla, Inc.," "Royal_Dutch_Shell," "Election," "Advertising," "Mortgage_loan," and "Letter." These often appear in complex or uncertain news contexts. This polarity-based interpretation confirms that the weekly model excels at capturing the cumulative and context-dependent effects of strategic, relational, and sector-wide news, enabling interpretable forecasts over longer temporal horizons.

Overall, the identified important entities in the daily and weekly prediction model representatives reveal distinct information preferences shaped by temporal granularity. The daily model relies predominantly on short-lived, content-specific triggers that drive immediate market reactions, such as financial disclosures, sentiment-related keywords, and economic terms. In contrast, the weekly model places greater emphasis on corporate entities and structural signals, reflecting slower information diffusion through organizational strategies, industry linkages, and policy developments. This distinction mirrors the pattern observed in the event-topic-heterogeneity analysis in Section 4.3, highlighting that while daily forecasts are driven by abrupt informational shocks, weekly forecasts capture cumulative, relational, and context-dependent patterns that evolve over time.

More broadly, this analysis demonstrates that our method delivers a level of interpretability that is difficult to achieve with existing sentiment- or embedding-based approaches. By leveraging a gradient-based attribution framework, our method makes entity-level informational heterogeneity explicit, revealing not only which entities matter on average, but also how the same entity can exert different directional effects across distinct event contexts. More importantly, the event-based representation enables direct traceability from model predictions back to specific events and their surrounding textual contexts, allowing predicted returns to be linked to concrete news realizations. This traceability clarifies why, and through which channels, news events and the entities embedded within them predict returns, positioning our method as a unified framework for prediction, interpretation, and mechanism-based analysis.

## 4.5 Entity-based Stock Comovement

Building on the entity-level interpretability analysis in Section 4.4, we further examine whether the most influential entities identified by the LLM act as common information hubs that trigger collective market reactions. The idea echoes the connected stocks per-spective of Anton and Polk (2014), who demonstrate that firms jointly held by the same investors tend to move together because shared ownership propagates correlated trading and attention. In a similar spirit, we explore an information-side counterpart of this mechanism: when multiple firms are recurrently linked to the same entities in news, they may become "information-connected" responding in tandem to new developments about those entities. Since entity-related information shocks dissipate quickly, in this analysis we restrict attention to only daily comovement patterns, aligning with the short-horizon nature of event-driven return adjustments.

Specifically, we focus on entity-induced comovement, where firms exposed to news involving the same key entity exhibit synchronous return patterns. Such comovement reflects the propagation of entity-specific information through economic linkages, supply-chain connections, or correlated investor sentiment, thereby translating the model's textual understanding into tangible cross-sectional dependencies in stock prices. In this sense, entity-driven comovement represents the information-side counterpart of ownership-driven connectedness, capturing how common attention and belief shocks transmit across firms that share salient informational anchors.

To assess this mechanism, we select the top-ranked entities from the daily prediction model as focal entities and identify the TOP-K stocks with the highest historical exposure to each focal entity, measured by their co-occurrence frequency in past event news. These stocks represent firms most consistently associated with a given entity in the news corpus, making them the most likely to react collectively when that entity reappears in subsequent events. We then evaluate whether these high-exposure portfolios exhibit stronger return comovement during event windows related to the same entity. The following subsection details the methodological procedure and empirical findings underlying this analysis.

### 4.5.1 Constructing Portfolios for Focal Entities

We begin by identifying the most influential entities in the daily prediction model (entities appear in Figure 4). For each focal entity, we then identify firms that historically display the strongest exposure, measured by the frequency of co-occurrence with that entity in event triplets. Exposure intensity is computed using a TF-IDF–style weighting scheme that adjusts for both within-firm frequency (term frequency) and across-firm rarity (inverse document frequency). Specifically, for firm $i$ and entity $e$, exposure is defined as

$$\text{Exposure}_{i,e} = \text{TF}_{i,e} \times \log\left(\frac{1 + N_{\text{docs}}}{1 + DF_e}\right), \tag{17}$$

where $\text{TF}_{i,e}$ is the number of times entity $e$ appears in firm $i$'s event history, $N_{\text{docs}}$ is the total number of firm-day observations, and $DF_e$ is the number of firms mentioning $e$ at least once. This weighting penalizes generic entities while emphasizing those that provide firm-specific information. Firms are then ranked by exposure, and the top $K = 15$ firms per entity are selected to form the TOP-$K$ exposure portfolio.

Next, to identify periods of collective response, we detect trading days on which a substantial fraction of firms within a TOP-K portfolio are jointly exposed to the same focal entity (or a semantically related entity). Specifically, each entity's shock day is defined as the trading day immediately following a news day on which at least 3 of the portfolio firms appear in events mentioning that entity or its closely related variants. Related entities are identified through co-occurrence networks and refined according to semantic or action-based relationships, including regulatory, partnership, and investment linkages. This approach isolates episodes of intensified collective attention to a given entity, when information shocks are most likely to propagate across firms. To ensure robustness, entities with fewer than five identified shock days are excluded from subsequent analysis. For instance, terms like "Bankruptcy," which almost exclusively corresponds to idiosyncratic firm-specific distress events, are filtered out at this stage to avoid conflating systematic comovement with isolated firm shocks.

### 4.5.2 Analysis of Comovements

Following the portfolio construction procedure, we obtain 84 focal entities, each associated with a portfolio of 15 stocks and a set of dates on which at least 3 of the portfolio firms are jointly exposed to the focal entity. We then analyze whether these portfolios exhibit stronger comovement in realized returns around the corresponding entity shock dates.

To quantify the degree of return comovement, we follow the approach of Huang, Huang, and Lin (2019). For each focal entity $e$ and its corresponding portfolio, we compute the average pairwise Pearson correlation of daily returns among the TOP-K constituent stocks within an event window of $[t-\tau, t+\tau]$, where $t$ denotes an entity shock day and $\tau$ represents the window length (typically 1–6 trading days). The same measure is computed for an equally sized non-event window as a baseline. To ensure comparability and avoid contamination from nearby shocks, we restrict the baseline period to 10 trading days before and after each event window, separated by a padding of 5 trading days from the event window boundaries.

Formally, let $\rho_e^{\text{event}}$ denote the mean pairwise correlation of portfolio returns during the event window and $\rho_e^{\text{base}}$ denote that during the baseline window. The comovement effect of entity $e$ is then defined as

$$\Delta \rho_e = \rho_e^{\text{event}} - \rho_e^{\text{base}}, \qquad (18)$$

which captures the increase in cross-firm correlation associated with the focal entity's appearance in contemporaneous news. In contrast to standard market-factor correlations, our measure reflects information-linked comovement: the extent to which firms historically tied to the same key entity respond synchronously to new entity-specific information shocks.

To further investigate heterogeneity in the strength of these comovement effects, we group the focal entities according to their importance scores derived in the previous subsection.

Given the long-tailed distribution of importance scores, we apply two grouping schemes: (1) an *equal-sized* grouping approach that evenly partitions entities into three groups; and (2) a *value-threshold* approach that classifies entities into *High*, *Medium*, and *Low* importance groups using thresholds of 0.0015 and 0.0010.[8] A summary of entities under each grouping method is reported in Table 9.

Table 9 about here

Table 10 about here

Table 10 summarizes the results of the entity-driven stock comovement analysis with a one-trading-day shift between the news day and the corresponding market reaction. First, across event windows, we observe that the average comovement effect ($\Delta \rho_e$) is positive in nearly all horizons, peaking around the $\pm 1$ window and gradually decaying thereafter. This pattern suggests that market responses to entity-related news are strongest immediately surrounding the information release and dissipate as the shock effect fades. Notably, many entities reach their highest comovement on day $\pm 1$ rather than at $t_0$, implying that the information is partially incorporated on the release day, consistent with the gradual diffusion of information documented in the literature.

Second, across importance groups, the *High*-importance entities consistently exhibit the strongest and most significant comovement effects, followed by the *Mid* group, whereas the *Low* group shows relatively muted and short-lived correlations. This monotonic pattern supports the interpretation that entities attracting higher model attention, and thus greater information relevance, tend to be associated with more synchronized market reactions among the relevant firms. Such results imply that our model's entity-level importance captures not only predictive relevance but also the intensity of shared information exposure across firms.

---

[8]The thresholds are determined based on the empirical support of the long-tailed distribution of importance scores. In the data, importance values range approximately from 0.0020 at the upper tail to 0.0005 at the lower tail. Accordingly, we partition entities into three economically meaningful importance regimes: *High* ($\geq 0.0015$), *Medium* ($[0.0010, 0.0015)$), and *Low* ($< 0.0010$).

Third, across grouping methods, both approaches produce similar temporal patterns; however, the value-threshold grouping, which applies a narrower cutoff on importance scores, isolates entities that exert a stronger informational influence. Under this specification, the high-importance group exhibits markedly stronger and more persistent comovement effects, further reinforcing the above conclusion that more salient entities serve as core information channels driving synchronized market reactions.

Table 11 about here

To provide a concrete illustration, Table 11 presents an example of the entity-driven comovement pattern for the focal entity *"Vaccine."* This entity belongs to the *High*-importance group under both grouping methods in Table 10, reflecting its strong model attribution and central role in driving collective market responses. For each identified shock day, we quantify its individual contribution to portfolio comovement through a *leave-one-out* procedure. Specifically, we recompute the post-event average pairwise correlation and the comovement effect ($\Delta\rho_e$) after excluding each event, and measure the resulting percentage decreases, referred to as "Pct of Post Corr. Drops" and "Pct of Corr. Difference Drops." These two metrics capture, respectively, how much a given event contributes to the overall level of return synchronization and to the incremental increase in correlations during the event window. The ten dates displayed in the table are manually selected from those exhibiting relatively large decreases in these measures, representing the most influential shocks linked to this entity.

Consistent with the aggregate evidence reported in Table 10, the corresponding portfolio of leading pharmaceutical and healthcare firms exhibits pronounced increases in cross-firm return correlations after major vaccine-related news. These news are largely tied to COVID-19 vaccine development milestones and associated public-policy actions. The most pronounced comovement occurred on May 19, 2020, when the U.S. Secretary of Health and Human Services announced a federal goal to secure 300 million vaccine doses by year-end.

This announcement marked a pivotal policy commitment that significantly aligned market expectations across the entire vaccine supply chain, producing the largest observed increase in pairwise correlations among portfolio firms. Overall, the results demonstrate that certain economically salient information shocks indeed trigger noticeable comovement within the portfolio. This pattern highlights how entity-level information serves as a conduit for cross-firm information diffusion. The case of *"Vaccine"* thus provides a clear, real-world validation of our model's interpretability: entities assigned higher importance scores not only enhance predictive accuracy but also correspond to substantive, market-relevant episodes of collective investor response. Moreover, this analysis offers methodological insights for studying comovement, showing how entity-level event identification and leave-one-out contribution analysis can reveal the underlying informational sources driving synchronous market behavior.

Figure 7 about here

To further illustrate how these firms become connected through the focal entity, Figure 7 visualizes the event triplets of the same portfolio (i.e., the top-15 firms with the highest *"Vaccine"* exposure) as a knowledge graph. The central node is *"Vaccine"*, the outer ring nodes correspond to the 15 focal firm entities, and the middle-layer nodes are other co-occurring entities appearing in the extracted triplets; an edge is drawn when two entities co-appear as a subject–object pair in at least one triplet. For visual clarity, we retain only triplets and nodes within two hops of *"Vaccine"* for each firm. According to the figure, the structural density of the network surrounding the *"Vaccine"* entity provides visual evidence of its role as a high-importance informational anchor that facilitates entity-induced comovement. The graph shows that the comovement observed in the TOP-K portfolio is not merely coincidental but is driven by the propagation of entity-specific shocks, such as regulatory milestones or policy commitments, through these documented economic and semantic linkages. Furthermore, the complexity of the edges within two hops of the central node underscores why *"Vaccine"* maintains a high importance score in the model, as it serves as a primary conduit for information diffusion that synchronizes investor beliefs and realized returns across the affected firms.

In summary, the evidence from both the aggregate analysis and the case study reveals that entity-level information plays a systematic role in shaping cross-firm return dynamics. Firms jointly exposed to salient entities react in a coordinated manner to new information, producing temporary but economically meaningful comovement in returns. These results suggest that the LLM-identified entities not only capture predictive textual signals but also correspond to real channels of information diffusion and correlated belief formation in financial markets. Conceptually, this mechanism represents an information-based counterpart to the ownership-driven connectedness documented by Anton and Polk (2014), highlighting that common exposure to news about the same entities can synchronize investor expectations just as shared ownership synchronizes trading flows. These findings position our SER representations as a promising tool for event studies. Because our SERs provide a standardized and economically meaningful decomposition of news into subject–action–object structures, they allow researchers to identify, group, and quantify firm-level exposures to specific informational triggers in a systematic way. This structured representation enables a scalable form of event classification and cross-firm impact tracing, offering a new pathway for studying information diffusion beyond traditional hand-coded or keyword-based approaches.

# 5 Conclusion

This paper introduces a powerful and interpretable event-driven framework for stock return prediction by utilizing large language models (LLMs) to extract structured event representation (SER) with attention-based deep learning. The approach not only achieves superior out-of-sample performance compared to existing textual methods but also bridges the gap between high-dimensional text data and interpretable financial analysis. By explaining how specific entities and event types influence market outcomes, our SER-based model offers insights into the economic mechanisms underlying stock return predictability. Our study highlights the critical value of structured and human-readable input features in enhancing both the accuracy and interpretability of text-based predictive models, paving the way for more trustworthy and mechanistically grounded financial forecasting.

While we focus on stock return predictability by newspaper data, the framework can be easily applied to other assets, such as options, bonds, and cryptocurrencies, and other text data, including firms' annual reports, social media, and earnings conference calls. In addition to asset returns, our framework can also be applied to many other topics in finance, such as corporate investment decisions, investors' trading motivations, and analysts' beliefs, which we leave for future research.

Table 1
Summary Statistics

|  | Daily | Weekly |
|---|---:|---:|
| Mean Cap. (000's) | 9,039,222 | 9,031,025 |
| Mean Ret. (%) | 0.0641 | 0.1862 |
| No. of Stocks | 59.86 | 213.55 |
| No. of News Articles | 170.86 | 1,221.44 |
| No. of News Articles / Stock | 2.85 | 2.07 |
| No. of Events | 1,222.84 | 8,778.34 |
| No. of Events / Stock | 20.43 | 14.55 |
| No. of Entities | 1,165.10 | 7,011.56 |
| No. of Entities/ Stock | 24.96 | 47.93 |
| No. of Actions | 460.48 | 2,006.71 |
| No. of Actions/ Stock | 14.71 | 26.45 |

This table reports summary statistics for the datasets used in the daily and weekly prediction tasks. "Mean Cap." denotes the average market capitalization (in USD) across all firms and periods. "Mean Ret." represents the mean daily or weekly realized return. "No. of Stocks" is the average number of traded firms with available return and news data per period. "No. of News Articles" refers to the total number of processed firm-level news items, while "No. of News Articles / Stock" gives the average number of news articles associated with each firm per period. "No. of Events" corresponds to the total number of structured event representations extracted from the news corpus, and "No. of Events / Stock" indicates the average number of events per firm per period. "No. of Entities" and "No. of Actions" represent the total counts of unique entities and actions identified in the extracted events, respectively, while "No. of Entities / Stock" and "No. of Actions / Stock" measure their averages per firm per period. All statistics are computed over the whole dataset window from January 2003 to December 2022.

**Table 2**
**Comparison of Portfolios Sorting Performance**

| Methods | | Daily Return Prediction | | | Weekly Return Prediction | | |
|---|---|---|---|---|---|---|---|
| | | AR | SR | $\alpha$ | AR | SR | $\alpha$ |
| Event-based | SER Model | 10.93% (3.31) | 0.78 | 10.58% (3.17) | 5.23% (3.03) | 0.63 | 4.21% (2.46) |
| Sentiment-based | Loughran McDonald | 2.04% (0.62) | 0.08 | 2.02% (0.61) | 0.15% (0.08) | $-0.13$ | 0.31% (0.19) |
| | RavenPack | 0.99% (1.11) | $-0.01$ | 1.01% (1.09) | 1.01% (1.72) | 0.01 | 0.99% (1.66) |
| Embedding-based | BERT | 2.12% (0.67) | 0.09 | 2.02% (0.60) | 3.56% (2.14) | 0.40 | 3.17% (1.92) |
| | FinBERT | 8.59% (2.66) | 0.61 | 7.81% (2.41) | 3.78% (2.31) | 0.44 | 3.22% (1.97) |

This table reports the annualized return (AR), Sharpe ratio (SR), and annualized Fama-French five-factor ($\alpha$) for various prediction methods in the daily and weekly stock return prediction tasks. For both AR and $\alpha$, the numbers in parentheses denote the corresponding $t$-statistics. The methods are grouped into three categories: event-based, sentiment-based, and embedding-based. Daily return prediction results are shown in columns 3–5, while weekly return prediction results are shown in columns 6–8. For each category, representative baseline models are reported alongside our proposed event-based models. Model specifications are described in Section 3.2. The sample period is from January 2008 to December 2022.

Table 3
Delayed Portfolio Performance

| Delayed Portfolio | | +1T | +2T | +3T | +4T | +5T | +6T | +7T | +8T | +9T |
|---|---|---|---|---|---|---|---|---|---|---|
| Daily (1T = 1 Day) | AR | 10.9% | −3.1% | 4.1% | 0.8% | 1.2% | 2.7% | 2.7% | −4.2% | −2.6% |
| | | (3.31) | (−1.01) | (1.26) | (0.25) | (0.38) | (0.78) | (0.64) | (−1.31) | (−0.87) |
| | SR | 0.78 | −0.35 | 0.25 | −0.01 | 0.02 | 0.13 | 0.10 | −0.42 | −0.31 |
| Weekly (1T = 1 Week) | AR | 5.23% | −0.07% | 1.43% | 2.07% | 3.48% | 2.05% | 1.38% | 1.81% | −0.27% |
| | | (3.03) | (−0.04) | (0.84) | (1.16) | (1.64) | (1.14) | (0.83) | (1.12) | (−0.17) |
| | SR | 0.63 | −0.17 | 0.07 | 0.15 | 0.30 | 0.15 | 0.06 | 0.13 | −0.21 |

This table reports the delayed portfolio performance at daily and weekly horizons. At the beginning of each period, long–short portfolios are formed based on the predicted return signals and then held with different implementation delays, ranging from one to nine trading intervals (columns +1T to +9T). For the daily case, 1T corresponds to one trading day, and for the weekly case, 1T corresponds to one trading week. Reported values include the annualized average return (AR), corresponding $t$-statistics in parentheses, and Sharpe ratio (SR). The sample period spans from January 2008 to December 2022.

Table 4
Fama-MacBeth Regression Analysis

|  | Methods | Daily Return Prediction | | | Weekly Return Prediction | | |
|---|---|---|---|---|---|---|---|
|  | Intercept | 0.469 (1.73) | 0.470 (1.72) | 0.462 (1.70) | 2.401 (2.14) | 2.451 (2.15) | 2.363 (2.11) |
| Event-based | SER Model | 0.222 (2.90) | 0.209 (2.68) | 0.223 (2.54) | 0.430 (2.38) | 0.493 (2.73) | 0.449 (2.53) |
| Sentiment-based | Loughran McDonald |  | 0.077 (0.98) |  |  | 0.129 (0.73) |  |
|  | RavenPack |  | −0.090 (−0.92) |  |  | −0.098 (−0.37) |  |
| Embedding-based | BERT |  |  | 0.016 (0.18) |  |  | −0.052 (−0.39) |
|  | FinBERT |  |  | 0.030 (0.33) |  |  | 0.165 (0.87) |
|  | Controls | Yes | Yes | Yes | Yes | Yes | Yes |
|  | Adj.$R^2$ (%) | 9.05 | 9.58 | 8.92 | 8.63 | 8.94 | 8.59 |

This table reports the out-of-sample Fama-MacBeth regression results for daily and weekly stock return prediction. The dependent variables are realized returns at the respective frequencies, and the independent variables correspond to predicted return signals from different models, grouped into three categories: event-based, sentiment-based, and embedding-based approaches. Model specifications are described in Section 3.2. All predictors are standardized to have a mean of zero and a unit standard deviation within each prediction window. Reported coefficients are the time-series averages of the cross-sectional regressions, with heteroskedasticity- and autocorrelation-consistent (HAC) standard errors computed using a lag length of 11 trading days. Control variables differ across tasks: for the daily prediction, we include the past 1-week, 2-week, and 1-month average realized daily returns of each stock. For the weekly prediction, we include the past 1-month, 3-month, and 6-month average realized weekly returns to capture short- to medium-horizon momentum and reversal effects. The sample period spans from January 2008 to December 2022.

Table 5
Top-20 Important Events (Daily)

| Rank | Event Representation | | | Imp. Score | Pos. Pct. | Neg. Pct. | Freq. |
|---|---|---|---|---|---|---|---|
| | Subject | Action | Object | | | | |
| 1 | National_Vision | rally | S&P_SmallCap_600 | 0.0110 | 100.00% | 0.00% | 79 |
| 2 | IBM | report | Revenue | 0.0089 | 100.00% | 0.00% | 24 |
| 3 | Revenue | raise | Percentage | 0.0087 | 100.00% | 0.00% | 25 |
| 4 | Harold_Hamm | offer | Stock | 0.0052 | 0.00% | 100.00% | 22 |
| 5 | Macy's | report | Sales | 0.0051 | 4.76% | 95.24% | 21 |
| 6 | Continental_Resources | receive | Take-private | 0.0043 | 0.00% | 100.00% | 23 |
| 7 | Walmart | raise | Wage | 0.0040 | 91.67% | 8.33% | 24 |
| 8 | Apple_Inc. | sell | IPhone | 0.0039 | 97.62% | 2.38% | 42 |
| 9 | Share | fell with | Percentage | 0.0037 | 96.43% | 3.57% | 28 |
| 10 | Netflix | add | Subscriber | 0.0033 | 94.59% | 5.41% | 37 |
| 11 | Kevin_Durant | score | Point | 0.0032 | 0.00% | 100.00% | 22 |
| 12 | Stock | fell with | Percentage | 0.0032 | 87.50% | 12.50% | 24 |
| 13 | Donald_Trump | sign | Executive_order | 0.0029 | 96.20% | 3.80% | 79 |
| 14 | European_Central_Bank | cut | Interest_rate | 0.0028 | 26.92% | 73.08% | 26 |
| 15 | Federal_Reserve_System | raise | Interest_rate | 0.0027 | 75.08% | 24.92% | 305 |
| 16 | Walmart | say | Sales | 0.0024 | 47.62% | 52.38% | 21 |
| 17 | Apple_Inc. | decline | Comment | 0.0022 | 95.56% | 4.44% | 45 |
| 18 | United_States | impose | Tariff | 0.0022 | 4.17% | 95.83% | 24 |
| 19 | Apple_Inc. | report | Revenue | 0.0021 | 93.10% | 6.90% | 29 |
| 20 | Walmart | report | Sales | 0.0020 | 28.21% | 71.79% | 39 |

This table reports the top 20 exact events ranked by their total (absolute) importance scores in the daily return prediction model. Event importance is computed using the gradient-based attribution procedure described in Section 4.1, where importance scores are first obtained for each occurrence of an event and then aggregated for events with identical (*subject-action-object*) representations. "Imp. Score" denotes the aggregated absolute importance, measuring the overall magnitude of each event's contribution to predicted returns. "Pos. Pct." and "Neg. Pct." report the proportions of positive and negative polarity importance, respectively, calculated as the absolute positive (or negative) contribution divided by the total absolute importance for that event. These polarity measures indicate whether the event tends to be return-enhancing or return-depressing on average. "Freq." denotes the number of times the exact event appears in the dataset.

## Table 6
## Top-20 Important Events (Weekly)

| Rank | Event Representation | | | Imp. Score | Pos. Pct. | Neg. Pct. | Freq. |
|---|---|---|---|---|---|---|---|
| | Subject | Action | Object | | | | |
| 1 | Investor | pour | United_States_dollar | 0.0055 | 75.00% | 25.00% | 20 |
| 2 | United_States_Department_of_the_Treasury | schedule | Finance | 0.0054 | 9.09% | 90.91% | 11 |
| 3 | United_States_House_of_Representatives | pass | Bill | 0.0054 | 20.00% | 80.00% | 15 |
| 4 | Investor | pull | United_States_dollar | 0.0051 | 32.00% | 68.00% | 25 |
| 5 | Google | decline | Comment | 0.0049 | 66.67% | 33.33% | 18 |
| 6 | Islamic_State_of_Iraq_and_the_Levant | claim | Responsibility | 0.0045 | 8.33% | 91.67% | 12 |
| 7 | Federal_Reserve_System | raise | Interest_rate | 0.0044 | 29.17% | 70.83% | 96 |
| 8 | Dow_Jones_Industrial_Average | raise | Point | 0.0042 | 73.33% | 26.67% | 15 |
| 9 | Standard_&_Poor's | cut | Credit_rating | 0.0041 | 36.36% | 63.64% | 11 |
| 10 | Apple_Inc. | fell with | Percentage | 0.0038 | 63.64% | 36.36% | 11 |
| 11 | Facebook | say | Advertising | 0.0036 | 45.45% | 54.55% | 11 |
| 12 | Donald_Trump | intercept | Tariff | 0.0036 | 36.36% | 63.64% | 11 |
| 13 | Epic_Games | boycott | Apple_Inc. | 0.0036 | 78.57% | 21.43% | 14 |
| 14 | Apple_Inc. | decline | Comment | 0.0035 | 85.00% | 15.00% | 20 |
| 15 | Donald_Trump | write | Twitter | 0.0034 | 18.18% | 81.82% | 11 |
| 16 | Donald_Trump | sign | Executive_order | 0.0033 | 78.13% | 21.88% | 32 |
| 17 | Carmelo_Anthony | score | Point | 0.0033 | 84.62% | 15.38% | 13 |
| 18 | Apple_Inc. | sell | IPhone | 0.0030 | 36.67% | 63.33% | 30 |
| 19 | Apple_Inc. | boycott | Samsung | 0.0029 | 18.18% | 81.82% | 11 |
| 20 | Mark_Zuckerberg | say | Facebook | 0.0028 | 58.33% | 41.67% | 12 |

This table reports the top 20 exact events ranked by their total (absolute) importance scores in the weekly return prediction model. Event importance is computed using the gradient-based attribution procedure described in Section 4.1, where importance scores are first obtained for each occurrence of an event and then aggregated for events with identical (*subject-action-object*) representations. "Imp. Score" denotes the aggregated absolute importance, measuring the overall magnitude of each event's contribution to predicted returns. "Pos. Pct." and "Neg. Pct." report the proportions of positive and negative polarity importance, respectively, calculated as the absolute positive (or negative) contribution divided by the total absolute importance for that event. These polarity measures indicate whether the event tends to be return-enhancing or return-depressing on average. "Freq." denotes the number of times the exact event appears in the dataset.

Table 7
Top-20 Important Event Topics

| Rank | Daily Model | | | | Weekly Model | | | |
|---|---|---|---|---|---|---|---|---|
| | Event Topics | Imp. Score | Pos. Pct. | Neg. Pct. | Event Topics | Imp. Score | Pos. Pct. | Neg. Pct. |
| 1 | Bankruptcy | 0.619 | 31.44% | 68.56% | Pharmaceutical Industry | 0.690 | 50.08% | 49.92% |
| 2 | Natural Gas Industry | 0.514 | 51.16% | 48.84% | Control Stakes | 0.681 | 53.16% | 46.84% |
| 3 | Control Stakes | 0.475 | 52.60% | 47.40% | Shareholder Payouts | 0.666 | 51.47% | 48.53% |
| 4 | Collaboration | 0.461 | 61.01% | 38.99% | Corporate Bond Risk | 0.665 | 45.96% | 54.04% |
| 5 | Key Role | 0.456 | 52.94% | 47.06% | Earning Forecast | 0.659 | 51.18% | 48.82% |
| 6 | Pharmaceutical Industry | 0.444 | 54.49% | 45.51% | Natural Gas Industry | 0.658 | 52.38% | 47.62% |
| 7 | Investment Banking | 0.441 | 60.39% | 39.61% | Corporate Governance | 0.654 | 44.63% | 55.37% |
| 8 | Commercial Airlines | 0.429 | 45.01% | 54.99% | Takeovers | 0.645 | 49.15% | 50.85% |
| 9 | Movie Industry | 0.427 | 50.32% | 49.68% | Retail | 0.638 | 49.91% | 50.09% |
| 10 | Oil Industry | 0.422 | 55.07% | 44.93% | Insurance Industry | 0.632 | 57.46% | 42.54% |
| 11 | Chinese Business Issues | 0.420 | 52.51% | 47.49% | Sale Performance | 0.630 | 48.99% | 51.01% |
| 12 | Global Exchange Strategies | 0.410 | 50.45% | 49.55% | Mutual Fund | 0.629 | 52.19% | 47.81% |
| 13 | Executive Appointments | 0.410 | 52.65% | 47.35% | Chinese Business Issues | 0.625 | 53.45% | 46.55% |
| 14 | Streaming Services | 0.409 | 52.45% | 47.55% | Media Broadcasting | 0.620 | 54.41% | 45.59% |
| 15 | Private/Public Sector | 0.406 | 56.71% | 43.29% | Credit Card | 0.620 | 56.92% | 43.08% |
| 16 | Media Broadcasting | 0.405 | 53.73% | 46.27% | Technology Company | 0.620 | 54.43% | 45.57% |
| 17 | Company Asset | 0.403 | 55.30% | 44.70% | Key Role | 0.619 | 51.30% | 48.70% |
| 18 | Aircraft Manufacturing | 0.402 | 50.95% | 49.05% | Earning Performance | 0.616 | 57.64% | 42.36% |
| 19 | Casinos | 0.402 | 62.62% | 37.38% | Auto Manufacturing | 0.616 | 59.81% | 40.19% |
| 20 | Auto Manufacturing | 0.401 | 54.32% | 45.68% | Global Exchange Strategies | 0.615 | 47.54% | 52.46% |

This table reports the top 20 event topics ranked by their total (absolute) importance scores for the daily and weekly return prediction models. Event-topic importance is computed using the gradient-based attribution method described in Section 4.1 and aggregated according to the LDA-based topic modeling procedure outlined in Section 4.2. "Imp. Score" denotes the absolute importance, measuring the overall magnitude of each topic's contribution to predicted returns. "Pos. Pct." and "Neg. Pct." denote the proportions of positive and negative polarity importance, respectively. These values are computed as the absolute positive (or negative) contribution divided by the total absolute importance, thereby indicating whether an entity is primarily associated with return-enhancing or return-depressing effects. The left panel summarizes results for the daily model, while the right panel presents those for the weekly model. Detailed examples of high-impact events associated with each event topic are provided in the Online Appendix OA.3 for reference.

Table 8
Top-20 Important Entities

| Rank | Daily Model | | | | Weekly Model | | | |
|---|---|---|---|---|---|---|---|---|
| | Entities | Imp. | Pos. Pct. | Neg. Pct. | Entities | Imp. Score | Pos. Pct. | Neg. Pct. |
| 1 | Bankruptcy | 0.486 | 6.76% | 93.24% | Tesla,_Inc. | 0.925 | 15.63% | 84.37% |
| 2 | Financial_statement | 0.305 | 99.87% | 0.13% | Consumer | 0.846 | 45.97% | 54.03% |
| 3 | Vaccine | 0.265 | 0.00% | 100.00% | United_Airlines | 0.768 | 4.77% | 95.23% |
| 4 | Cost | 0.239 | 100.00% | 0.00% | Royal_Dutch_Shell | 0.758 | 6.66% | 93.34% |
| 5 | California | 0.235 | 100.00% | 0.00% | Retail | 0.748 | 26.04% | 73.96% |
| 6 | Bond | 0.233 | 6.02% | 93.98% | Mortgage_loan | 0.723 | 18.86% | 81.14% |
| 7 | Price | 0.228 | 99.77% | 0.23% | Letter | 0.689 | 0.70% | 99.30% |
| 8 | Challenge | 0.225 | 100.00% | 0.00% | Deal | 0.684 | 98.60% | 1.40% |
| 9 | Revenue | 0.223 | 89.00% | 11.00% | Airline | 0.679 | 65.14% | 34.86% |
| 10 | Data | 0.216 | 100.00% | 0.00% | Election | 0.665 | 2.74% | 97.26% |
| 11 | Sales | 0.213 | 14.53% | 85.47% | Microsoft | 0.652 | 92.45% | 7.55% |
| 12 | United_Kingdom | 0.210 | 100.00% | 0.00% | Advertising | 0.644 | 6.40% | 93.60% |
| 13 | Washington,_D.C. | 0.202 | 100.00% | 0.00% | ExxonMobil | 0.624 | 8.06% | 91.94% |
| 14 | Share | 0.196 | 97.16% | 2.84% | Earnings_per_share | 0.624 | 1.98% | 98.02% |
| 15 | Fine | 0.195 | 0.00% | 100.00% | Board_of_directors | 0.573 | 13.20% | 86.80% |
| 16 | Finance | 0.192 | 100.00% | 0.00% | Saudi_Arabia | 0.570 | 16.14% | 83.86% |
| 17 | Comment | 0.189 | 99.57% | 0.43% | Washington,_D.C. | 0.558 | 44.33% | 55.67% |
| 18 | Chairman | 0.186 | 98.20% | 1.80% | Person | 0.547 | 65.00% | 35.00% |
| 19 | City | 0.183 | 100.00% | 0.00% | Verizon_Communications | 0.543 | 92.44% | 7.56% |
| 20 | Asset | 0.180 | 99.86% | 0.14% | Exchange-traded_fund | 0.543 | 39.20% | 60.80% |

This table reports the top 20 entities ranked by their total (absolute) importance scores for both the daily and weekly return prediction models. Entity-level importance is derived from the gradient-based attribution method described in Section 4.1, aggregated across all occurrences of each entity within the representative dataset. "Imp. Score" denotes the absolute importance, capturing the overall magnitude of each entity's contribution to predicted returns. "Pos. Pct." and "Neg. Pct." denote the proportions of positive and negative polarity importance, respectively. These values are computed as the absolute positive (or negative) contribution divided by the total absolute importance, thereby indicating whether an entity is primarily associated with return-enhancing or return-depressing effects. The left panel summarizes results for the daily model, while the right panel presents those for the weekly model. Detailed examples of high-impact events associated with each entity are provided in the Online Appendix OA.3 for reference.

## Table 9
## Entity Group Summary for Stock Comovement Analysis

| Group (Total=84) | Equal Grouped | | Value Grouped | |
| --- | --- | --- | --- | --- |
| | No. | Entities | No. | Entities |
| High | 28 | Vaccine, California, Bond, Data, Fine, Comment, City, Boeing, European_Commission, Terrorism, New_York_City, Euro, Oracle_Corporation, Statement, Decision-making, Tariff, Pfizer, General_Motors, North_Korea, Alphabet_Inc., Donation, Deutsche_Bank, Federal_Reserve, Amazon, Game, Obama_administration | 8 | Vaccine, California, Bond, Data, Fine, Comment, City, Boeing |
| Mid | 28 | France, Intel, Europe, European_Central_Bank, Election, BlackRock, Alibaba_Group, Barack_Obama, Xi_Jinping, Starbucks, Target_Corporation, Walmart, Serena_Williams, Uber, Ford_Motor_Company, ExxonMobil, Federal_Aviation_Administration, Clinical_trial, New_York_Knicks, John_Kerry, The_Walt_Disney_Company, Food_and_Drug_Administration, Citigroup, United_States_Department_of_the_Treasury, Letter, Verizon_Communications, Wells_Fargo, Law, Morgan_Stanley | 12 | European_Commission, Terrorism, New_York_City, Euro, Oracle_Corporation, Statement, Decision-making, Tariff, Pfizer, Chief_executive_officer, General_Motors, North_Korea, Apple_Inc., Performance, Alphabet_Inc. |
| Low | 28 | Iran, United_States_Congress, Market, Mutual_fund, Coca-Cola, Bank, Settlement, Hillary_Clinton, Aircraft, Republican_Party, Journalist, Exchange-traded_fund, Barclays, New_York_Yankees, New_York_Mets, Airline, BP, Volkswagen, Product, Islamic_State_of_Iraq_and_the_Levant, Vladimir_Putin, Chris_Christie, Book, Mark_Zuckerberg, Janet_Yellen, Smartphone, Theresa_May, Joe_Biden, Family, Angela_Merkel | 64 | Donation, Deutsche_Bank, Federal_Reserve, Amazon, Game, Obama_administration, France, Intel, Europe, European_Central_Bank, Election, Russia, BlackRock, Alibaba_Group, Barack_Obama, Xi_Jinping, Starbucks, Target_Corporation, Walmart, Serena_Williams, Uber, Ford_Motor_Company, ExxonMobil, Federal_Aviation_Administration, Clinical_trial, New_York_Knicks, John_Kerry, The_Walt_Disney_Company, Food_and_Drug_Administration, Citigroup, United_States_Department_of_the_Treasury, Letter, Verizon_Communications, Wells_Fargo, Law, Morgan_Stanley, Iran, United_States_Congress, Market, Mutual_fund, Coca-Cola, Bank, Settlement, Hillary_Clinton, Aircraft, Journalist, Exchange-traded_fund, Barclays, New_York_Yankees, New_York_Mets, Airline, BP, Volkswagen, Product, Islamic_State_of_Iraq_and_the_Levant, Vladimir_Putin, Chris_Christie, Book, Mark_Zuckerberg, Janet_Yellen, Smartphone, Theresa_May, Joe_Biden, Family, Angela_Merkel |

This table summarizes the grouping of focal entities used in the stock comovement analysis. Entities are grouped according to their importance scores derived from the attribution analysis in Section 4.4. The *Equal Grouped* columns divide entities evenly into three groups of approximately equal size, while the *Value Grouped* columns classify entities into *High*, *Mid*, and *Low* importance groups based on thresholds of 0.0015 and 0.0010. The total sample includes 84 focal entities identified through the portfolio construction procedure described in Section 4.5. These groupings are used in subsequent analyses to examine heterogeneity in information-linked comovement effects across entities of varying importance.

Table 10
**Entity Driven Stock Comovement Analysis**

| Group | Event Window | Equal Grouped | | | | Value Grouped | | | |
|---|---|---|---|---|---|---|---|---|---|
| | | Average | Pct. of Mean>0 | Pct. of $p<.05$ | Pct of $p<.01$ | Average | Pct. of Mean>0 | Pct. of $p<.05$ | Pct. of $p<.01$ |
| High | ±0d | 0.023 | 57.10% | 42.90% | 32.10% | 0.025 | 75.00% | 50.00% | 37.50% |
| | ±1d | 0.029 | 71.40% | 42.90% | 32.10% | 0.063 | 87.50% | 75.00% | 75.00% |
| | ±2d | 0.017 | 50.00% | 42.90% | 35.70% | 0.045 | 75.00% | 50.00% | 50.00% |
| | ±3d | 0.011 | 57.10% | 46.40% | 35.70% | 0.045 | 87.50% | 75.00% | 50.00% |
| | ±4d | 0.013 | 57.10% | 42.90% | 39.30% | 0.054 | 87.50% | 75.00% | 75.00% |
| | ±5d | 0.001 | 50.00% | 35.70% | 32.10% | 0.045 | 87.50% | 75.00% | 75.00% |
| | ±6d | 0.004 | 53.60% | 28.60% | 28.60% | 0.047 | 75.00% | 62.50% | 62.50% |
| Mid | ±0d | −0.030 | 42.90% | 25.00% | 17.90% | 0.034 | 41.70% | 41.70% | 33.30% |
| | ±1d | −0.011 | 39.30% | 28.60% | 28.60% | 0.012 | 66.70% | 33.30% | 16.70% |
| | ±2d | −0.020 | 32.10% | 17.90% | 17.90% | 0.009 | 50.00% | 50.00% | 33.30% |
| | ±3d | −0.016 | 35.70% | 28.60% | 17.90% | 0.007 | 50.00% | 41.70% | 33.30% |
| | ±4d | −0.016 | 32.10% | 21.40% | 21.40% | 0.007 | 50.00% | 33.30% | 33.30% |
| | ±5d | −0.021 | 28.60% | 17.90% | 17.90% | −0.014 | 41.70% | 25.00% | 16.70% |
| | ±6d | −0.018 | 32.10% | 25.00% | 21.40% | −0.019 | 41.70% | 8.30% | 8.30% |
| Low | ±0d | −0.007 | 53.60% | 21.40% | 21.40% | −0.013 | 50.00% | 25.00% | 20.30% |
| | ±1d | −0.003 | 53.60% | 28.60% | 25.00% | −0.003 | 48.40% | 28.10% | 25.00% |
| | ±2d | −0.023 | 32.10% | 14.30% | 10.70% | −0.019 | 31.20% | 17.20% | 15.60% |
| | ±3d | −0.022 | 25.00% | 7.10% | 3.60% | −0.019 | 31.20% | 18.80% | 12.50% |
| | ±4d | −0.022 | 28.60% | 14.30% | 10.70% | −0.019 | 31.20% | 18.80% | 15.60% |
| | ±5d | −0.020 | 25.00% | 17.90% | 17.90% | −0.020 | 26.60% | 17.20% | 17.20% |
| | ±6d | −0.012 | 28.60% | 21.40% | 14.30% | −0.014 | 32.80% | 23.40% | 18.80% |

This table reports the results of the entity-driven stock comovement analysis with all focal entities (n=84). For each focal entity, the comovement effect ($\Delta\rho_e$) is computed as the difference between the average pairwise correlation of portfolio returns in the event window and that in the baseline window. The percentages indicate the fraction of entities exhibiting positive comovement ($\Delta\rho_e > 0$) and those with statistically significant effects at the 5% and 1% levels. Results are presented separately under the *Equal Grouped* and *Value Grouped* classification schemes described in Table 9. Results are based on comovement analysis with a one-trading-day shift between the news day and the corresponding market reaction, across event windows ranging from ±0d to ±6d.

**Table 11**
**A Sample of Entity Driven Stock Comovement (Focal Entity: Vaccine)**

| Shock Dates | Pct. of Post Corr. Drops | Pct. of Corr. Difference Drops | Main Events |
|---|---|---|---|
| Oct 02 2014 | 2.16% | 8.73% | The Food and Drug Administration approved a **vaccine** for the first time for the deadly type of meningitis that affected the students. |
| May 19 2020 | 18.69% | 84.26% | On Sunday, Health and Human Services Secretary Alex Azar said in a CBS interview that the Trump administration has set a goal for 300 million doses of a **vaccine** to be available to people in the U.S. by year-end. |
| Jul 23 2020 | 0.55% | 10.20% | Some pharmaceutical companies pledged to members of Congress that they wouldn't seek a profit from the Covid-19 **vaccines**. |
| Dec 04 2020 | 1.09% | 2.50% | Britain on Wednesday granted emergency approval to the coronavirus **vaccine** from pharmaceutical giant Pfizer and the German company BioNTech, becoming the first Western country to authorize mass inoculations that could lead to an end of the pandemic. |
| Feb 08 2021 | 1.26% | 2.04% | Biden promised to get those **doses of hope** out. |
| Feb 25 2021 | 2.18% | 9.57% | Maryland expects to receive shipments of the yet-to-be cleared Johnson & Johnson **vaccine** as soon as next week, Gov. Larry Hogan (R) said Tuesday. |
| Feb 26 2021 | 2.86% | 11.50% | Moderna Inc. said it has made the initial batch of doses of a new Covid-19 **vaccine** designed to better protect people against a new strain of the coronavirus. |
| Mar 03 2021 | 0.97% | 0.21% | The U.K. government's strategy of stretching out limited **vaccine supplies** by delaying a second shot of vaccine by up to 12 weeks after the first to protect more people from the disease. Some 600 million doses of the nation's first two authorized **vaccines**, made by Pfizer-BioNTech and Moderna, are scheduled to be delivered over the next four months. |
| Apr 16 2021 | 0.78% | 6.84% | J&J said this week it would delay the rollout of **its vaccine** in Europe because of potential safety concerns. Denmark on Wednesday said it was suspending indefinitely use of the **vaccine**. |
| Oct 14 2021 | 0.56% | 4.31% | A Food and Drug Administration review released Tuesday found that a booster shot of the Moderna coronavirus **vaccine** enhanced virus-fighting antibodies in people who had received the standard two-dose regimen at least six months earlier. A unit of a Japanese company backed by Philip Morris International plans to launch the first plant-based Covid-19 **vaccine** that is potentially cheaper and easier to transport and store than conventional jabs. |

This table reports representative news events associated with the focal entity "Vaccine" that influence stock comovement within its corresponding portfolio. For each entity shock day, two metrics are reported: "Pct of Post Corr. Drops," which measures the percentage decrease in the average post-event pairwise correlations when the event is excluded, and "Pct of Corr. Difference Drops," which quantifies the reduction in the comovement effect ($\Delta \rho_e$) under the leave-one-out evaluation. Both measures are computed within an event window of $\pm 1$ trading day around each shock date. Based on these measures, we manually select 10 dates (total = 33) exhibiting relatively large declines in comovement when omitted, representing the most influential shocks linked to this entity. The "Main Events" column summarizes the contexts of the key events drawn from contemporaneous news coverage. The sample period is from January 2008 to December 2021.

* The corresponding portfolio includes: Altimmune Inc, Moderna Inc, Novavax Inc, Emergent Biosolutions Inc, Pfizer Inc, Tenet Healthcare Corp, Johnson & Johnson, Tyson Foods Inc, Walgreens Boots Alliance Inc, Philip Morris International Inc, Six Flags Entertainment Corp, Listed Funds Trust, Baxter International Inc, Inovio Pharmaceuticals Inc, Merck & Co Inc New.

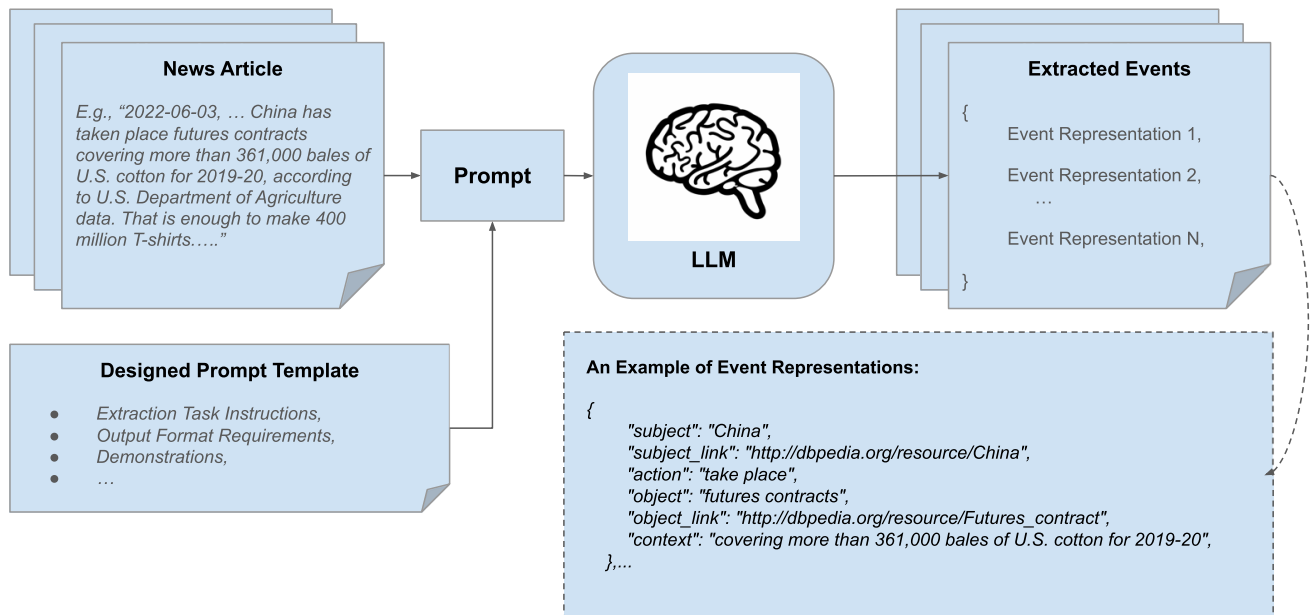

**Figure 1**
**Event Extraction Pipeline**
This figure illustrates the pipeline for extracting events using a large language model (LLM). Each news article is transformed into a structured input using a carefully designed prompt template that specifies task instructions, output format, and examples. The LLM processes this prompt and generates extracted events in the form of structured representations. The box on the bottom right provides an example of a structured event representation, demonstrating the type of output produced by the pipeline. We post the full prompt template and a representative sample output in the Online Appendix OA.2.

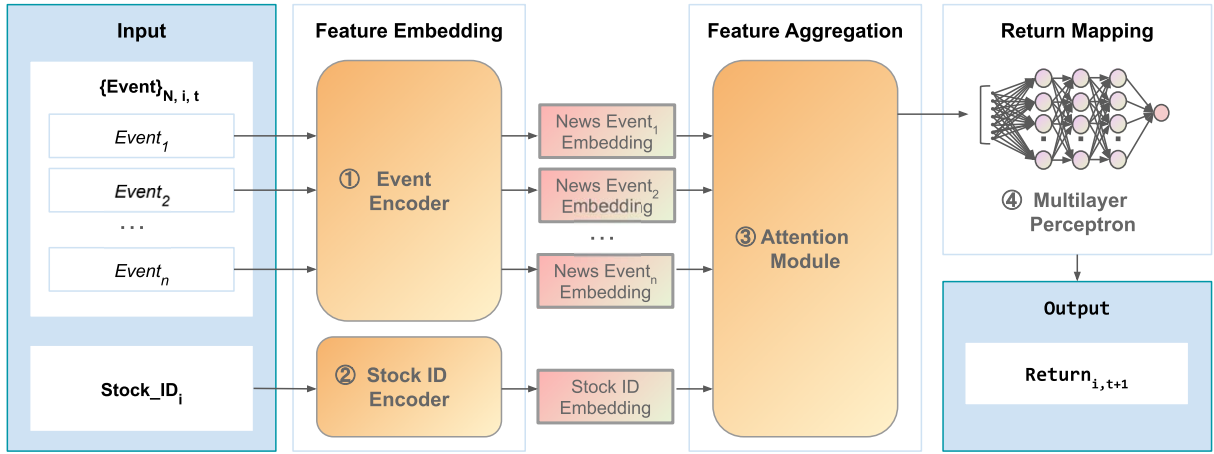

**Figure 2**
**The Framework of Event-driven Return Prediction**
This figure presents the framework of how to estimate the expected return based on LLM-extracted events. Accordingly, the blue boxes represent the input (the event set and stock ID) and output (expected return) of the model. The process of mapping input to output includes three main steps (represented by transparent boxes in the figure), namely feature embedding, feature aggregation, and return mapping. In the feature embedding step, event representation and stock ID are converted into vector features, respectively. Then, in the feature aggregation step, they are further aggregated into one-dimensional vectors through the attention mechanism. Finally, in the return mapping step, a multi-layer perceptron is used to map them to expected returns. The ①-④ in the figure mark the key modules of each step, whose details are explained in Section 2.2.2-2.2.4.

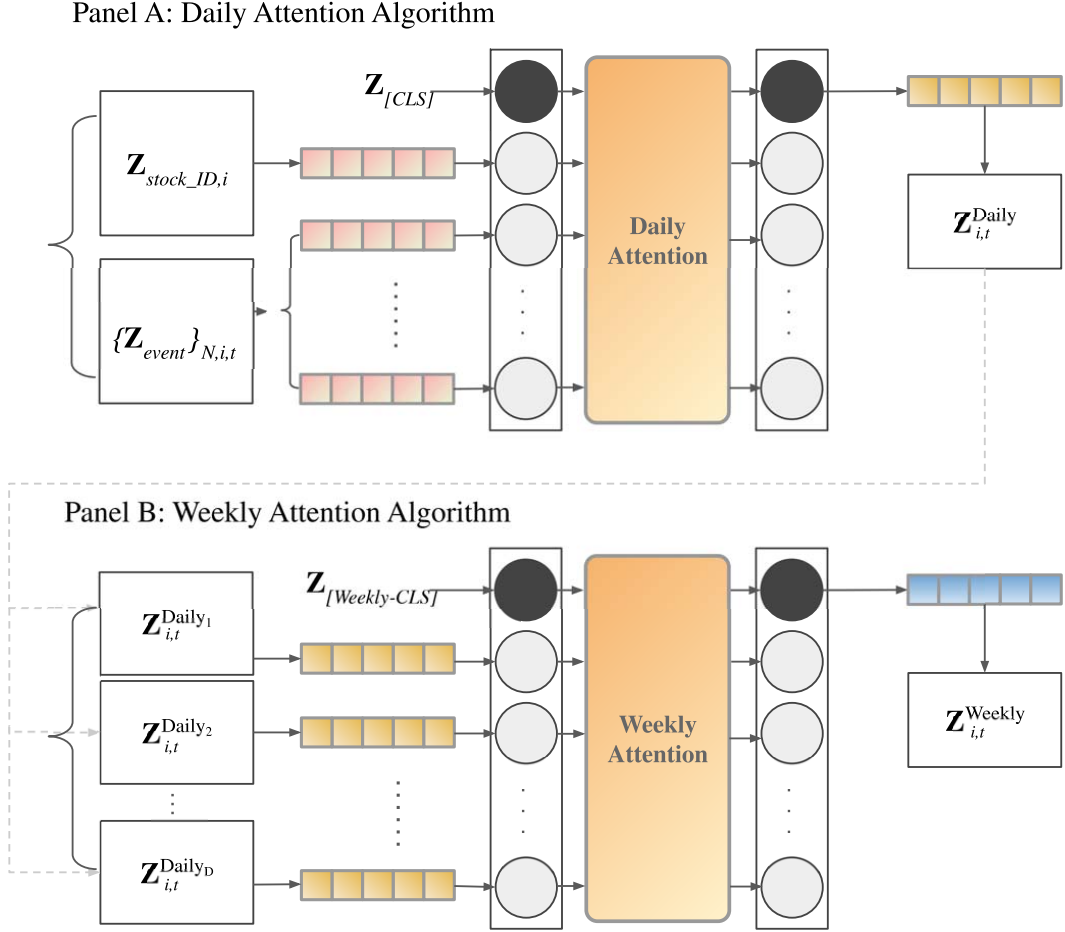

**Figure 3**
**The Framework of Feature Aggregation**
This figure illustrates the two feature aggregation schemes described in Section 2.2.3. Panel A presents the daily attention mechanism, which applies multi-head self-attention to the set of same-day event embeddings $\{\mathbf{z}_{\text{event}}\}_{N,i,t}$, together with the stock-identifier embedding $z_{\text{stock\_ID},i}$ and a learnable summary token $\mathbf{z}_{[\text{CLS}]}$. This module captures cross-event interactions as well as event–stock contextual dependencies, enabling the model to learn heterogeneous event importance within a trading day and to aggregate all information into a single daily representation $\mathbf{Z}_{i,t}^{\text{Daily}}$. Panel B illustrates the weekly attention mechanism, which performs cross-day aggregation by applying self-attention over the sequence of daily representations $\mathbf{Z}_{i,t}^{\text{Daily}_1}, \mathbf{Z}_{i,t}^{\text{Daily}_2}, \ldots, \mathbf{Z}_{i,t}^{\text{Daily}_D}$ within a week, together with a weekly-level summary token $\mathbf{z}_{[\text{Weekly-CLS}]}$. This hierarchical design explicitly models inter-day temporal dependencies and yields the weekly representation $\mathbf{Z}_{i,t}^{\text{Weekly}}$.

**Figure 4**
**Entity Importance Word Cloud for Daily Return Prediction**
This figure displays word clouds of entity importance for daily return prediction. The top panel shows the most influential entities based on absolute attribution scores aggregated across the dataset. The bottom panels separate entities by polarity, with the left panel presenting those contributing positively to predicted returns and the right panel showing those contributing negatively.

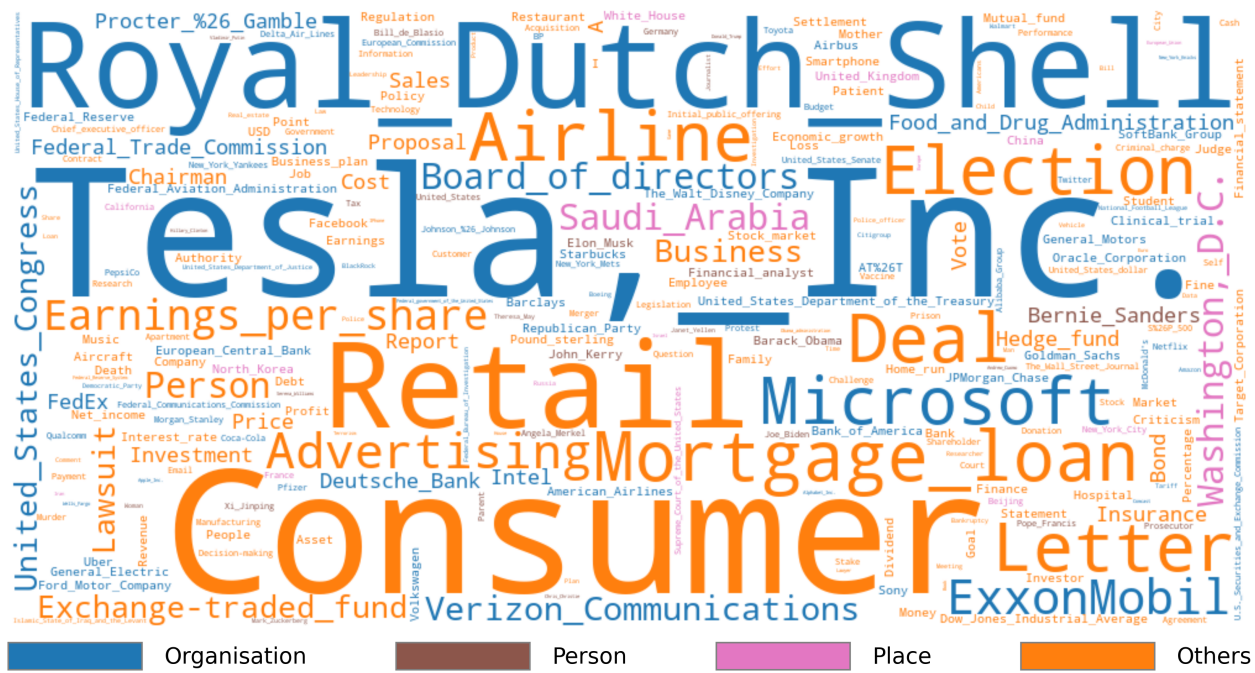

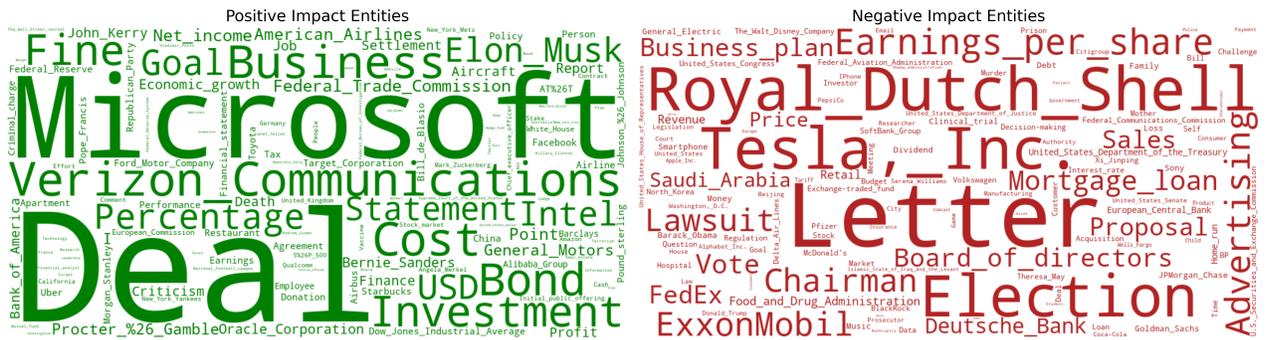

**Figure 5**
**Entity Importance Word Cloud for Weekly Return Prediction**
This figure displays word clouds of entity importance for weekly return prediction. The top panel shows the most influential entities based on absolute attribution scores aggregated across the dataset. The bottom panels separate entities by polarity, with the left panel presenting those contributing positively to predicted returns and the right panel showing those contributing negatively.

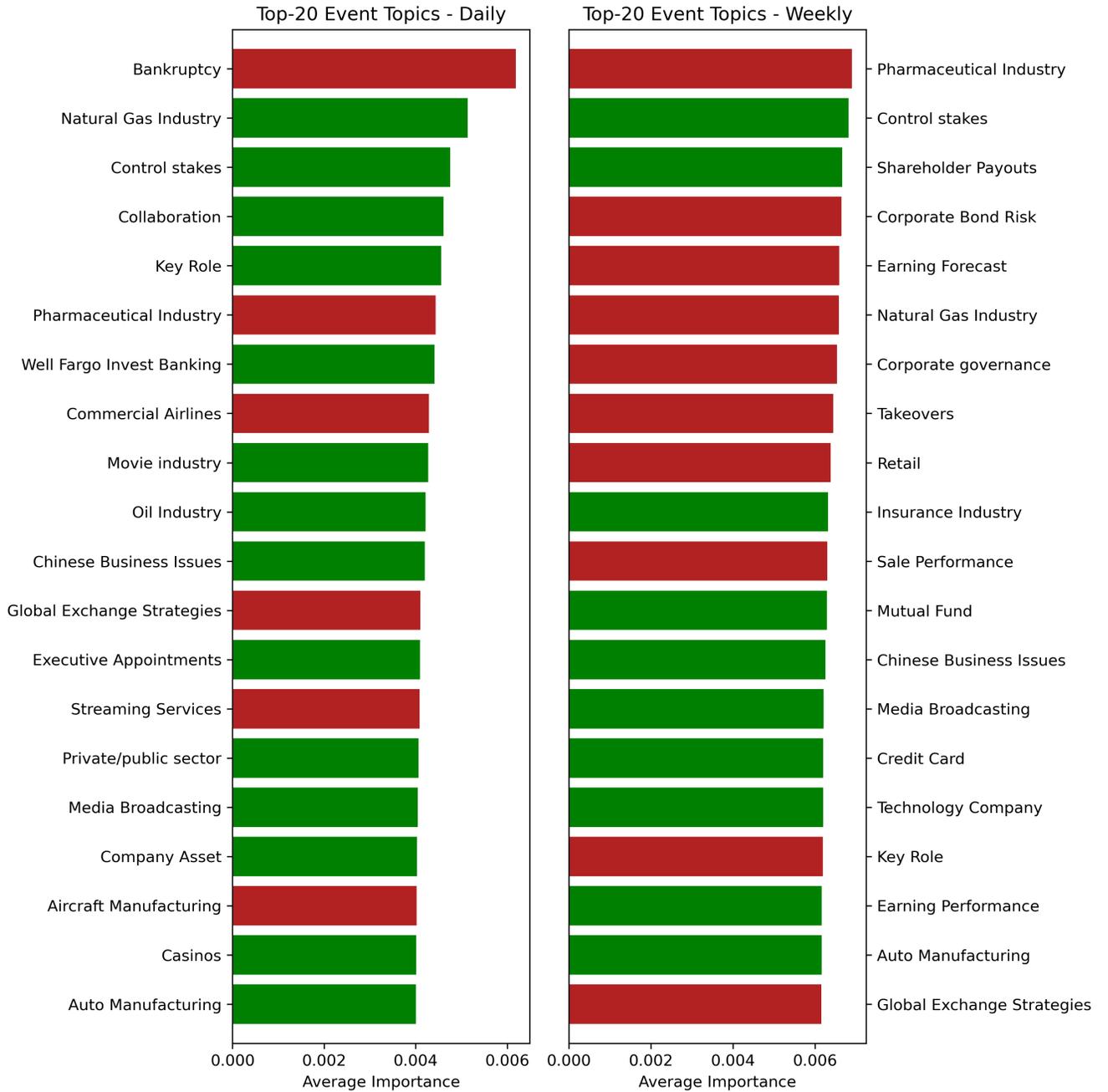

**Figure 6**
**Comparison of Top 20 Event Topics**
This figure compares the average importance scores of top 20 event topics that appear in the daily model (left panel) and weekly model (right panel). For both panels, the $y$-axis shows the list of topics sorted by their average importance scores, while the $x$-axis shows the value of the corresponding average importance scores. Each bar is colored by the dominant importance-polarity share of events within the corresponding topic: green indicates that positive events account for a larger percentage, while red indicates that negative events account for a larger percentage.

**Figure 7**
**A Sample of Entity Driven Stock Comovement (Focal Entity: Vaccine)**
This figure visualizes entity-driven comovement using "Vaccine" as the focal entity. We build an entity graph from the news-event triplets of the top-15 firms with the highest "Vaccine" exposure in our representative dataset, retaining only triplets (and nodes) within two hops of "Vaccine" for each firm. The central rectangular node is "Vaccine" (red). The outer ring shows the 15 focal firm entities (deep blue). Middle-layer nodes are other co-occurring entities in the extracted triplets (light blue). An edge indicates that the two entities co-appear as a subject–object pair in at least one extracted event triplet from the focal firms' news. This graph shows how "Vaccine" and related entities serve as a shared informational anchor that connects and influences the focal firms.

# Online Appendix for

# "Structured Event Representation and Stock Return Predictability"

This document supplements the paper "Structured Event Representation and Stock Return Predictability". It provides additional results and robustness analyses which are not displayed in the published text.

## OA.1 Constructing Dataset for Event-Driven Return Prediction

This appendix outlines the processing procedures used to construct the news–event–return datasets employed in the empirical return predicting.

### OA.1.1 News Articles Preprocessing

1. **Sample collection.** Given the news data collected primarily from ProQuest and complemented with Factiva, we begin by retaining all news articles published between January 1, 2003 and December 31, 2023.

2. **Duplicate removal.** For each publication date, we compute cosine similarity across article texts and drop an article when its similarity with an earlier article exceeds 0.9.

3. **Event extraction.** For every remaining article, we apply a pretrained large language model (LLM), the OpenAI `gpt-3.5-turbo-0125` model, using a standardized prompt to extract the structured event representations (SERs). The details of this step can be found in Section OA.2.

### OA.1.2 CRSP Return Panel Construction

1. **For daily return panel,** we download the daily stock returns for all CRSP-listed firms during the same sample window.

2. **For weekly return panel,** they are aggregated from daily return panel using the standard compounding formula over calendar weeks by the following formulation.

$$R_{\text{week}} = \exp\left(\sum_{i=1}^{5} \ln(1+r_i)\right) - 1, \tag{A.1}$$

where $r_i$ denotes the daily return on trading day $i$ within a given week, and the summation is taken over all valid trading days in that week.

### OA.1.3 Linking Articles to Firms

1. **Fuzzy matching of firm names.** We employ a text-based fuzzy matching algorithm to identify mentions of CRSP firms in each article. A firm match is retained only if its name appears at least twice.

2. **Assignment of firm identifiers.** Each article is linked to the firm generating the highest similarity score. Articles containing ambiguous or overly common expressions are excluded to reduce false matches.

3. **Merging with returns.** Articles are merged with stock returns at the daily and weekly frequencies based on publication timing:

   (1) Daily panel: articles on date $t$ are merged with daily returns on $t+1$.

   (2) Weekly panel: articles in week $t$ are merged with weekly returns in week $t+1$.

   For each stock–date observation, we collect all SERs extracted from all linked articles.

### OA.1.4 Event Processing

1. **Remove duplicated events.** Within each stock–date observation, event representations identical in the format of (*subject-action-object*) appearing across multiple articles are collapsed into a single event.

2. **Action normalization.** To reduce lexical redundancy, we lemmatize all action terms and vectorize them using a pretrained BERT model (`bert-base-uncased`). We then cluster their vector representations using an unsupervised algorithm (DBSCAN). All terms within a cluster are replaced by the cluster center.

3. **Entity indexing.** Each unique subject or object expression is assigned a unique entity identifier based on their DBpedia URLs, while each unique action expression is assigned a unique action identifier. In this way, each SER is transformed into an indexed triplet (*subject token, action token, object token*).

Finally, based on the above processing steps, we can construct two panels for the prediction tasks, with each event list containing the standardized SER token triplets associated with that stock-period.:

(1) Daily panel: (Stock, Date, Articles, Next-Date Return, Event List)

(2) Weekly panel: (Stock, Week, Date, Articles, Next-Week Return, Event List)

## OA.2 Using LLM to Extract Events from News Articles

This appendix provides additional details on the large language model (LLM)–based event extraction procedure used in our main analysis. As discussed in the main text, for each news article we employ an LLM agent to extract the key event representations. Each event is encoded as a structured subject–action–object tuple, supplemented with DBpedia links for both entities and a short context sentence that preserves the local semantic meaning of the event.

[Tables OA.1, OA.2 about here]

Table OA.1 presents the prompt templates used to instruct the LLM to identify salient events from the article and return them in a standardized JSON format. Table OA.2 provides a representative example of a news article together with the extracted events generated under this specification.

## OA.3 Supplementary Tables for Entity- and Event-Level Importance Analyses

This appendix provides supplementary tables that report the detailed results of the entity- and event-level interpretability analyses described in the main text. Tables OA.3 and OA.4 present the **top 20 entities** ranked by their absolute importance (ABS Imp.) scores for the daily and weekly prediction models, respectively.

[Tables OA.3, OA.4, OA.5, and OA.6 about here]

Tables OA.5 and OA.6 list the **top 20 event topics** identified through the LDA-based topic modeling procedure and ranked by their average normalized importance scores. For each table, we report the absolute importance of each entity or topic, the proportions of positive and negative polarity contributions (Pos. Pct. and Neg. Pct.), and representative examples of high-impact news events associated with them.

These tables provide comprehensive evidence underlying the visual analyses in the main text, offering fine-grained insight into which entities and event types most strongly influence the model's predictions across temporal horizons. Together, they complement the main results by illustrating the interpretability patterns at both the *entity* and *event* levels, and by clarifying the polarity-driven differences between short-term (daily) and longer-term (weekly) predictive mechanisms.

## OA.4 Event Topic Modeling

This section introduces the details of the event topic modeling procedure used to categorize extracted events into latent thematic groups. The purpose of this analysis is to capture broader semantic structures underlying news events and to examine how entity importance and market responses vary across different thematic domains. Each event is represented by a *context field*, which is a sentence extracted from the original news article that contains the identified event, providing rich textual information for topic clustering.

We employ a latent Dirichlet allocation (LDA) model to uncover the latent topic distributions among these event contexts. LDA is a probabilistic generative model that represents each document (in this case, an event sentence) as a mixture of topics, with each topic characterized by a distribution over words. This approach effectively captures the notion that news narratives often involve multiple intertwined themes, such as policy intervention, financial performance, or corporate behavior.

A key hyperparameter of LDA is the number of topics. To determine this systematically, we follow the model selection procedure of Bybee, Kelly, and Su (2023), which evaluates model fit across alternative topic numbers using the Bayes factor to balance explanatory power and model complexity. As shown in Figure OA.1, the Bayes factor increases steadily before stabilizing around 250 topics, which we adopt as the optimal number for event-level clustering.

Figure OA.1 about here

Table OA.7 about here

Table OA.7 presents representative examples of the resulting topic distributions. Each topic is labeled based on the dominant semantic patterns of its top-ranked keywords to enhance interpretability. These topics provide an interpretable and compact representation of the event corpus, forming the basis for subsequent analyses on how different topics of news events contribute to expected return prediction in the main text.

Table OA.1
The Prompt Templates for LLM-Extracted News Events

---

**SYSTEM PROMPT:**
You are an expert agent specialized in analyzing news texts. Your task is to identify the most salient events from a given news text as requested by the user prompt. You must generate the output in a JSON format containing a list with JSON objects having the following keys: "subject", "subject_link", "action", "object", "object_link" and "context".
- The "subject" key must contain the text of the extracted subject entity.
- The "subject_link" key must contain the DBpedia link for the subject entity.
- The "action" key must contain the verb or action that connects the subject to the object.
- The "object" key must represent the text of the entity which is the object of the action.
- The "object_link" key must contain the DBpedia link for the object entity.
- The "context" key must contain additional context or information about the event.
Ensure that both the subject and object are named entities present in DBpedia, and that both the subject and object are nouns. Extract only the most salient events from the news text.

- - - - - - - - - - - - - - - - - - - - - - - - - - - - - - - - - - - - - - - - - - - - - - - - - - - - - - - - -

**USER PROMPT:**
Based on the following example, extract salient events from the provided text.
{Examples}
For the following specification, extract salient events as in the provided example.
{News publication Date, News Article}

---

This table presents the prompt templates used for event extraction in our pipeline. The system prompt instructs the LLM to identify salient events from each news article and to return them in a structured JSON format, including the subject and object entities. Both subject and object are required to correspond to DBpedia-recognized named entities. The user prompt provides the article text and requests event extraction following the specified template.

Table OA.2
Event Extraction Example

> **INPUT:** 2017-04-20, President Trump signed an executive order on Tuesday that could lead to significant changes in the H-1B immigrant worker program that is heavily used by the technology industry.
> While tech has often clashed with Mr. Trump, at least on this issue, many executives are cautiously optimistic. Why? They are not the biggest users of H-1B visas – that distinction goes to outsourcing companies – and some hope that changes to the program will help them bring highly educated workers to the United States.
> The tempered comments from technology executives stood in contrast to their opposition to earlier orders by Mr. Trump banning travelers from a handful of predominantly Muslim countries. Immigrants have risen to the pinnacle of the industry, leading icons like Google and Microsoft, and many rank-and-file workers from abroad have become entrepreneurs behind the most successful start-ups.
> ......

**LLM OUTPUT:**

```
[
  {
    "subject": "President Trump",
    "subject_link": "http://dbpedia.org/resource/Donald_Trump",
    "action": "signed",
    "object": "executive order",
    "object_link": "http://dbpedia.org/resource/Executive_order",
    "context": "President Trump signed an executive order on Tuesday that could
    ↪  lead to significant changes in the H-1B immigrant worker program that is
    ↪  heavily used by the technology industry.",
  },
  {
    "subject": "executive order",
    "subject_link": "http://dbpedia.org/resource/Executive_order",
    "action": "could lead to significant changes in",
    "object": "H-1B immigrant worker program",
    "object_link": "http://dbpedia.org/resource/H-1B_visa",
    "context": "President Trump signed an executive order on Tuesday that could
    ↪  lead to significant changes in the H-1B immigrant worker program that is
    ↪  heavily used by the technology industry.",
  },
```

```
{
  "subject": "technology industry",
  "subject_link": "http://dbpedia.org/resource/Technology_industry",
  "action": "heavily used by",
  "object": "H-1B immigrant worker program",
  "object_link": "http://dbpedia.org/resource/H-1B_visa",
  "context": "President Trump signed an executive order on Tuesday that could
    lead to significant changes in the H-1B immigrant worker program that is
    heavily used by the technology industry.",
},
{
  "subject": "technology executives",
  "subject_link": "http://dbpedia.org/resource/Technology_executive",
  "action": "stood in contrast to",
  "object": "opposition to earlier orders",
  "object_link": "http://dbpedia.org/resource/Executive_order",
  "context": "The tempered comments from technology executives stood in
    contrast to their opposition to earlier orders by Mr. Trump banning
    travelers from predominantly Muslim countries.",
},

]
```

This table provides an illustrative example of our event-extraction procedure using the prompt template in Table OA.1. The input panel shows the original news article paragraphs and its publication date supplied to the LLM. The output panel reports the extracted events in JSON format. Each event includes the subject and object entities with their corresponding DBpedia links, the action verb, and the associated context sentence.

Table OA.3
Top-20 Important Entities for Daily Model

| Rank | Entities | ABS Imp. | Pos. Pct. | Neg. Pct. | Top Important Events |
|---|---|---|---|---|---|
| 1 | Bankruptcy | 0.486 | 6.76% | 93.24% | (Chesapeake_Energy-hire-Bankruptcy), (Chesapeake_Energy-prepare to apply for-Bankruptcy), (AMC_Theatres-warn-Bankruptcy), (Hertz_Global_Holdings-file for-Bankruptcy), (VIP_Cinema-file for-Bankruptcy), (Sabine_Oil_%26_Gas_Corporation-file for-Bankruptcy) |
| 2 | Financial Statement | 0.305 | 99.87% | 0.13% | (Chesapeake_Energy-report-Financial_statement), (VF_Corporation-sink-Financial_statement), (Google-report-Financial_statement), (The_Hertz_Corporation-identify-Financial_statement), (The_Hertz_Corporation-restate-Financial_statement), (Honeywell-report-Financial_statement) |
| 3 | Vaccine | 0.265 | 0.00% | 100.00% | (Vaccine-could weigh in-Virus), (Vaccine-arrive-Port_city), (Moderna-describe-Vaccine), (Inovio_Pharmaceuticals-say-Vaccine), (Moderna-plan on develop-Vaccine), (Inovio_Pharmaceuticals-plan on develop-Vaccine) |
| 4 | Cost | 0.239 | 100.00% | 0.00% | (University-suffer-Cost), (3D_printing-take-Cost), (United_States_dollar-raise-Cost), (Robert_Clarke-say-Cost), (Sysco-would be grateful for-Cost), (Chesapeake_Energy-cut-Cost) |
| 5 | California | 0.235 | 100.00% | 0.00% | (California-boycott-Trump_administration), (Michael_Stamer-say-California), (California-boycott-Company), (California-use-Water_conservation), (California-order-Indoor_dining), (California-charge-Sticker) |
| 6 | Bond | 0.233 | 6.02% | 93.98% | (Qi_Junwen-say-Bond), (Shandong_Ruyi-say-Bond), (AGI-believe-Bond), (European_Central_Bank-extend-Bond), (Industrial_and_Commercial_Bank_of_China-issue-Bond), (Morningstar,_Inc.-state-Bond) |
| 7 | Price | 0.228 | 99.77% | 0.23% | (Netflix-raise-Price), (Netflix-raise-Price), (Biogen-raise-Price), (Eli_Lilly_and_Company-raise-Price), (Pfizer-raise-Price), (Starbucks-announce-Price) |
| 8 | Challenge | 0.225 | 100.00% | 0.00% | (Sears-face-Challenge), (Doug_Lawler-face-Challenge), (IBM-know-Challenge), (John_Kim-say-Challenge), (William_Ackman-issue-Challenge), (Philip_Lowe-warn-Challenge) |

Table OA.3 (continued)

| Rank | Entities | ABS Imp. | Pos. Pct. | Neg. Pct. | Top Important Events |
|------|----------|----------|-----------|-----------|----------------------|
| 9 | Revenue | 0.223 | 89.00% | 11.00% | (Chesapeake_Energy-report-Revenue), (AMC_Entertainment-say-Revenue), (AMC_Entertainment-bring charge-Revenue), (Revenue-rise-Percentage), (Nvidia-post-Revenue), (AMC_Entertainment-document in-Revenue) |
| 10 | Data | 0.216 | 100.00% | 0.00% | (Freddie_Mac-release-Data), (Caterpillar_Inc.-release-Data), (CME_Group-move-Data), (Eli_Lilly_and_Company-expect-Data), (Google-break out-Data), (Fiberlink_Communications-shatter-Data) |
| 11 | Sales | 0.213 | 14.53% | 85.47% | (Lululemon_Athletica-report-Sales), (SABMiller-report-Sales), (The_Hershey_Company-revenue edge up-Sales), (Nvidia-report-Sales), (Tailored_Brands-decline-Sales), (Macy%27s-struggle with-Sales) |
| 12 | United Kingdom | 0.210 | 100.00% | 0.00% | (Stitch_Fix-say-United_Kingdom), (Auditor-have to disclose-United_Kingdom), (Starbucks-say-United_Kingdom), (European_Union_nationality_law-be seat-United_Kingdom), (Margaret_Thatcher-make-United_Kingdom), (United_Kingdom-stand up to-Nazi_Germany) |
| 13 | Washington, D.C. | 0.202 | 100.00% | 0.00% | (Washington,_D.C.-refer-Luxury_goods), (Washington,_D.C.-have be-Medicare_Advantage), (Washington,_D.C.-experience-Rain), (Eric_Rosengren-come to-Washington,_D.C.), (Jake_Tapper-acknowledge-Washington,_D.C.), (Washington,_D.C.-invest-United_States_dollar) |
| 14 | Share | 0.196 | 97.16% | 2.84% | (Baozun_Inc.-close-Share), (CFRA_Research-estimate-Share), (Tyson-fell with-Share), (Leonard_Riggio-sell-Share), (Publicis_Groupe-acquire-Share), (Publicis_Groupe-buy-Share) |
| 15 | Fine | 0.195 | 0.00% | 100.00% | (Fine-could amount to-United_States_dollar), (Commonwealth_Bank-agree to shave-Fine), (Ross_Stores-agree to shave-Fine), (Lucille_Griffin-leave out-Fine), (Consumer_Financial_Protection_Bureau-impose-Fine), (Pacific_Gas_and_Electric_Company-agree to shave-Fine) |
| 16 | Finance | 0.192 | 100.00% | 0.00% | (Virtu_Financial-seek-Finance), (Lyft-raise-Finance), (Chesapeake_Energy-raise-Finance), (Robert_Peck-estimate-Finance), (Larry_Silverstein-try to quash-Finance), (Cash_flow-turn-Finance) |

Table OA.3 (continued)

| Rank | Entities | ABS Imp. | Pos. Pct. | Neg. Pct. | Top Important Events |
|------|----------|----------|-----------|-----------|----------------------|
| 17 | Comment | 0.189 | 99.57% | 0.43% | (Email-return-Comment), (Chipotle_Mexican_Grill-be not available to-Comment), (Fox_News-decline-Comment), (Brent_Thill-say-Comment), (Mylan-decline-Comment), (Aubrey_McClendon-decline-Comment) |
| 18 | Chairman | 0.186 | 98.20% | 1.80% | (Howard_Schultz-be-Chairman), (Dov_Charney-fire-Chairman), (Janet_Yellen-nominate-Chairman), (Ginni_Rometty-remain-Chairman), (Dennis_J._Wilson-step down from-Chairman), (John_Donahoe-expect to command-Chairman) |
| 19 | City | 0.183 | 100.00% | 0.00% | (Richard_Ayoade-visit-City), (City-engage in-Competition), (Make_Music_New_York-outlive-City), (Housing_court-authorize-City), (Los_Angeles_Dodgers-buzz back at-City), (City-agree-Million) |
| 20 | Asset | 0.180 | 99.86% | 0.14% | (Chesapeake_Energy-sell-Asset), (Chesapeake_Energy-announce-Asset), (Aubrey_McClendon-leave behind-Asset), (Robert_Wood-say-Asset), (Greenland_Group-plan to add-Asset), (Chevron_Corporation-buy-Asset) |

This table lists the top-20 entities ranked by their absolute importance (ABS Imp.) scores, derived from the gradient-based attribution method, based on the representative daily prediction model. "Pos. Pct." and "Neg. Pct." denote the proportions of each entity's polarity importance that are positive or negative, respectively, which are computed as the ratio of the absolute positive (or negative) contribution to the total absolute importance. The final column ("Top Important Events") provides representative examples of events in which each entity attains the highest local attribution scores. Each event is represented in a (*subject-action-object*) triplet.

Table OA.4
Top-20 Important Entities for Weekly Model

| Rank | Entities | ABS Imp. | Pos. Pct. | Neg. Pct. | Top Important Events |
|---|---|---|---|---|---|
| 1 | Tesla, Inc. | 0.925 | 15.63% | 84.37% | (Tesla,_Inc.-announce-Debt_security), (Elon_Musk-open source-Tesla,_Inc.), (R._W._Baird-face-Tesla,_Inc.), (Tesla,_Inc.-extend-Vehicle), (New_Jersey-charge-Tesla,_Inc.), (Jefferies_Financial_Group-upgrade-Tesla,_Inc.) |
| 2 | Consumer | 0.846 | 45.97% | 54.03% | (Consumer-obtain-Service), (Kroger-look to promote-Consumer), (Consumer-make shift to-Online_shopping), (Consumer-base-Purchase_decision), (Mexico-push for-Consumer), (Consumer-pay-Debt) |
| 3 | United Airlines | 0.768 | 4.77% | 95.23% | (United_Airlines-behave-Trend), (United_Airlines-oppose-Gulf_Carriers), (United_Airlines-boycott-Gulf_Airlines), (United_Airlines-lobby-Government), (United_Airlines-access-Washington,_D.C.) |
| 4 | Royal Dutch Shell | 0.758 | 6.66% | 93.34% | (Royal_Dutch_Shell-recession-Wave_glider), (Obama_administration-conditional grant of approval-Royal_Dutch_Shell), (Royal_Dutch_Shell-seek-Drilling), (Royal_Dutch_Shell-correct-Problem_solving), (Royal_Dutch_Shell-lead-Negotiation), (Royal_Dutch_Shell-pay-Cash) |
| 5 | Retail | 0.748 | 26.04% | 73.96% | (Retail-may miss-Federal_Reserve_System), (Campbell_Soup_Company-say-Retail), (Walmart-reopen-Retail), (Dollar_General-close-Retail), (Massmart-open-Retail), (Dollar_General-introduce-Retail) |
| 6 | Mortgage Loan | 0.723 | 18.86% | 81.14% | (Frank_Nothaft-say-Mortgage_loan), (Mortgage_loan-decline-2017), (Freddie_Mac-report-Mortgage_loan), (Fannie_Mae-buy-Mortgage_loan), (Citigroup-purchase-Mortgage_loan), (Mortgage_loan-reach-Interest_rate) |

Table OA.4 (continued)

| Rank | Entities | ABS Imp. | Pos. Pct. | Neg. Pct. | Top Important Events |
|---|---|---|---|---|---|
| 7 | Letter | 0.689 | 0.70% | 99.30% | (Dean_Moore-send-Letter), (Fred_Kameny-write-Letter), (James_Welch-write-Letter), (Zoe's_Kitchen-receive-Letter), (East_Brunswick,_New_Jersey-write-Letter), (Investment_Company_Institute-send-Letter) |
| 8 | Deal | 0.684 | 98.60% | 1.40% | (Kamala_Harris-ink-Deal), (Hewlett_Packard_Enterprise-announce-Deal), (Novavax-sign-Deal), (Valeant_Pharmaceuticals-maintain-Deal), (United_States_Congress-boycott-Deal), (Shahid_Khan-express-Deal) |
| 9 | Airline | 0.679 | 65.14% | 34.86% | (Airline-update-Loyalty_program), (Airline-agree-Layoff), (Supreme_Court_of_the_United_States-pattern-Airline), (Airline-comply with-Order), (Airline-avoid-Syria), (Airline-say-Trade_policy) |
| 10 | Election | 0.665 | 2.74% | 97.26% | (Italy-hold-Election), (Italy-hold-Election), (Doug_Oberhelman-say-Election), (Omnicom_Group-say-Election), (Kathy_Carter-announce-Election), (Michel_Temer-decide-Election) |
| 11 | Microsoft | 0.652 | 92.45% | 7.55% | (Microsoft-say-Windows_10), (Microsoft-announce-Windows_10), (Microsoft-bundle-Offering), (Microsoft-announce-HoloLens), (Microsoft-curb-General_internist), (Jerry_Shen-say-Microsoft) |
| 12 | Advertising | 0.644 | 6.40% | 93.60% | (Budweiser-run-Advertising), (Advertising-feature-Stephen_Curry), (YouTube-announce-Advertising), (Pharmaceutical_industry-spend-Advertising), (Verizon_Communications-agree-Advertising), (Viacom-expect-Advertising) |
| 13 | ExxonMobil | 0.624 | 8.06% | 91.94% | (ExxonMobil-oppose-Kyoto_Protocol), (ExxonMobil-sell-Bond), (ExxonMobil-reiterate-Capital_expenditure), (ExxonMobil-replace-Oil_and_gas_reserves), (ExxonMobil-describe-Oil_and_gas_production), (ExxonMobil-receive-Reprimand) |

Table OA.4 (continued)

| Rank | Entities | ABS Imp. | Pos. Pct. | Neg. Pct. | Top Important Events |
|------|----------|----------|-----------|-----------|----------------------|
| 14 | Earnings Per Share | 0.624 | 1.98% | 98.02% | (Financial_analyst-document in-Earnings_per_share), (Campbell_Soup_Company-cut-Earnings_per_share), (Mondelez_International-report-Earnings_per_share), (UnitedHealth_Group-guide for-Earnings_per_share), (AutoZone-expect-Earnings_per_share), (Adobe_Inc.-increase-Earnings_per_share) |
| 15 | Board of Directors | 0.573 | 13.20% | 86.80% | (Ron_Doerfler-serve-Board_of_directors), (ExxonMobil-welcome-Board_of_directors), (Arconic-be not observe-Board_of_directors), (Nasdaq-propose-Board_of_directors), (ExxonMobil-concede-Board_of_directors), (Bill_Rogers-serve-Board_of_directors) |
| 16 | Saudi Arabia | 0.570 | 16.14% | 83.86% | (Saudi_Arabia-consider sell-Share), (Saudi_Arabia-strategy now that it have have-Saudi_Aramco), (Boeing-secure-Saudi_Arabia), (Saudi_Arabia-tell-OPEC), (Saudi_Arabia-soothe-Oil_price), (Lockheed_Martin-say-Saudi_Arabia) |
| 17 | Washington, D.C. | 0.558 | 44.33% | 55.67% | (Washington,_D.C.-refer-Luxury_goods), (Andy_Najar-return-Washington,_D.C.), (Washington,_D.C.-would have be freeze in-Legislature), (Washington,_D.C.-avoid-Bureaucracy), (Washington,_D.C.-have be-Medicare_Advantage), (Washington,_D.C.-be-Residential_area) |
| 18 | Person | 0.547 | 65.00% | 35.00% | (The_Sun-pay-Person), (Kuwait-sentence-Person), (Person-tell-Person), (Person-pepper-Conversation), (Person-ask-Person), (Fox_News-curb-Person) |
| 19 | Verizon Communications | 0.543 | 92.44% | 7.56% | (Verizon_Communications-provide-Disney%2B), (Verizon_Communications-pull out of talk with-Satellite_communications), (Verizon_Communications-increase-Data_allowance), (Verizon_Communications-raise-Price), (Verizon_Communications-confirm-Data_breach), (Verizon_Communications-say-Increase) |

Table OA.4 (continued)

| Rank | Entities | ABS Imp. | Pos. Pct. | Neg. Pct. | Top Important Events |
|------|----------|----------|-----------|-----------|----------------------|
| 20 | Exchange-traded Fund | 0.543 | 39.20% | 60.80% | (U.S._Securities_and_Exchange_Commission-propose-Exchange-traded_fund), (Exchange-traded_fund-require to report on-Net_asset_value), (Legg_Mason-branch out aggressively into-Exchange-traded_fund), (Deutsche_Asset_%26_Wealth_Management-say-Exchange-traded_fund), (Philip_Blancato-share-Exchange-traded_fund), (Deutsche_Bank-transfer-Exchange-traded_fund) |

This table lists the top-20 entities ranked by their absolute importance (ABS Imp.) scores, derived from the gradient-based attribution method, based on the representative weekly prediction model. "Pos. Pct." and "Neg. Pct." denote the proportions of each entity's polarity importance that are positive or negative, respectively, which are computed as the ratio of the absolute positive (or negative) contribution to the total absolute importance. The final column ("Top Important Events") provides representative examples of events in which each entity attains the highest local attribution scores. Each event is represented in a (*subject-action-object*) triplet.

**Table OA.5**
**Top-20 Important Event Topics for Daily Model**

| Rank | Topics | ABS Imp. | Pos. Pct. | Neg. Pct. | Top Important Events |
|---|---|---|---|---|---|
| 1 | Bankruptcy | 0.619 | 31.44% | 68.56% | (Greensill-file-Insolvency), (Penn_Virginia-file for-Chapter_11,_Title_11,_United_States_Code), (Ron_Bloom-move on to-Assistant_to_the_President_for_Manufacturing_Policy), (Tusk_Energy_Services-file for-Chapter_11,_Title_11,_United_States_Code), (Pacific_Gas_and_Electric_Company-win-Court_approval), (SkyMall-win-Court_approval) |
| 2 | Natural Gas Industry | 0.514 | 51.16% | 48.84% | (Chesapeake_Energy-wrestle with-Debt), (Doug_Lawler-officially take over as chief executive on-Chesapeake_Energy), (Chesapeake_Energy-raise-United_States_dollar), (Doug_Lawler-face-Challenge), (Chesapeake_Energy-eliminate-Dividend), (Oasis_Petroleum-file for-Chapter_11,_Title_11,_United_States_Code) |
| 3 | Control Stakes | 0.475 | 52.60% | 47.40% | (Shandong_Ruyi-buy-SMCP), (The_Hertz_Corporation-buy-Preferred_stock), (Canyon_Partners-purchase-Debt), (Mudrick_Capital_Management-commit-Debt_financing), (Square_Inc-raise-Financial_capital), (Grenade-be buy by-Lion_Capital_LLP) |
| 4 | Collaboration | 0.461 | 61.01% | 38.99% | (Hudson's_Bay_Company-say-Saks.com), (Saks_Fifth_Avenue-split off-Saks.com), (National_Vision-rally-S%26P_SmallCap_600), (Jan_Singer-launch-Bra), (L_Brands-prereported-Revenue), (Yng-Ru_Chen-join-Tattly) |
| 5 | Key Role | 0.456 | 52.94% | 47.06% | (Amy_Howe-face-Chief_executive_officer), (Aaron_Cooper-take over as-Chief_executive_officer), (Acies_Acquisition_Corp.-lead by-Jim_Murren), (Joseph_Bae-grant-Stock_award), (Adam_Aron-whip-Meme_stock), (Scott_Nuttall-grant-Stock_award) |

Table OA.5 (continued)

| Rank | Topics | ABS Imp. | Pos. Pct. | Neg. Pct. | Top Important Events |
|---|---|---|---|---|---|
| 6 | Pharmaceutical Industry | 0.444 | 54.49% | 45.51% | (Andrea_Fletcher-move on to-Roblox), (Synergy_Pharmaceuticals-sell-Asset), (Inovio_Pharmaceuticals-rush to develop-Vaccine), (Rigel_Pharmaceuticals-rally-Eli_Lilly_and_Company), (Bausch_Health-change-Name_change), (Linden_Care-dispense-Horizon_Pharma) |
| 7 | Investment Banking | 0.441 | 60.39% | 39.61% | (AMC_Entertainment_Holdings-say-Membership), (AMC_Entertainment-bring charge-Movie_theater), (AMC_Entertainment-disclose-Cash_burn_rate), (Spirit_AeroSystems-be simply not available-Fuselage), (AMC_Entertainment-bring charge-Movie_theater), (AMC_Entertainment-rally-Share_price) |
| 8 | Commercial Airlines | 0.429 | 45.01% | 54.99% | (Air_Berlin-file for-Insolvency), (Primera_Air-saw-Insolvency), (Monarch_Airlines-declare-Insolvency), (Sun_Country_Airlines-own-Passenger_aircraft), (Lufthansa-fly-Passenger_aircraft), (Air_Berlin-file for-Insolvency) |
| 9 | Movie Industry | 0.427 | 50.32% | 49.68% | (Marvel_Studios-move on to-Thor:_Ragnarok), (Walt_Disney_Studios-set or shift-Release_date), (AMC_Entertainment_Holdings-drive-Ticket), (Newark_Teachers_Union-decline to sign off on-Grant), (Avatar-have-Release_date), (AMC_Entertainment-close-Cinema) |
| 10 | Oil Industry | 0.422 | 55.07% | 44.93% | (National_Iranian_Oil_Company-broach-West_Karoon_field), (Chesapeake_Energy-raise-Production_outlook), (Eni-start pump-Oil), (Chesapeake_Energy-raise-Billion), (Chevron_Corporation-reach-Oil), (ExxonMobil-indefinitely postpone-Oil_and_gas_assets) |
| 11 | Chinese Business Issues | 0.420 | 52.51% | 47.49% | (Shandong_Ruyi-say-Bond), (AMC_Entertainment-acquisitions of-Cinema), (Carmike_Cinemas-reach a deal-Iron_Man_3), (Chesapeake_Energy-sell-Asset), (Square_Inc-say it would send-Afterpay), (PayPal-agree to fully acquire-Paidy) |

Table OA.5 (continued)

| Rank | Topics | ABS Imp. | Pos. Pct. | Neg. Pct. | Top Important Events |
|------|--------|----------|-----------|-----------|----------------------|
| 12 | Global Exchange Strategies | 0.410 | 50.45% | 49.55% | (Bank-move on to-Electronic_trading_platform), (AMC_Entertainment-launch-Effort), (Intercontinental_Exchange-consider alter-EBay), (WW_International-gain-Goldman_Sachs), (Justin_Trudeau-fatal beat in-Charge), (Santa_Claus-sulk off-North_Pole) |
| 13 | Executive Appointments | 0.410 | 52.65% | 47.35% | (Doug_Lawler-massage-Profit), (Doug_Lawler-join-Chesapeake_Energy), (Doug_Lawler-have improve-Financial_discipline), (Chesapeake_Energy-hire-Doug_Lawler), (Air_Berlin-later replace-Thomas_Winkelmann), (Anthony_Wood-say-Streaming_media) |
| 14 | Streaming Services | 0.409 | 52.45% | 47.55% | (Mark_Ghilarducci-rail-Telecommunications_industry), (International_Monetary_Fund-set to be decide in-Special_Drawing_Rights), (LightSquared-file for-Chapter_11,_Title_11,_United_States_Code), (Comcast-acquire-Contingent_Network_Services), (AMC_Entertainment-launch-Private_placement), (Federal_Communications_Commission-reclassify-High_speed_Internet) |
| 15 | Private/Public Sector | 0.406 | 56.71% | 43.29% | (Private_equity-have sell-Company), (John_Waldron-say-Private_investment), (Tenet_Healthcare-acquire-Aspen_Healthcare), (Michael_Dell-fatal beat in-Hedge_fund), (AMC_Entertainment-spend-Month), (Les_Edgar-buy-TVR) |
| 16 | Media Broadcasting | 0.405 | 53.73% | 46.27% | (Dish_Network-have not strike-Viacom), (Arsenio_Hall-announce-Cancellation), (ESPN-underpin-Valuation), (Warner_Bros.-postpone-Tenet), (AMC_Entertainment-announce-Agreement), (Fox_Corporation-buy-Bento_Box_Entertainment) |
| 17 | Company Asset | 0.403 | 55.30% | 44.70% | (Pfizer-broach-Merger), (Bausch_Health-serve as-Bidder), (Intelsat-say-Asset_sale), (Intelsat-explore-Asset_sale), (Beijing-hint-Share_sale), (Geraint_Thomas-be buffet by-Warren_Barguil) |

Table OA.5 (continued)

| Rank | Topics | ABS Imp. | Pos. Pct. | Neg. Pct. | Top Important Events |
|------|--------|----------|-----------|-----------|----------------------|
| 18 | Aircraft Manufacturing | 0.402 | 50.95% | 49.05% | (Nordam_Group-file for-Chapter_11,_Title_11,_United_States_Code), (SpiceJet-start import-Jet_fuel), (Boeing-move-Aircraft_manufacturing), (Airbus-underpin-Airbus_A350_XWB), (James_Taiclet-say-Statement), (Boeing-emerge from-Aircraft_manufacturer) |
| 19 | Casinos | 0.402 | 62.62% | 37.38% | (Jim_Murren-leave for-MGM_Resorts_International), (Jim_Murren-say-Baseball), (Caesars_Entertainment_Inc.-pull bid for-Gambling_license), (Arthur_Compton-shin-Light), (Jim_Murren-seize-Parking), (Genting_Group-announce-Gambling_license) |
| 20 | Auto Manufacturing | 0.401 | 54.32% | 45.68% | (Tesla,_Inc.-need to pick up-Pace), (Tesla_Motors-set to invoke-Gigafactory), (Goldman_Sachs_Asset_Management-begin disclose-Private_equity), (Huang_Weiguo-lead-Contractor), (The_Hertz_Corporation-file-Bankruptcy), (Retail-accelerate purchase-Tesla,_Inc.) |

This table lists the top 20 event topics ranked by their average normalized importance scores for the daily return prediction model. The importance scores are derived from the gradient-based attribution method and aggregated at the topic level according to the LDA-based topic modeling procedure. "ABS Imp." denotes the absolute importance of each event topic, capturing the overall magnitude of its contribution to predicted returns. "Pos. Pct." and "Neg. Pct." represent the proportions of positive and negative polarity importance, respectively, which are calculated as the absolute positive (or negative) contribution divided by the total absolute importance, indicating whether each topic is predominantly associated with return-enhancing or return-depressing effects. The final column ("Top Important Events") provides representative examples of high-impact events within each topic that exhibit the strongest local attribution scores in the daily model.

Table OA.6
Top-20 Important Event Topics for Weekly Model

| Rank | Topics | ABS Imp. | Pos. Pct. | Neg. Pct. | Top Important Events |
|------|--------|----------|-----------|-----------|----------------------|
| 1 | Pharmaceutical Industry | 0.690 | 50.08% | 49.92% | (Teva_Pharmaceutical_Industries-use-Copaxone), (Abdel_Fattah_el-Sisi-ask-Terrorism), (Kite-expect-Biologics_License_Application), (Valeant_Pharmaceuticals-reveal-Subpoena), (Osteoporosis-be-Women's_health), (Joseph_Papa-curb-Company) |
| 2 | Control Stakes | 0.681 | 53.16% | 46.84% | (Elliott_Management_Corporation-propose-Debt_refinancing), (Analog_Devices-in talk to raise-Maxim_Integrated), (Analog_Devices-acquisition of-Maxim_Integrated), (Pentair-agree to fully acquire-Erico_Global), (Signet-pay-Cash), (Remington_Outdoor_Company-acquire by-Cerberus_Capital_Management) |
| 3 | Shareholder Payouts | 0.666 | 51.47% | 48.53% | (Goldman_Sachs-want to block-Return_on_equity), (David_Solomon-tell-Shareholder), (AMC_Entertainment-explore-Fundraising), (Goldman_Sachs-reset-Return_on_equity), (Goldman_Sachs-repurchase-Preferred_stock), (RTI_International-receive-Alcoa) |
| 4 | Corporate Bond Risk | 0.665 | 45.96% | 54.04% | (Prudential_Plc-deduct-Corporate_bond), (Moody's-cut-Credit_rating), (United_States-pepper-Credit_rating), (Moody's_Investors_Service-warn-Tesla,_Inc.), (Thomson_Reuters-note-Issuance), (Moody's-higher-Credit_rating) |
| 5 | Earning Forecast | 0.659 | 51.18% | 48.82% | (Amazon-spoon-Profit), (Macy's-document in-Earnings_per_share), (SABMiller-report-Sales), (Best_Buy-increase-Profit), (Financial_analyst-forecast-Financial_forecast), (Activision-tell-Mobile_phone) |

Table OA.6 (continued)

| Rank | Topics | ABS Imp. | Pos. Pct. | Neg. Pct. | Top Important Events |
|---|---|---|---|---|---|
| 6 | Natural Gas Industry | 0.658 | 52.38% | 47.62% | (Royal_Dutch_Shell-convert-Drilling_rig), (ConocoPhillips-divest-Alternative_energy), (Canpotex-be-Distribution), (General_Electric-spend-Acquisition), (Exxon-add-Natural_gas), (Continental_Resources-plan to ban-Oil_production) |
| 7 | Corporate Governance | 0.654 | 44.63% | 55.37% | (Microsoft-recommend-Shareholder), (Board_of_Trustees-determine-Fund), (Martin_Sorrell-head of-WPP_plc), (David_Einhorn-propose-Candidate), (Muriel_Siebert-mourn the pass of-Muriel_Siebert), (MSCI_Inc.-reject-ValueAct_Capital) |
| 8 | Takeovers | 0.645 | 49.15% | 50.85% | (Pacific_Gas_and_Electric_Company-reject-Proposal), (Wrapports-disclose-Ownership), (Government_of_Australia-reject-GrainCorp), (HP_Inc.-reject-Acquisition), (Xerox-say-Board_of_directors), (MKS_Instruments-curb-Proposal) |
| 9 | Retail | 0.638 | 49.91% | 50.09% | (Walmart-immersion-Sale_of_goods), (REI-decide to stop use-Black_Friday), (Zact-sell through-Best_Buy), (LaToya_Evans-release-Statement), (Best_Buy-plan to reopen-Retail_store), (Kevin_Johnson-write-Letter) |
| 10 | Insurance Industry | 0.632 | 57.46% | 42.54% | (American_International_Group-use-Chartis), (Conor_O'Clery-say-Hello), (Johnson_Controls-hire-Centerview_Partners), (Mondelez-sell-Vegemite), (Blackstone_Group-eschew-Balance_sheet), (The_Bank_of_New_York_Mellon-hit-Invesco) |
| 11 | Sale Performance | 0.630 | 48.99% | 51.01% | (Pacific_Gas_and_Electric_Company-record-Revenue), (Tim_Touchette-deeply worry-Booking), (WPP_plc-report-Revenue), (Amazon-expect-Revenue), (Casey's_General_Stores_Inc.-report-Profit), (United_Continental_Holdings-expect-Unit_revenue) |

Table OA.6 (continued)

| Rank | Topics | ABS Imp. | Pos. Pct. | Neg. Pct. | Top Important Events |
|------|--------|----------|-----------|-----------|----------------------|
| 12 | Mutual Fund | 0.629 | 52.19% | 47.81% | (Bruce_R._Berkowitz-manage-Mutual_fund), (VanEck-take in-Asset), (Legg_Mason-hire-Executive_officer), (Financial_adviser-use-BlackRock), (The_Bank_of_New_York_Mellon-be-Custodian_bank), (Investor-put-United_States_dollar) |
| 13 | Chinese Business Issues | 0.625 | 53.45% | 46.55% | (Dalian_Wanda_Group-obey-Buyout), (Kakao-agree to buy-Daum_Communications), (Suncity_Group_Holdings-suspend-Share_trading), (Alibaba_Group-agree to buy-Stake), (Jay_Penske-purchase-Variety), (Royal_Caribbean_Cruises-agree to buy-Stake) |
| 14 | Media Broadcasting | 0.620 | 54.41% | 45.59% | (Time_Warner-negotiate-NBA_TV), (Univision-make-Digital_asset), (NBCUniversal-sell-Advertising), (Nickelodeon-sever tie with-Dan_Schneider), (David_Madden-add to-Portfolio), (AT%26T-combination of-Discovery_Inc.) |
| 15 | Credit Card | 0.620 | 56.92% | 43.08% | (PayPal-make-Loan), (United_States_Court_of_Appeals_for_the_Second_Circuit-reverse-Trial_judge), (American_Express-expand-Credit_card), (Evan_Chesler-say-Fee), (Citigroup-appoint-David_Chubak), (Tesla,_Inc.-suspend-Bitcoin) |
| 16 | Technology Company | 0.620 | 54.43% | 45.57% | (VMware-use-Product), (Amazon-build and sell-Consumer_behaviour), (Seamless-agree to collaborate-Grubhub), (Markit-boycott-CoreOne_Technologies), (Montage_International-begin raise-One_Medical), (Jonathan_Mayer-report-Tracking_technology) |
| 17 | Key Role | 0.619 | 51.30% | 48.70% | (Heinz-appoint-Executive_officer), (Raj_Subramaniam-say-Growth), (The_Hershey_Company-plan to retire in-David_Tacka), (Scott_Brickman-recruit-Andrew_Kerin), (Jorge_Gomez-be take over-Chief_financial_officer), (Diane_Tryneski-say-Search_engine) |

Table OA.6 (continued)

| Rank | Topics | ABS Imp. | Pos. Pct. | Neg. Pct. | Top Important Events |
|---|---|---|---|---|---|
| 18 | Earning Performance | 0.616 | 57.64% | 42.36% | (United_States-indicate-Murder_rate), (People-retire-Retirement), (Kohl%27s-test-Feature), (Buffalo_Wild_Wings-saw-Stock), (Kohl%27s-open-Retail), (Zillow-trade-Stock_market) |
| 19 | Auto Manufacturing | 0.616 | 59.81% | 40.19% | (General_Motors-recall-Vehicle), (Fiat_Chrysler_Automobiles-give up-Fiat_500), (BYD-become-Electric_car), (Family_name-litigation against-General_Motors), (New_Jersey-charge-Tesla,_Inc.), (Henry_Ford_Health_System-include-Medical_clinic) |
| 20 | Global Exchange Strategies | 0.615 | 47.54% | 52.46% | (Xpeng-make debut on-Hong_Kong_Stock_Exchange), (Cloudera-struggle as a list company-Offer), (Euronext-include-Stock_exchange), (Hong_Kong_Exchanges_and_Clearing-tell-Launch_ceremony), (Chinese_company-decline-Percentage), (IntercontinentalExchange-employ-Wash_trade) |

This table reports the top 20 event topics ranked by their average normalized importance scores for the weekly return prediction model. The scores are computed using the same gradient-based attribution framework as in the daily model, aggregated at the topic level according to the LDA-derived topic classifications. "ABS Imp." indicates the absolute importance of each event topic, measuring its average contribution to predicted weekly returns. "Pos. Pct." and "Neg. Pct." denote the proportions of positive and negative polarity importance, which are computed as the ratio of absolute positive (or negative) contributions to total absolute importance, reflecting whether the topic tends to have favorable or adverse return implications. The final column ("Top Important Events") lists representative examples of news events under each topic with the highest local attribution scores, illustrating how these event types influence the model's longer-horizon predictions.

Table OA.7
Topic Distributions

| Index | Topic Label | Key Terms |
|---|---|---|
| 1 | International Visits | leave, french, trip, race, france, australia, zealand, new_zealand, horse, paris, american, finally, hollande, say, pilot, president, derby, reveal, probably, win, would_leave, run, triple, visit, come |
| 2 | Performing Arts | play, theater, director, broadway, show, star, direct, production, musical, cast, film, actor, comedy, stage, garden, writer, producer, david, artistic, write, madison, revival, performance, theatre, artistic_director |
| 3 | Legal Consultancy Reports | release, firm, report, law, represent, show, law_firm, say, partner, data, llp, lawyer, consult, dead, find, accord, tuesday, privately, monday, audit, practice, report_release, federation, hold, wednesday |
| 4 | Agriculture | food, meat, olympic, chicken, gold, medal, like, make, plant, tyson, farm, beef, burger, cross, eat, product, use, olympics, farmer, ingredient, include, fry, blue, milk, mcdonald |
| 5 | Student Loan | past, past_year, loan, past_month, decade, past_decade, student, past_week, exxon, mobil, exxon_mobil, osborne, student_loan, george, george_osborne, debt, borrower, corp, couple, spend, payment, mobil_corp, least, chancellor, percent_past |
| 6 | SEC | commission, security, exchange, investigate, security_exchange, exchange_commission, sec, whether, disclose, practice, probe, company, say, trade, regulator, agency, file, federal, examine, martin, investigation, allege, report, also, charge |
| 7 | Religion Affairs | set, brown, pope, church, francis, catholic, increasingly, breast, cardinal, pope_francis, stage, roman, cancer, record, also, breast_cancer, bishop, vatican, become, highlight, visit, priest, mr_brown, set_new, van |
| 8 | Long/Short Term | long, term, take, short, place, long_term, take_place, short_term, effect, step, run, long_run, say, take_effect, long_time, report, block, late, period, toward, take_long, end, near, decade, fall |
| 9 | Investment Banking | stanley, morgan, morgan_stanley, debt, ever, row, raise, issue, birth, trillion, bond, give, american, ever_since, gorman, senior, gen, total, joseph, give_birth, cautious, rice, estimate, take, say |
| 10 | Economic Reports | report, show, department, rise, sale, data, august, february, labor, fell, january, home, consumer, say, march, index, manager, rate, july, labor_department, reuters, adjust, december, april, commerce |

Table OA.7 (continued)

| Index | Topic Label | Key Terms |
|---|---|---|
| 11 | Computers | use, access, system, gain, say, information, hacker, tool, pull, computer, control, internet, network, one, software, government, website, email, could_use, target, know, security, intend, give, company |
| 12 | Trade War | tariff, steel, import, china, american, export, good, chinese, impose, aluminum, trade, produce, make, producer, mexico, product, industry, production, plan, metal, threaten, parenthood, material, duty, trump |
| 13 | Chinese Business Issues | company, group, deal, ltd, chinese, buy, china, alibaba, acquire, hold, site, commerce, inc, sell, corp, online, base, technology, giant, e_commerce, strike, holding, reach, own, softbank |
| 14 | Policy and Program Initiatives | number, fall, reduce, say, top, priority, keep, grow, large, canadian, last_fall, fill, dimon, last, struggle, small, gap, large_number, count, head, number_people, people, grow_number, mr_dimon, include |
| 15 | Technology Company | service, company, data, customer, technology, provide, digital, software, base, use, inc, provider, platform, cloud, online, internet, help, business, center, system, say, analytics, storage, product, focus |
| 16 | Climate Change Issues | change, climate, policy, say, climate_change, make, issue, address, impact, make_change, rule, paris, say_change, include, name, effect, way, forth, global, significant, mind, propose, course, approach, call |
| 17 | Corporate Bond Risk | debt, bond, rating, risk, investor, credit, warn, moody, downgrade, default, financial, issue, company, grade, junk, service, corporate, credit_rating, investment, fame, sell, hall, government, payment, hall_fame |
| 18 | Treasury Talks | tuesday, say, treasury, say_tuesday, meet, secretary, picture, press, sony, department, treasury_department, treasury_secretary, steven, mnuchin, lew, hearing, issue, steven_mnuchin, u_treasury, office, jacob, entertainment, tuesday_would, sony_picture, tuesday_say |
| 19 | Corporate Cost Projections | wednesday, say, soon, mid, say_wednesday, low, equipment, cost, also, company, single, say_would, also_say, begin, late, expect, low_cost, digit, caterpillar, single_digit, ireland, wednesday_would, company_say, march, maker |
| 20 | Micro Chips | use, technology, system, chip, design, computer, company, program, task, allow, process, say, force, intel, develop, base, start, also, build, work, task_force, machine, create, material, software |
| 21 | Wide Range | world, wide, range, put, together, world_wide, internal, alex, wide_range, service, say, rodriguez, barclays, put_together, center, internal_revenue, revenue_service, revenue, broad, owner, barclays_center, part, include, occasion, use |

Table OA.7 (continued)

| Index | Topic Label | Key Terms |
|---|---|---|
| 22 | Company Spokes Person | comment, decline, decline_comment, respond, request, spokesman, confirm, spokeswoman, afternoon, immediately, liberty, request_comment, respond_request, detail, company, make, civil_liberty, specific, civil, decline_say, whether, representative, union, spokesman_decline, american |
| 23 | Disease Research | study, test, patient, researcher, find, disease, risk, drug, cancer, blood, use, research, treatment, health, brain, say, heart, show, cause, human, result, doctor, treat, virus, people |
| 24 | Major Concerns | put, raise, without, pressure, concern, question, ability, say, face, walker, already, place, also, worry, significant, scott, raise_concern, risk, future, raise_question, prospect, doubt, even, increase, signal |
| 25 | Trade Agreements | agreement, trade, deal, world, reach, cup, sign, world_cup, american, canada, free, negotiate, pacific, partnership, reach_agreement, north, trans, free_trade, north_american, trade_deal, mexico, trade_agreement, pact, town, final |
| 26 | Stock Market Performance | share, stock, percent, fell, rise, index, gain, market, close, point, trading, dow, jones, high, record, monday, tuesday, drop, wednesday, thursday, average, price, low, dow_jones, rally |
| 27 | Trump | trump, mr_trump, say, email, call, criticism, remark, speak, lack, president, critic, tweet, tell, describe, campaign, speech, make, trump_say, write, quickly, attack, also, response, parker, promise |
| 28 | Biographies | write, book, john, story, novel, life, author, mccain, publish, monsanto, read, memoir, farmer, writer, seed, crop, john_mccain, write_book, love, james, david, queen, history, describe, elizabeth |
| 29 | Mortgages | loan, financial, mortgage, crisis, bank, financial_crisis, lender, home, government, finance, back, housing, security, borrower, england, lending, freddie, fannie, sell, new_england, buyer, mac, side, west_side, corp |
| 30 | Carolina | spend, north, south, america, carolina, north_america, north_carolina, south_carolina, town, money, state, plan_spend, spend_year, american, mile, spend_time, small, florida, much, across, hour, plain, africa, would_spend, spend_much |
| 31 | Mining Industry | gold, mining, rio, copper, metal, miner, iron, wynn, project, commodity, ore, say, coal, glencore, operation, brazil, price, company, iron_ore, world, tinto, rio_tinto, production, bhp, tonne |
| 32 | Media Broadcasting | television, warner, network, channel, cable, time_warner, broadcast, cbs, deal, medium, inc, fox, station, little, comcast, show, include, sport, broadcaster, corp, viewer, sky, also, pay, entertainment |

Table OA.7 (continued)

| Index | Topic Label | Key Terms |
|---|---|---|
| 33 | Job Cuts | job, cut, mail, say, mcconnell, mitch, eliminate, number, king, mitch_mcconnell, e_mail, majority, leader, cut_job, martin, new_job, say_would, plan, promise, workforce, luther, job_cut, voice, leader_mitch, majority_leader |
| 34 | National Security and Foreign Threats | intelligence, attack, investigator, agency, russian, kill, use, report, terrorist, chemical, committee, weapon, say, hack, intelligence_agency, document, american, people, evidence, spy, information, chemical_weapon, russia, congressional, suspect |
| 35 | Community Engagement | another, community, almost, local, idea, commit, say, outline, eli, cambridge, set, lilly, broad, resident, crime, yet_another, brennan, work, group, violent, across, center, eli_lilly, powell, part |
| 36 | Bankruptcy | general, attorney, attorney_general, file, bankruptcy, state, protection, office, court, inspector, chapter, suit, file_bankruptcy, state_attorney, inspector_general, seek, bankruptcy_protection, eric, general_eric, sue, company, case, say, general_office, patent |
| 37 | Oil Industry | oil, price, barrel, production, crude, oil_price, energy, shell, company, cut, output, export, producer, oil_company, gas, dutch, opec, global, crude_oil, low, supply, barrel_day, brent, drill, fall |
| 38 | Basketball Culture | game, team, season, coach, play, win, player, playoff, basketball, lead, score, n_f, f_l, ranger, victory, sunday, goal, last, championship, beat, regular, washington, final, tournament, head |
| 39 | Earning Performance | end, earlier, value, earlier_year, market, end_year, earlier_month, march, period, june, share, trade, roughly, accord, year_end, earnings, market_value, lose, year_earlier, company, september, expect, stock, april, compare |
| 40 | Cultural Events | sell, july, september, begin, weekend, late, start, sharply, author, last, say, best, book, science, copy, best_sell, rapid, last_weekend, drop, scalia, fiction, record, report, last_july, begin_sell |
| 41 | Mobile device Industry | google, device, mobile, search, phone, use, system, user, app, android, software, apps, apple, microsoft, smartphone, version, operate, window, tablet, also, smartphones, feature, call, web, company |
| 42 | Taxes | tax, corporate, income, pay, rate, company, tax_rate, revenue, low, benefit, hill, income_tax, corporate_tax, capitol, cut, break, high, capitol_hill, raise, profit, government, plan, credit, reduce, return |

Table OA.7 (continued)

| Index | Topic Label | Key Terms |
|---|---|---|
| 43 | Capital Management | research, capital, firm, say, group, management, llc, development, foundation, analyst, donation, make, funding, director, research_firm, base, capital_management, accord, research_development, investment, project, lead, partner, market, manage |
| 44 | Gender and Age Demographics | old, year_old, woman, young, age, men, girl, man, son, boy, young_woman, become, daughter, find, name, old_son, black, child, female, old_man, home, mother, old_daughter, ross, month_old |
| 45 | Cash Flow | still, ago, cash, decade, year_ago, almost, flow, free, decade_ago, cash_flow, say, week_ago, month_ago, positive, generate, free_cash, almost_year, say_still, hernandez, turn, entirely, hold, several, even, receive |
| 46 | Baseball Game | run, hit, inning, game, yankee, home, pitch, season, mets, homer, field, home_run, base, start, single, lead, allow, score, pitcher, win, leave, ball, double, walk, throw |
| 47 | UN Affairs | nation, class, united, united_nation, middle, council, middle_class, world, security, resolution, security_council, u_n, call, international, substance, power, abuse, eastern, country, general, work, refugee, yoga, lead, member |
| 48 | Personal Reflections | say, life, think, something, know, tell, people, friend, like, see, really, live, ask, come, everyone, ever, anything, lot, work, want, need, always, happen, often, understand |
| 49 | Immigration | away, pass, immigration, border, immigrant, pass_away, peacefully, yorkers, new_yorkers, illegal, law, legislation, lawmaker, cross, country, age, congress, citizenship, home, mexico, family, migrant, bill, child, custom |
| 50 | Hedge Funds | fund, hedge, hedge_fund, manager, investor, money, investment, spent, raise, invest, fund_manager, pension, capital, return, bet, management, pension_fund, firm, portfolio, help, spent_year, investment_fund, run, include, fund_raise |
| 51 | Global Exchange Strategies | list, exchange, effort, stock, stock_exchange, company, newly, nyse, nasdaq, bet, trade, sport, york_stock, part, create, include, hong_kong, london, trading, buy, address, top, part_effort, plan, legal |
| 52 | Mental Health | say, much, thing, want, leave, know, never, like, love, always, really, keep, good, behind, tell, someone, think, lot, whole, mental, place, royal, actually, bad, make |
| 53 | Sports Cultural Life | series, red, world, hour, within, rock, telephone, boston, sox, world_series, yankee, red_sox, stadium, cross, interview, scott, within_hour, bull, fan, roll, flag, rick, telephone_interview, call, hatch |

Table OA.7 (continued)

| Index | Topic Label | Key Terms |
|---|---|---|
| 54 | Currency Market | dollar, currency, side, foreign, low, exchange, east, euro, yen, east_side, u_dollar, upper, strong, value, rise, fedex, foreign_exchange, hit, fell, bitcoin, effect, fall, yuan, lower, parcel |
| 55 | Timeline Recall | last, last_year, last_week, last_month, summer, say_last, last_summer, announce, minute, announce_last, begin, spring, since_last, last_minute, decade, private, december, sector, last_time, last_spring, part, late, follow, last_day, last_decade |
| 56 | Auto Manufacturing | car, ford, vehicle, motor, electric, general, tesla, general_motor, motor_co, battery, sale, say, truck, model, maker, sell, production, factory, auto, ford_motor, plan, electric_vehicle, company, plant, build |
| 57 | Space Program | administration, space, launch, drone, nasa, rocket, federal, plan, test, trump_administration, satellite, trump, aviation, use, earth, ahead, mission, commercial, federal_aviation, aviation_administration, spacex, fly, flight, station, orbit |
| 58 | Wages and Employment Conditions | employee, worker, pay, wage, hour, work, job, minimum, company, employer, lay, benefit, say, raise, cut, staff, salary, increase, minimum_wage, labor, leave, workforce, bonus, louis, offer |
| 59 | Tobacco Industry | safety, crash, board, cigarette, say, accident, system, cause, tobacco, pilot, national, plane, investigator, transportation, train, ronald, report, issue, flight, control, boeing, safety_board, involve, reagan, e_cigarette |
| 60 | Commemorations | church, hold, park, service, saturday, memorial, street, avenue, funeral, national, celebrate, center, anniversary, washington, sunday, attend, home, friday, march, honor, west, hill, manhattan, birthday, cemetery |
| 61 | News conference | news, conference, say, agency, news_conference, american, report, african, medium, african_american, say_news, release, news_medium, tell, press, defend, monday, news_release, news_agency, agency_say, speak, hold, wednesday, organization, news_corp |
| 62 | Investment Banking | group, goldman, inc, sachs, goldman_sachs, group_inc, sachs_group, banker, investment, firm, former, include, lead, solomon, work, group_say, bank, partner, financial, allen, inc_say, trading, lloyd, banking, head |
| 63 | Letter | letter, attention, park, write, draw, send, room, write_letter, central, christmas, sign, attract, central_park, send_letter, turn, gray, open, resident, call, vincent, living, tree, receive, urge, express |
| 64 | Social Innovation | create, opportunity, work, character, create_new, kind, describe, also, would_create, jeremy, call, experience, version, present, american, society, self, corbyn, mind, base, blake, world, rather, khosrowshahi, seek |

Table OA.7 (continued)

| Index | Topic Label | Key Terms |
|---|---|---|
| 65 | Economic Growth Record High | high, level, end, low, high_level, january, record, low_level, hit, high_end, rise, last, reach, june, april, say, level_since, level_year, demand, quality, information, see, last_week, decade, current |
| 66 | Secretarial and Diplomatic Actions | state, john, secretary, secretary_state, kerry, john_kerry, former, kennedy, say, department, state_john, meet, state_department, williams, john_f, washington, mile, mcdonnell, robert, governor, tillerson, president, mr_kerry, assistant, f_kennedy |
| 67 | Game Scoring | point, score, game, rebound, score_point, lead, point_rebound, season, assist, add, high, guard, james, minute, shot, knicks, win, net, anthony, finish, second, victory, pointer, career, high_point |
| 68 | Problems | problem, find, great, say, address, solve, fix, solution, help, struggle, dollar, multibillion, issue, multibillion_dollar, work, cause, even, bad, inequality, face, debate, never, acknowledge, serious, place |
| 69 | Pensions | account, able, far, say, longer, use, wing, creation, intention, possible, take, saving, work, retirement, reach, hold, without, money, go_far, twitter, would_able, right, fact, individual, even |
| 70 | Economic Growth and Recovery | economy, growth, economic, demand, market, investor, global, china, analyst, say, strong, slow, grow, amid, concern, sign, trade, remain, recovery, continue, financial, weak, expect, economist, sector |
| 71 | California | los, angeles, los_angeles, california, calif, southern, west, san, antonio, base, southern_california, san_antonio, county, city, bureau, move, protection, consumer_financial, coast, hill, home, financial_protection, protection_bureau, consumer, oakland |
| 72 | Israeli-Palestinian Conflict | israel, israeli, palestinian, netanyahu, peace, jewish, benjamin, west, gaza, fire, say, hamas, benjamin_netanyahu, cease, bank, jerusalem, west_bank, authority, mr_netanyahu, abbas, leader, jew, cease_fire, minister_benjamin, recognition |
| 73 | Commercial Airlines | airline, flight, air, airport, travel, american, passenger, delta, international, fly, say, plane, carrier, southwest, inc, united, line, pilot, american_airline, air_line, international_airport, seat, delta_air, airway, route |
| 74 | Executive Appointments | executive, chief, chief_executive, director, say, former, chairman, executive_director, executive_say, company, manage, manage_director, ceo, chairman_chief, become, former_chief, group, name, replace, john, new_chief, richard, lead, appoint, nadella |

Table OA.7 (continued)

| Index | Topic Label | Key Terms |
|---|---|---|
| 75 | Company Stakeholding | stake, own, take, percent, company, advantage, cook, tim, sell, take_advantage, percent_stake, orleans, new_orleans, business, international, control, hold, family, machine, tim_cook, ownership, apple, share, corp, international_business |
| 76 | Chinese Politics Issues | chinese, china, beijing, government, jinping, xi_jinping, party, larry, leader, president, office, communist, foreign, box, chinese_government, liu, president_xi, hong_kong, box_office, country, communist_party, wang, mr_xi, authority, hogan |
| 77 | National Statement | statement, say, say_statement, decision, issue, spokesman, important, issue_statement, release, defense, carter, step, welcome, monday, statement_say, secretary, write, jim, support, today, strongly, bnp, financial, defense_secretary, tuesday |
| 78 | Acquired Investment | america, bank, special, bank_america, counsel, lynch, robert, merrill, corp, mueller, library, merrill_lynch, general, interest, america_corp, lauren, ralph, general_counsel, branch, special_counsel, net, say, america_merrill, head, neutrality |
| 79 | Venture Capital | venture, walk, capital, away, back, dog, animal, venture_capital, turn, walk_away, renew, refuse, firm, raise, bear, cat, partner, lead, start, corner, take, capital_firm, toward, head, funding |
| 80 | Urban Real Estate | building, foot, square, build, hotel, space, home, property, square_foot, office, apartment, house, city, street, park, manhattan, floor, construction, lease, tower, room, developer, plan, project, center |
| 81 | Pandemic Vaccines and Drug Trials | vaccine, covid, pfizer, trial, say, shot, coronavirus, moderna, dos, dose, receive, biontech, effective, covid_vaccine, test, clinical, inc, booster, study, clinical_trial, people, result, health, use, variant |
| 82 | Product & Service Launches | clear, launch, make, say, early, make_clear, mark, baker, happen, highly, hughes, stephen, launch_new, guy, field, think, plan, spencer, way, seem, colbert, profitable, unlikely, service, waste |
| 83 | Bush | financial, service, george, bush, housing, financial_service, w_bush, affordable, president, george_w, jeb, affordable_housing, financial_time, unit, washington, jeb_bush, community, president_george, mr_bush, adviser, george_washington, development, former, service_committee, urban |
| 84 | Time Windows | early, morning, late, tuesday, monday, thursday, wednesday, afternoon, friday, rest, even, say, stage, march, hour, cutting, begin, early_year, action, sunday, wednesday_morning, monday_morning, june, thursday_morning, tuesday_morning |

Table OA.7 (continued)

| Index | Topic Label | Key Terms |
|---|---|---|
| 85 | Government Reports | official, say, official_say, least, people, government, kill, city, senior, local, state, u_official, government_official, top, city_official, least_people, include, administration_official, accord, administration, health_official, state_official, bomb, kill_least, meet |
| 86 | Socioeconomic Inequality | money, charge, murder, income, low, degree, count, raise, low_income, story, raise_money, tell, make, assault, use, make_money, arrest, put, degree_murder, schultz, male, attempt, family, tell_story, include |
| 87 | Commodities | try, come, determine, back, say, keep, also, discuss, prove, come_back, still, avoid, whether, possible, impact, future, persuade, help, try_get, figure, make, stop, determine_whether, try_make, prevent |
| 88 | Musk&Tesla | office, county, attorney, musk, district, say, elon, elon_musk, district_attorney, tesla, thompson, schneiderman, eric, manhattan, mr_musk, sheriff, attorney_office, brooklyn, yorker, new_yorker, eric_schneiderman, office_say, prosecutor, u_attorney, work |
| 89 | Weapon Control | gun, sure, make, super, bowl, make_sure, tough, super_bowl, control, check, say, background, rifle, carry, violence, weapon, ban, keep, gun_control, background_check, bob, firearm, owner, gun_violence, shoot |
| 90 | Strategic Initiatives and Future Investments | move, come, forward, dollar, billion, billion_dollar, step, toward, move_forward, plan, say, ahead, quickly, come_year, come_week, closer, come_month, effort, go_forward, come_day, company, put, bolster, say_move, fast |
| 91 | Cognitive Pathways | way, lot, say, beyond, find, make, look, work, find_way, pave, make_way, part, well, along, pave_way, help, transform, bed, easy, new_way, think, good, learn, give, look_way |
| 92 | White House | house, white, white_house, golden, press, state, never, say, golden_state, president, black, warrior, house_official, office, house_say, staff, secretary, pichai, state_warrior, house_press, press_secretary, sundar, sundar_pichai, tell, adviser |
| 93 | Sports Actions | back, coffee, hand, end, line, foot, onto, front, ball, defensive, bar, run, receiver, table, wide, room, eye, yard, behind, inch, toward, man, put, inside, door |
| 94 | Rail | train, association, say, program, group, station, institute, line, project, rail, transit, bus, subway, transportation, system, industry, national, policy, director, american, metro, trade, track, railroad, provide |

Table OA.7 (continued)

| Index | Topic Label | Key Terms |
|---|---|---|
| 95 | Invest Management | management, investment, head, say, client, wealth, global, manager, portfolio, wealth_management, strategy, asset, banking, team, firm, investment_management, adviser, business, division, david, financial, equity, strategist, advise, responsibility |
| 96 | Corporate Settlements | pay, agree, settlement, settle, agree_pay, fine, fee, claim, company, part, penalty, reach, deal, agreement, admit, charge, resolve, allegation, government, lawsuit, also, case, pay_settle, without, would_pay |
| 97 | Medical Care and Patient Services | medical, hospital, center, patient, health, doctor, care, aside, say, medicine, clinic, medical_center, treatment, set, set_aside, cancer, surgery, condition, emergency, physician, child, director, study, device, take |
| 98 | Japan | minister, prime, prime_minister, government, say, abe, finance, foreign, cameron, country, foreign_minister, finance_minister, david, former, parliament, electronics, david_cameron, japan, british, shinzo, shinzo_abe, charter, minister_say, meet, cabinet |
| 99 | Economic Uncertainty | question, whether, ask, answer, remain, say, many, light, affect, benefit, come, question_whether, still, unemployment, rate, trend, answer_question, continue, likely, green, wonder, response, give, economist, ask_whether |
| 100 | Casinos | different, express, casino, resort, vega, la_vega, choose, mgm, type, say, key, vegas, view, gamble, tillerson, support, las, reason, la_vegas, fight, part, las_vega, portion, however, show |
| 101 | Earnings Losses | loss, add, quarter, company, report, say, total, big, share, subscriber, november, netflix, net, suffer, expect, add_new, sell, post, estimate, say_add, customer, record, would_add, drop, period |
| 102 | Corporate Governance | board, shareholder, member, director, company, chairman, vote, investor, ceo, activist, board_member, serve, management, icahn, committee, independent, activist_investor, board_director, elliott, appoint, also, seat, proposal, proxy, replace |
| 103 | Court Rulings | court, appeal, judge, decision, order, authority, rule, circuit, say, federal, court_appeal, state, ruling, metropolitan, plan, district, housing, project, transportation, issue, development, panel, columbia, challenge, case |
| 104 | Global Banking | bank, financial, deutsche, capital, deutsche_bank, regulator, banking, big, investment, authority, barclays, investment_bank, large, risk, big_bank, lender, regulation, system, conduct, bank_say, citigroup, libor, london, u_k, institution |

Table OA.7 (continued)

| Index | Topic Label | Key Terms |
|---|---|---|
| 105 | Convertible/Preferred | good, young, davis, say, political, adviser, economic, delivery, harm, argue, cause, common, frustration, food, reputation, people, express, chance, word, blow, give, make, food_delivery, top, deliver |
| 106 | Collaboration | join, bring, work, together, brand, vision, wilson, company, back, force, say, also, work_together, rally, group, come, would_join, national, join_force, hurt, bring_back, anticipation, bring_new, bring_together, tomorrow |
| 107 | National Security Disclosures | national, government, security, say, national_security, spokesman, program, agency, support, spokesman_say, foreign, cohen, adviser, council, snowden, u_government, policy, modi, michael, surveillance, edward, organization, nonprofit, security_agency, work |
| 108 | President Election | trump, president, mr_trump, donald, donald_trump, president_trump, president_donald, elect, campaign, donald_j, trump_say, j_trump, president_elect, election, criticize, policy, tweet, elect_donald, carson, trump_campaign, call, presidential, promise, republican, meet |
| 109 | Revenue Growth | focus, growth, shift, grow, say, company, consumer, strategy, economic, brand, market, part, expand, rapidly, product, big, sale, nike, industry, international, accelerate, fast, boost, improve, area |
| 110 | Legal Disputes | case, say, evidence, document, involve, issue, argue, court, claim, legal, report, lawyer, paper, accord, review, make, find, cite, hear, show, dispute, also, rule, fraud, suggest |
| 111 | Key Role | chief, executive, chief_executive, officer, say, company, financial, chief_financial, financial_officer, operate, president, chief_operate, operate_officer, succeed, step, john, president_chief, become, leave, role, name, chairman, former, executive_officer, ceo |
| 112 | Tech Hiring | help, hire, company, tech, work, firm, big, also, recently, build, say, relatively, tech_company, bring, would_help, employee, could_help, attract, team, recruit, staff, worker, talent, start, fortune |
| 113 | Treasury Yield Fluctuations | every, note, yield, treasury, every_year, every_day, point, gold, rise, almost, basis, single, ounce, year_treasury, high, percentage, treasury_note, yield_year, fell, price, nearly, tide, percentage_point, benchmark, nearly_every |
| 114 | Couriers | amazon, inc, corp, com, amazon_com, delivery, com_inc, pandemic, coronavirus, company, service, prime, coronavirus_pandemic, include, corp_say, customer, package, warehouse, online, seller, amazon_say, fedex, ship, sell, order |

Table OA.7 (continued)

| Index | Topic Label | Key Terms |
|---|---|---|
| 115 | Industry Lavor Relation & Safety | union, worker, auto, recall, maker, bloomberg, labor, factory, michael, auto_maker, switch, strike, say, safety, plant, contract, represent, company, barra, vehicle, car, chrysler, part, ignition, mary |
| 116 | Russia | russia, russian, military, putin, ukraine, president, vladimir, moscow, mr_putin, ukrainian, force, soviet, nato, eastern, defense, gen, vladimir_putin, war, president_vladimir, western, country, government, ally, europe, say |
| 117 | US Senate | senate, vote, democrat, house, committee, republican, bill, leader, senator, majority, state, control, approve, member, chairman, congress, legislation, democratic, hold, tuesday, hearing, lawmaker, support, sen, representative |
| 118 | Survey Result | percent, accord, increase, average, survey, rise, estimate, rate, total, find, percent_year, compare, sale, rise_percent, report, data, percentage, percent_percent, american, population, almost, annual, drop, point, half |
| 119 | Soft Drink Industry | decide, whether, water, drink, island, decide_whether, beverage, staten, staten_island, consider, coke, food, pepsico, say, bottle, cola, rousseff, soda, greenberg, coca, coca_cola, drinking, president, leave, consider_whether |
| 120 | Sexual Misconduct | sexual, allegation, editor, magazine, assault, accuse, woman, report, former, abuse, newspaper, harassment, accusation, article, weinstein, sexually, publish, sexual_assault, sexual_harassment, misconduct, resign, allege, claim, employee, write |
| 121 | Shareholder Payouts | share, return, dividend, cash, stock, buy, company, pay, shareholder, say, investor, plan, buyback, capital, back, increase, raise, acquisition, also, say_would, use, debt, proceeds, buy_back, deal |
| 122 | Real Estate Development | real, johnson, estate, real_estate, johnson_johnson, property, latin, america, say, latin_america, agent, commercial, investment, mr_johnson, trust, johnson_say, real_time, estate_investment, vaccine, laden, broker, estate_agent, developer, sell, real_life |
| 123 | Trade Commission | let, deny, future, say, never, trade, bay, collins, tampa, commission, tampa_bay, trade_commission, boy, scout, maduro, terry, commodity, adam, president, manager, sign, future_trade, let_go, federal_trade, gay |
| 124 | Corporate Restructuring | plan, back, say, company, push, scale, say_plan, operation, merger, expand, part, chemical, effort, large, expansion, scale_back, go_back, also, company_plan, test, push_back, deal, manufacturing, invest, build |
| 125 | Interest Rates and Market Expectations | rate, interest, interest_rate, bond, investor, yield, raise, low, federal, reserve, high, rise, feed, increase, federal_reserve, market, fed, inflation, government, rate_increase, expect, short, term, raise_rate, treasury |

Table OA.7 (continued)

| Index | Topic Label | Key Terms |
|---|---|---|
| 126 | Accountability and Executive Actions | take, action, step, say, would_take, part, take_step, office, murdoch, control, take_part, take_action, seriously, take_office, say_take, take_control, charge, threat, job, take_year, bannon, could_take, pen, since_take, responsibility |
| 127 | Emerging Markets and Investment Risks | half, emerge, market, concern, amount, interest, potential, emerge_market, express, conflict, say, factor, half_year, jose, express_concern, due, free, conflict_interest, century, cut, san_jose, significant, year_half, half_century, edge |
| 128 | Lawsuits | lawsuit, file, claim, seek, legal, introduce, complaint, legislation, say, federal, court, accuse, graham, company, allege, suit, file_lawsuit, challenge, action, sue, without, lawyer, state, lawsuit_file, argue |
| 129 | Retail | store, retailer, online, sale, mart, wal, wal_mart, inc, chain, buy, retail, customer, walmart, shop, grocery, open, sell, shopper, say, item, brand, order, holiday, macy, shopping |
| 130 | Conference Call | call, within, say, analyst, tell, conference, investor, conference_call, company, phone, executive, earnings, earnings_call, call_mr, phone_call, chief, within_year, tuesday, tell_investor, thursday, company_call, call_analyst, say_conference, within_day, tell_analyst |
| 131 | Negative Interviews | say, interview, hard, make, even, enough, say_interview, little, difficult, work, nothing, wrong, ahead, hit, keep, find, recent, break, good, progress, line, come, expect, bit, choice |
| 132 | Market Motives | give, say, look, reason, position, like, move, opportunity, prepare, lose, would_give, away, also, buyer, look_like, potential, rather, stand, much, money, good, give_u, control, prefer, chance |
| 133 | Energy Supply | power, energy, plant, emission, fuel, carbon, coal, reduce, solar, gas, utility, electricity, nuclear, clean, use, wind, project, power_plant, renewable, company, electric, cut, grid, plan, greenhouse |
| 134 | Consumer Marketplace | product, sell, group, company, app, consumer, save, store, say, also, advocacy, brand, apple, business, include, app_store, developer, advocacy_group, noble, barnes, new_product, use, purchase, buy, remove |
| 135 | Social Media Platforms | facebook, user, video, post, content, platform, zuckerberg, twitter, app, say, youtube, mark, site, instagram, share, page, message, online, reality, company, mark_zuckerberg, use, social, virtual, account |
| 136 | Global Market & Major Players | market, large, world, global, world_large, big, company, share, become, stock, market_share, china, stock_market, country, future, investor, ever, sale, global_market, grow, world_big, capital, volatility, europe, enter |

Table OA.7 (continued)

| Index | Topic Label | Key Terms |
|---|---|---|
| 137 | Turkey | post, twitter, washington, frank, write, washington_post, erdogan, say, turkish, blog, turkey, attempt, dodd, president, dodd_frank, message, schumer, charles, bag, blog_post, tweet, recep, tayyip, recep_tayyip, tayyip_erdogan |
| 138 | Public Figures | many, history, high, profile, american, way, high_profile, find, see, like, others, many_year, say_many, bad, seem, even, say, become, include, modern, come, people, great, eager, make |
| 139 | Weather Conditions | role, play, know, weather, best, stay, temperature, also, best_know, winter, cold, like, part, character, eye, play_role, snow, heat, degree, lead, condition, keep, warm, important, also_know |
| 140 | Disease Control | wife, survive, son, daughter, grandchild, control, center, dakota, ebola, death, prevention, disease, child, mary, disease_control, center_disease, survive_wife, control_prevention, great, interest, confirm, ann, sister, west, thomas |
| 141 | Electronics | game, lose, full, full_time, win, streak, washington, play, console, sony, video, nintendo, job, video_game, end, videogame, playstation, microsoft, lose_job, xbox, title, mobile, player, big, activision |
| 142 | Bay Area Housing | san, francisco, san_francisco, area, silver, washington, base, apartment, perry, rent, spring, height, bay, move, buy, bedroom, home, house, chicago, city, live, bay_area, sell, sterling, sheldon |
| 143 | EU | european, union, european_union, commission, country, britain, europe, european_commission, france, leave, resource, member, vote, brexit, germany, u_k, italy, parliament, language, bloc, human, brussels, leader, human_resource, jean |
| 144 | Public Opinions | people, look, say, matter, familiar, accord, want, people_familiar, person, accord_people, familiar_matter, say_people, many, forward, situation, work, talk, like, employ, look_forward, say_look, chamber, people_say, many_people, young |
| 145 | Media Disclosures | public, news, offering, initial, fox, public_offering, fox_news, initial_public, company, make, go_public, network, make_public, raise, plan, relation, private, nbc, public_company, service, myers, interest, public_service, debut, anchor |
| 146 | Pharmaceutical Industry | drug, pharmaceutical, company, inc, valeant, pfizer, generic, allergan, plc, deal, patent, mylan, pharmaceutical_company, treatment, maker, merck, astrazeneca, abbvie, medicine, international, pharmacy, biotech, cancer, base, mental |

Table OA.7 (continued)

| Index | Topic Label | Key Terms |
|---|---|---|
| 147 | Positive Sentiment | say, face, like, challenge, feel, good, think, spokeswoman, seem, felt, still, important, spokeswoman_say, would_like, feel_like, concerned, really, especially, see, come, make, say_think, idea, campbell, great |
| 148 | Senior Executives | president, vice, vice_president, senior, say, executive, senior_vice, appoint, executive_vice, marketing, former, development, manager, national, operation, chairman, association, mike, adviser, vice_chairman, serve, global, senior_executive, penny, company |
| 149 | Publishing | book, property, eventually, publish, intellectual, multiple, club, publisher, intellectual_property, title, secret, author, also, kushner, sam, include, salary, new_book, american, read, publishing, print, become, land, jar |
| 150 | District Court Proceedings | judge, district, stop, lawyer, trial, court, defense, district_court, public, chuck, federal, say, health, case, u_district, witness, district_judge, public_health, federal_district, rule, panel, order, hagel, ask, testify |
| 151 | Investment Banking | well, fargo, well_fargo, holding, inc, know, say, intelligence, artificial, holding_inc, artificial_intelligence, well_know, bank, citigroup, fargo_co, customer, corp, sale, perform, account, include, unit, citigroup_inc, scandal, work_well |
| 152 | World Trade | china, pound, country, world, india, trade, south, japan, bottom, pre, big, hong_kong, line, korea, include, brazil, world_trade, south_korea, export, vietnam, bottom_line, taiwan, top, like, tax |
| 153 | Natural Disasters | damage, hurricane, storm, fire, coast, puerto, rico, flood, puerto_rico, cause, island, area, rescue, taylor, mile, home, water, resident, people, wind, wave, power, florida, hit, ship |
| 154 | Emerging Market Innovation | car, drive, electric, vehicle, general, south, self, motor, africa, general_electric, race, toyota, autonomous, self_drive, driver, technology, south_africa, african, test, uber, corp, electric_co, south_african, drive_car, company |
| 155 | Streaming Services | service, stream, network, verizon, offer, wireless, mobile, customer, video, inc, netflix, plan, communication, pay, content, launch, subscriber, carrier, cable, internet, stream_service, free, company, sprint, music |
| 156 | Federal Reserve | recent, federal, likely, reserve, federal_reserve, thursday, recent_year, say, say_thursday, bank, recent_month, government, capital, spending, recent_day, federal_government, would_likely, recent_week, reserve_bank, stress, test, china, plan, capital_spending, increase |

Table OA.7 (continued)

| Index | Topic Label | Key Terms |
|---|---|---|
| 157 | Federal Communications Commission | ask, commission, federal, agency, review, steal, rule, consider, delay, communication, say, process, vehicle, prosecutor, whether, government, also, federal_communication, block, communication_commission, independent, fcc, unclear, chairman, wheeler |
| 158 | Public Health Initiatives | need, organization, say, help, heavily, invest, provide, willing, volunteer, rely, support, FALSE, world, nonprofit, health, would_need, make, say_need, much, program, work, world_health, health_organization, community, invest_heavily |
| 159 | NY Politics | york, new_york, city, york_city, cuomo, state, york_time, andrew, gov, andrew_cuomo, mr_cuomo, york_state, gov_andrew, york_base, cuomo_say, governor, office, york_say, upstate, move, monday, york_gov, city_new, nixon, run |
| 160 | Monetary Policy | bank, central, central_bank, policy, european, euro, bond, monetary, european_central, japan, stimulus, inflation, ecb, government, economy, rate, eurozone, monetary_policy, market, program, economic, draghi, ease, money, purchase |
| 161 | Veterans Affairs | veteran, affair, retire, city, finance, stone, department, progress, kansa, correction, former, kansa_city, inmate, roll, wait, jail, current, williams, campaign, prison, veteran_affair, diamond, campaign_finance, john, pre |
| 162 | Criminal Trials | court, charge, supreme, supreme_court, federal, judge, guilty, prison, sentence, prosecutor, plead, case, plead_guilty, convict, trial, criminal, fraud, former, face, rule, count, justice, state, arrest, appeal |
| 163 | Sports Drafting | pick, draft, poor, standard, round, harris, money, santa, navy, joseph, rich, laundering, overall, money_laundering, select, better, top, word, draft_pick, david, picture, round_pick, read, standard_poor, team |
| 164 | Financial Results | cent, per, share, quarter, earlier, per_cent, year_earlier, profit, net, report, cent_share, compare, rise, period, income, loss, end, fell, revenue, net_income, share_year, per_share, earnings, post, net_profit |
| 165 | Control Stakes | deal, private, buy, equity, company, investment, firm, investor, agree, stake, group, sell, private_equity, value, capital, acquire, inc, raise, include, partner, equity_firm, acquisition, debt, base, agree_buy |
| 166 | Comparative Growth & Time Multiples | nearly, double, half, nearly_year, dozen, far, decade, nearly_decade, nearly_half, accord, nearly_percent, average, remain, triple, size, number, nearly_double, half_dozen, bell, haram, boko_haram, boko, far_year, nearly_month, anderson |

Table OA.7 (continued)

| Index | Topic Label | Key Terms |
|---|---|---|
| 167 | Governmental Health Services | see, want, say, service, human, cruise, health, say_see, also, say_want, human_service, available, health_human, line, department, ship, guard, want_see, come, department_health, coast, shutdown, could_see, carnival, government |
| 168 | Regulatory Policy | regulator, industry, rule, across, say, country, protect, company, regulation, standard, also, across_country, mean, privacy, competition, set, consumer, propose, practice, regulatory, federal, financial, issue, argue, benefit |
| 169 | Film and Literary Honors | award, prize, best, receive, win, winner, honor, tony, grand, oscar, jury, film, academy, actor, grand_jury, actress, include, nobel, nomination, star, performance, pulitzer, national, nominate, medal |
| 170 | Economic Performance | last, last_year, last_month, last_week, revenue, annual, generate, previous, late, sale, gross, late_last, product, total, domestic, compare, record, say, report, estimate, company, end, gross_domestic, domestic_product, reach |
| 171 | Culinary Culture | restaurant, wine, film, chef, make, movie, bar, dinner, like, food, cook, include, serve, star, kitchen, direct, taste, daniel, feature, menu, meal, offer, lunch, style, drink |
| 172 | Stock Prices | price, cost, cut, high, low, increase, rise, say, stock, raise, share, share_price, stock_price, company, reduce, cut_cost, high_price, estimate, low_price, fall, raise_price, saving, much, consumer, average |
| 173 | Arts | art, museum, work, artist, exhibition, gallery, show, design, paint, collection, painting, american, piece, modern, sculpture, include, contemporary, century, create, auction, sell, curator, feature, portrait, cola |
| 174 | Small Business | business, small, company, expand, small_business, core, model, sell, grow, say, part, size, group, large, big, owner, help, business_model, also, unit, run, build, spin, consumer, include |
| 175 | Future Outlook & Planning | next, next_year, expect, next_month, next_week, plan, say, tell, reporter, tell_reporter, early, say_expect, gate, early_next, decade, generation, begin, end, next_day, within, next_generation, next_decade, start, come, next_season |
| 176 | Startup and Development Programs | start, ago, year_ago, aim, found, begin, company, month_ago, start_year, twice, since_start, week_ago, jump, atlantic, say, start_new, move, city, would_start, worth, work, start_work, say_start, plan_start, company_start |
| 177 | Performing Projects | several, include, complete, work, project, ballet, several_year, company, century, mixed, also, village, several_month, recently, dancer, several_time, american, review, major, city_ballet, theater, dance, various, several_week, late |

Table OA.7 (continued)

| Index | Topic Label | Key Terms |
|---|---|---|
| 178 | Economic Openings | open, door, hope, development, morris, international, location, economic, restaurant, open_new, philip, say, economic_development, open_door, american, close, plan, plan_open, philip_morris, center, say_hope, hotel, latin, would_open, office |
| 179 | Company Founder | company, founder, executive, chief, say, chief_executive, co_founder, jeff, ceo, say_company, bezos, chairman, top, founder_chief, jeff_bezos, remain, amazon, combine, found, active, company_would, investment, chief_investment, company_say, top_executive |
| 180 | Smartmobile Launches | apple, iphone, samsung, screen, device, introduce, phone, model, ipad, iphones, apple_say, feature, watch, smartphone, app, display, use, camera, apple_inc, launch, design, inch, galaxy, also, product |
| 181 | Tennis Events | open, williams, match, round, win, title, tournament, final, double, tennis, seed, play, top, beat, tour, championship, champion, grand, serena, player, victory, digit, slam, world, serena_williams |
| 182 | Earning Forecast | quarter, earnings, report, result, expect, sale, revenue, forecast, analyst, fiscal, share, company, expectation, profit, full, growth, full_year, quarterly, estimate, fiscal_year, guidance, rise, fell, end, percent |
| 183 | FBI | investigation, federal, conduct, bureau, club, government, say, cooperate, authority, mill, golf, river, internal, general, hudson, probe, bureau_investigation, federal_bureau, lead, criminal, review, open, agency, federal_government, inquiry |
| 184 | NYC Politics | city, mayor, blasio, de_blasio, bill, council, bill_de, mr_de, mayor_bill, say, city_council, member, hall, packard, hewlett, hewlett_packard, administration, blasio_say, council_member, city_hall, plan, resident, public, campaign, commissioner |
| 185 | University Affairs | university, college, school, student, professor, study, graduate, science, degree, law, harvard, valley, silicon, silicon_valley, education, institute, columbia, program, state, campus, business, california, earn, receive, state_university |
| 186 | Asia Affairs | civil, asia, war, europe, talk, region, right, unveil, civil_right, across, civil_war, end, pacific, alliance, green, dame, walgreens, africa, notre, southeast, notre_dame, part, boot, peace, asia_pacific |
| 187 | Electoral Politics | party, election, political, vote, voter, campaign, republican, leader, conservative, democratic, candidate, support, national, opposition, presidential, poll, president, group, win, lead, liberal, government, seat, victory, politics |

Table OA.7 (continued)

| Index | Topic Label | Key Terms |
|---|---|---|
| 188 | Natural Gas Industry | gas, natural, energy, oil, natural_gas, pipeline, oil_gas, resource, corp, company, comey, james, petroleum, project, director, shale, natural_resource, inc, texas, field, station, chevron, export, well, drilling |
| 189 | National Security | security, send, message, publicly, email, say, homeland, homeland_security, text, department, trade, service, receive, company, publicly_trade, letter, harley, department_homeland, text_message, neil, force, guard, cyber, agency, davidson |
| 190 | Consumer Brands | agent, free, home, others, beer, free_agent, brand, inbev, return, pet, brewer, dog, ben, sign, ab_inbev, spirit, craft, become, busch, desire, van, owner, diageo, anheuser, among_others |
| 191 | Obama | obama, president, administration, president_obama, mr_obama, obama_administration, barack, barack_obama, trump, president_barack, trump_administration, congress, executive, order, policy, obama_say, sign, executive_order, speech, american, bill, address, action, former, clinton |
| 192 | Big Tech Giants | inc, google, yard, touchdown, alphabet, facebook, apple, company, food, whole, giant, parent, alphabet_inc, apple_inc, whole_food, tech, pass, yard_touchdown, inc_say, run, throw, advertising, facebook_inc, include, amazon |
| 193 | Fashion Cultural Life | show, wear, fashion, movie, designer, like, design, film, series, black, dress, star, collection, make, look, white, color, brand, shirt, netflix, men, style, line, create, feature |
| 194 | Insider Trading | trading, information, provide, london, insider, trade, firm, stock, share, exchange, insider_trading, cohen, detail, trader, sac, confidential, tip, activity, provide_information, irving, provide_detail, capital, martoma, mr_cohen, information_officer |
| 195 | Iraq | force, islamic, state, islamic_state, attack, group, syria, iraq, military, syrian, kill, government, american, fight, militant, rebel, fighter, city, afghanistan, army, taliban, terrorist, iraqi, troop, war |
| 196 | Corporate Announcements | announce, plan, manager, announce_plan, general, announce_would, general_manager, october, monday, easy, company, deal, tuesday, thursday, wednesday, late, make_easy, announce_new, company_announce, sign, also, make, hampshire, new_hampshire, recently |
| 197 | Contracts & Later Developments | later, contract, worth, year_later, sign, later_year, year_contract, month_later, day_later, holder, fairfax, suicide, hour, expire, date, county, evening, extend, week_later, later_month, hour_later, commit, fairfax_county, contract_worth, renew |

Table OA.7 (continued)

| Index | Topic Label | Key Terms |
|-------|-------------|-----------|
| 198 | Food Industry | continue, say, monday, supply, chain, company, say_monday, ministry, regulatory, restaurant, demand, meet, supply_chain, mcdonald, filing, would_continue, say_would, work, food, also, spokesman, operate, fast, company_say, foreign |
| 199 | State Politics | follow, state, unite, west, nation, unite_state, virginia, texas, behind, american, spot, native, indian, west_virginia, ohio, manchester, large, austin, move, intend, unite_nation, nation_large, point, place, follow_year |
| 200 | Expert Views | even, say, expert, chance, big, learn, even_though, become, ticket, expert_say, often, generation, lesson, live, give, interested, brick, keep, struggle, work, chen, many, much, suggest, idea |
| 201 | WSJ Political Coverage | street, wall, wall_street, journal, street_journal, publish, report, biden, journal_report, news, joe, review, mr_biden, corp, joe_biden, tell, paper, malloy, speak, study_publish, president_biden, interview, recently, news_corp, accord |
| 202 | Environment | social, medium, social_medium, environmental, agency, protection, environmental_protection, group, front, protection_agency, security, virginia, program, northern, medium_company, social_security, runner, service, education, use, network, center, northern_virginia, front_runner, pruitt |
| 203 | Corrections/Amplifications | home, name, article, good, news, incorrectly, bad, refer, depot, home_depot, write, give, say, page, penney, column, j_c, marijuana, edition, c_penney, use, misstate, news_article, misspell, good_news |
| 204 | Urban Mobility | long, island, truck, model, car, production, long_island, vehicle, sport, road, drive, driver, opera, mile, utility, tesla, city, bike, pickup, metropolitan, sport_utility, utility_vehicle, speed, metropolitan_opera, highway |
| 205 | Middle East Politics | saudi, country, arabia, saudi_arabia, muslim, government, arab, travel, egypt, ban, macron, president, lead, support, kingdom, emmanuel, terrorism, qatar, international, protest, emmanuel_macron, united, gulf, iran, egyptian |
| 206 | Political Engagement | recent, closely, wait, watch, say, work, recent_year, recent_week, winter, hold, compensation, romney, see, work_closely, coordinator, offensive, industry, successful, pass, mitt, mitt_romney, recent_month, mourns, federation, watched |
| 207 | Athlete Injuries | injury, hop, miss, leave, round, right, knee, season, return, back, surgery, suffer, hip, game, shoulder, leg, sustain, hand, end, practice, break, ankle, head, foot, play |

| Index | Topic Label | Key Terms |
|---|---|---|
| 208 | Human Rights Affairs | right, human, say, human_right, group, activist, advocate, spanish, deputy, freedom, n_c, director, right_group, international, issue, loeb, speech, far, record, track, thing, direction, protect, government, call |
| 209 | Environmental Conservation | balance, water, sheet, find, fish, balance_sheet, like, river, specie, use, animal, clean, lake, land, scientist, bird, sea, tree, wildlife, body, near, plant, forest, discover, almost |
| 210 | Sports Highlights | jackson, diego, final, san, warren, san_diego, star, elizabeth, conference, eastern, kelly, michael, murray, eastern_conference, elizabeth_warren, jordan, play, hole, phil, team, chicago, western, coach, hand, former |
| 211 | Movie Industry | disney, christie, fox, chris, walt, movie, jersey, studio, walt_disney, film, century, new_jersey, star, lee, chris_christie, kraft, heinz, century_fox, gov, war, mr_christie, park, entertainment, disney_co, star_war |
| 212 | Athlete Performance | morgan, j_p, p_morgan, career, season, rank, chase, finish, morgan_chase, record, brady, straight, top, tom, chase_co, quarterback, start, save, make, high, win, holiday, shot, henrik, lundqvist |
| 213 | Insurance Industry | asset, insurance, management, plc, sell, asset_management, company, group, manager, insurer, life, unit, asset_manager, business, holding, say, property, base, include, sale, investment, insurance_company, u_k, group_plc, value |
| 214 | Mutual Fund | fund, investor, venture, stock, joint, investment, etf, joint_venture, mutual, bond, blackrock, money, trade, asset, mutual_fund, index, manage, exchange, pension, manager, exchange_trade, accord, invest, vanguard, portfolio |
| 215 | Government Budgets | budget, house, cut, tax, spending, bill, government, congress, republican, speaker, state, ryan, representative, federal, propose, plan, lawmaker, increase, rep, funding, proposal, fiscal, include, paul, local |
| 216 | Shooting | police, man, kill, officer, arrest, shot, shoot, say, police_officer, death, people, fire, men, suspect, find, dead, authority, police_say, black, wound, attack, identify, gun, car, body |
| 217 | Disease Impact | die, home, age, cause, bear, cancer, heart, percentage, hospital, point, april, march, july, percentage_point, edward, june, death, manhattan, whose, complication, retire, oct, washington, james, sunday |
| 218 | Aircraft Manufacturing | boeing, jet, aircraft, plane, order, airbus, max, engine, say, production, air, system, commercial, boeing_co, ground, aerospace, airline, maker, technology, delivery, part, defense, deliver, fleet, fly |
| 219 | Sale Performance | sale, quarter, revenue, profit, expect, growth, decline, say, company, report, rise, margin, increase, strong, record, post, operating, grow, high, boost, business, operate, drop, gain, volume |

Table OA.7 (continued)

| Index | Topic Label | Key Terms |
|---|---|---|
| 220 | Implementation Goals | goal, minute, score, receive, approval, win, bid, second, period, final, set, tie, give, kick, achieve, field, regulatory, score_goal, make, minute_second, lead, regulatory_approval, leave, game, field_goal |
| 221 | Nancy Pelosi | begin, become, best, option, bring, say, camp, training, work, nancy, team, back, pelosi, life, known, upon, leader, nancy_pelosi, late, spring, would_become, recently, exercise, part, also |
| 222 | Germany | woman, germany, german, men, gay, marriage, sex, merkel, angela, patent, couple, chancellor, angela_merkel, right, group, france, christian, religious, support, ban, men_woman, sex_marriage, berlin, chancellor_angela, m_merkel |
| 223 | Credit Card | credit, card, payment, bank, credit_card, suisse, credit_suisse, consumer, visa, fee, customer, account, transaction, use, merchant, mastercard, swiss, charge, pay, accept, american, say, ubs, purchase, line |
| 224 | Sports Events | team, league, player, major, football, national, sport, game, baseball, soccer, coach, play, club, fan, season, champion, major_league, american, college, premier, field, basketball, nfl, former, professional |
| 225 | Takeovers | offer, say, company, believe, share, reject, value, takeover, bid, approach, proposal, deal, yesterday, base, say_believe, rival, pursue, make, accept, propose, company_say, say_yesterday, group, shareholder, potential |
| 226 | TV & Comedy Programming | night, show, host, sunday, late, saturday, monday, friday, chase, thursday, behind, wednesday, jpmorgan, jpmorgan_chase, monday_night, tuesday, live, schedule, tuesday_night, comedy, episode, wednesday_night, saturday_night, stand, talk |
| 227 | Executive Transitions | say, smith, david, steve, really, director, clark, person, relationship, robert, jamie, paul, kevin, work, peter, mr_smith, daniel, brian, jonathan, head, smith_say, job, suspend, dimon, resign |
| 228 | Justice Department | department, justice, justice_department, state, state_department, antitrust, say, criminal, concern, defense, case, department_say, review, quality, write, investigation, transportation, department_justice, defense_department, law, kennedy, system, official, federal, hold |
| 229 | Texas | around, world, turn, around_world, cruz, campaign, ted, radio, country, ted_cruz, around_country, turn_around, texas, mr_cruz, say, around_time, globe, begin, city, around_globe, build, run, station, around_p, sen_ted |
| 230 | Private/Public Sector | partner, develop, private, firm, company, engineer, equity, say, sector, fixed, help, income, manage, base, vast, majority, lead, develop_new, private_equity, fixed_income, software, private_sector, technology, business, focus |

Table OA.7 (continued)

| Index | Topic Label | Key Terms |
|---|---|---|
| 231 | Modern Music Industry | music, song, album, band, record, middle, perform, release, singer, play, east, middle_east, jazz, pop, rock, sound, sing, include, like, feature, dance, star, show, musician, concert |
| 232 | Presidential Campaigns | clinton, campaign, republican, democratic, candidate, presidential, sander, senator, primary, democrat, state, governor, hillary, run, hillary_clinton, former, race, sen, mrs, election, mrs_clinton, trump, bernie, paul, voter |
| 233 | School Affairs | school, student, high, high_school, teacher, education, child, public, district, parent, public_school, test, class, college, charter, teach, program, grade, attend, elementary, system, girl, alone, graduate, school_district |
| 234 | Software Industry | close, microsoft, cloud, compute, corp, reopen, amazon, deal, wood, oracle, cloud_compute, service, say, microsoft_corp, software, business, without, expect, operate, office, company, web, tiger, window, tiger_wood |
| 235 | Pharmaceutical Regulation | drug, food, prince, administration, approve, county, food_drug, drug_administration, use, fda, approval, george, prince_george, say, william, prescription, treatment, aid, opioid, prince_william, bashar, gilead, al_assad, george_county, bashar_al |
| 236 | Data Privacy | data, authority, information, company, collect, personal, breach, say, privacy, consumer, use, protection, competition, disclose, customer, fail, fine, sue, user, online, identity, give, antitrust, google, gather |
| 237 | Health Insurance | health, care, health_care, insurance, act, bill, state, plan, law, republican, affordable, insurer, coverage, senator, care_act, affordable_care, rubio, exchange, health_insurance, sen, democrat, federal, medicaid, sponsor, repeal |
| 238 | Monetary Policy Committees | member, jersey, new_jersey, committee, staff, yellen, port, say, feed, authority, janet, chairman, chairwoman, chair, congress, governor, staff_member, policy, port_authority, janet_yellen, tell, fed, powell, former, york_new |
| 239 | North Korea | north, korea, south, north_korea, korean, kim, south_korea, missile, south_korean, american, north_korean, nuclear, jong, test, kim_jong, leader, jong_un, pyongyang, north_american, military, mr_kim, seoul, say, weapon, launch |
| 240 | Music Performance | music, festival, clinton, dance, perform, work, performance, present, hillary, concert, orchestra, opera, hall, film, hillary_clinton, play, theater, program, premiere, composer, symphony, center, feature, season, piece |
| 241 | Ride-Hailing Services | law, uber, driver, enforcement, law_enforcement, state, ride, buffett, say, company, lyft, berkshire, california, violate, hail, federal, inc, hathaway, warren, technology, rule, service, ride_hail, require, license |

Table OA.7 (continued)

| Index | Topic Label | Key Terms |
|---|---|---|
| 242 | State-Owned Finance | state, united, united_state, million, dollar, million_dollar, hundred, own, state_own, hundred_million, american, country, government, ten, ten_million, postal, state_would, service, across, state_government, washington, include, california, postal_service, canada |
| 243 | Police/Crime | police, fire, officer, department, say, police_department, incident, county, report, crime, police_officer, miller, force, chief, commissioner, arrest, protester, york_police, detective, use, nixon, city, police_say, firefighter, call |
| 244 | Veteran Service | less, thousand, war, world, serve, army, world_war, war_ii, hundred, ten, ten_thousand, force, hundred_thousand, navy, less_year, become, marine, expensive, air, veteran, people, far, american, less_month, likely |
| 245 | Decision Making | make, decision, say, sense, would_make, also, available, make_decision, say_make, company_make, use, make_sense, also_make, mistake, difference, could_make, view, effort, much, common, make_big, decision_make, product, point, commitment |
| 246 | Company Asset | meeting, annual, session, say, sale, explore, hold, discuss, auction, trust, plan, jeff, strategic, simon, attend, sotheby, annual_meeting, schedule, shareholder, realty, process, possibility, jeff_session, seller, potential |
| 247 | Family Roles | family, child, mother, father, husband, parent, love, wife, young, son, friend, brother, daughter, sister, bear, life, live, survive, death, member, adult, age, work, beloved, leave |
| 248 | Fee Regulations | require, allow, rule, measure, limit, leader, say, propose, government, would_allow, would_require, secure, also, deadline, fee, fail, promise, opposition, proposal, state, plan, within, include, requirement, force |
| 249 | Nuclear Affairs | iran, nuclear, sanction, deal, impose, iranian, agreement, theresa, foreign, lift, talk, theresa_may, international, weapon, russia, program, ban, restriction, country, tehran, gamble, american, negotiation, power, procter |
| 250 | Beach Life | beach, ready, tour, sit, front, finally, stand, window, palm, bar, faith, murphy, come, susan, fla, home, collins, palm_beach, guide, even, good, hand, look, france, miami |

This table reports representative examples from the 250 latent topics generated by the latent Dirichlet allocation (LDA) model applied to the event-level context sentences. Each topic is characterized by its most probable key terms, which collectively describe the underlying thematic structure of the event corpus. The topic labels are manually assigned based on the dominant semantic patterns of these terms to enhance interpretability.

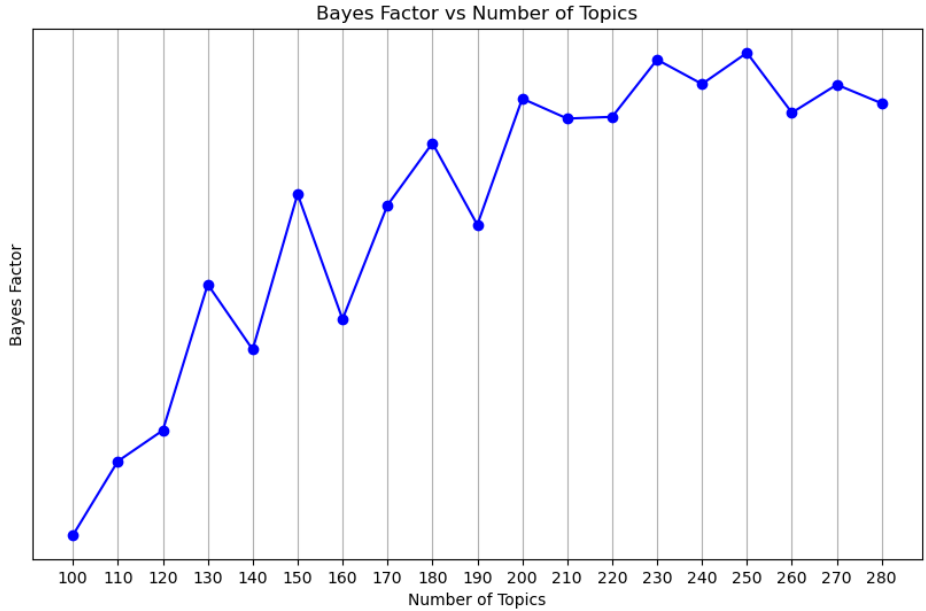

**Figure OA.1**
**Topic Volume Optimization for Event Topic Modeling**
This figure illustrates the optimization procedure for determining the number of topics in the event-level latent Dirichlet allocation (LDA) model. Each event is represented by its context sentence extracted through the LLM-based preprocessing pipeline, and LDA is applied to cluster these event contexts into coherent thematic topics. The Bayes factor, following the model selection procedure of Bybee, Kelly, and Su (2023), measures the relative model fit across different topic counts while penalizing overfitting. The figure plots the Bayes factor values for topic numbers ranging from 100 to 280, showing that model fit improves substantially up to around 250 topics and stabilizes thereafter. Accordingly, we select 250 topics as the optimal volume for event topic modeling.